\documentclass[fleqn,usenatbib]{mnras}

\usepackage{newtxtext,newtxmath}
\usepackage{pdflscape}

\usepackage[T1]{fontenc}
\usepackage{ae,aecompl}

\usepackage{graphicx}    
\usepackage{amsmath}    
\usepackage{amssymb}    
\usepackage{subfig} 

\title[Radio polarimetric study of Seyfert galaxies vs starburst galaxies]{A Radio Polarimetric Study to Disentangle AGN Activity and Star-Formation in Seyfert Galaxies}

\author[Sebastian et al.]{
Biny Sebastian,$^{1}$\thanks{E-mail: biny@ncra.tifr.res.in}
P. Kharb,$^{1}$
C. P. O' Dea,$^{2}$
J. F. Gallimore,$^{3}$
and S. A. Baum$^{2}$
\\
$^{1}$National Centre for Radio Astrophysics (NCRA) - Tata Institute of Fundamental Research (TIFR), S. P. Pune University Campus,\\ Post Bag 3, Ganeshkhind, Pune 411007, India\\
$^{2}$Physics and Astronomy, University of Manitoba, Winnipeg, Canada \\
$^{3}$Department of Physics and Astronomy, Bucknell University, Lewisburg, PA 17837, USA \\}

\pubyear{2020}

\begin{document}
\label{firstpage}
\pagerange{\pageref{firstpage}--\pageref{lastpage}}
\maketitle

\begin{abstract}
To understand the origin of radio emission in radio-quiet AGN and differentiate between the contributions from star formation, AGN accretion, and jets, we have observed a nearby sample of Seyfert galaxies along with a comparison sample of starburst galaxies using the EVLA in full-polarization mode in the B-array configuration. The radio morphologies of the Seyfert galaxies show lobe/bubble-like features or prominent cores in radio emission whereas the starburst galaxies show radio emission spatially coincident with the star-forming regions seen in optical images. There is tentative evidence that Seyferts tend to show more polarized structures than starburst galaxies at the resolution of our observations. We find that unlike a sample of Seyfert galaxies hosting kilo-parsec scale radio (KSR) emission, starburst galaxies with superwinds do not show radio-excess compared to the radio-FIR correlation. This suggests that shock acceleration is not adequate to explain the excess radio emission seen in Seyferts and hence most likely have a jet-related origin. We also find that the [O~{\small III}] luminosity of the Seyferts is correlated with the off-nuclear radio emission from the lobes, whereas it is not well correlated with the total emission which also includes the core. This suggests strong jet-medium interaction, which in turn limits the jet/lobe extents in Seyferts. We find that the power contribution of AGN jet, AGN accretion, and star formation is more or less comparable in our sample of Seyfert galaxies. We also find indications of episodic AGN activity in many of our Seyfert galaxies.
\end{abstract}

\begin{keywords}
galaxies: Seyfert --- galaxies: jets --- galaxies: sample --- radio continuum: galaxies --- galaxies: active
\end{keywords}


\section{Introduction}

 It is known that only a fraction ($\sim10\%$) of active galaxies host powerful radio jets which in some cases extend up to several Mpcs in size \citep{begelman1984}. The question of why some active galaxies host powerful jets and others not, is yet to be settled. While the origin of radio emission in radio-loud active galactic nuclei (AGN) with the radio-loudness parameter, R$\gg$10, \citep[R being defined as the ratio of the radio flux density at 5~GHz to the optical nuclear flux density in the B-band;][]{Kellermann89} is decidedly attributed to jets powered by the central engine, this is not true for radio-quiet AGN. Although many studies have contested this RL/RQ bimodality and have argued in favour of a continuous  distribution \citep{lacy2001,wals2005,white2007,bonchi2013}, several studies have suggested mechanisms other than jets to explain the radio emission seen in radio-quiet AGN. Several possibilities that can lead to radio emission in radio-quiet AGN that have been discussed in the literature include star formation-driven superwinds, winds driven by the AGN, the coronal activity of the accretion disk, or jets which are intrinsically low powered (see \cite{panessa2019} and references therein).

Radio synchrotron emission seen in starburst galaxies owes its origin to supernovae and stellar winds which give rise to the relativistic electron population. This electron population combined with the large-scale magnetic fields that thread the galaxy gives rise to the observed radio synchrotron emission in these systems. This emission is diffuse and present at the scales of the galaxy and beyond. The polarized emission from these radio halos of starburst galaxies often shows X-shaped morphology \citep{beck2015,krause2020}.

Ultrafast outflows observed in X-rays \citep{tombesi2010,pounds2014} and broad absorption lines in several quasars \citep{everett2007} are thought to be signatures of AGN driven winds. The supersonic velocities of these radiatively/thermally driven winds result in shock acceleration of particles which in turn leads to synchrotron emission \citep{zakamska2014,hwang2018,nims2015}. Yet another mechanism proposed for the origin of the radio emission is the free-free emission mechanism from the optically thin plasma in the wind \citep{Blustin09}. The spectral slope of the radio emission is expected to be $\alpha=-0.1$  (using $S_\nu\propto\nu^\alpha$) in this case. However, in most of the Seyfert galaxies, we find steeper spectral indices in the larger scale emission diminishing the importance of this mechanism.
Magnetohydrodynamically launched winds are an alternative to the radiation-driven winds and have been proposed to explain outflows observed in the X-ray \citep{fukumura2015}. It has also been proposed in the literature that the radio emission is coronal in origin as suggested by the correlation between the radio and X-ray luminosity observed in these systems \citep{panessa2007,Laor08}. However, such coronal emission  might be confined to the core region with the kpc-scale radio emission being unrelated to it. More work needs to be done in this context.

Seyfert galaxies are usually classified as radio-quiet in nature. Yet, when nuclear optical emission is properly accounted for, either by using high-resolution optical imaging from the {Hubble Space Telescope} (HST) or by removing galactic contribution, $>50\%$ of Seyfert galaxies wind up as ``radio-loud'' (RL) sources with $R>10$ \citep{HoPeng01,Kharb14a}. High angular resolution studies like those by \cite{Ulvestad81,Kukula95,Thean00,Thean01} show compact nuclear structures and sometimes evidence for multiple components, suggesting the presence of low-power jets. Moreover, kiloparsec-scale radio structures (KSR) of extents $\sim1-10$ kpc, have been detected in a large fraction of them in complete samples \citep{Baum93,colbert1996b,gallimore06}.

However, the debate on the origin of KSRs is yet to be settled. While studies like \cite{Baum93} suggested KSRs are starburst-wind-driven, later studies have favored an AGN-jet-driven origin \citep[e.g.,][]{colbert1996b}. The observed misalignment between the KSRs and their host galaxy minor axes \citep[e.g.,][]{gallimore06}; the presence of relic lobe emission around KSRs \citep[e.g.,][]{kharb2016}; and the presence of parsec-scale radio jets leading into lobes \citep[e.g.,][]{Kharb14b}, favor an AGN-jet-driven scenario. 

\cite{gallimore06} suggest that a low-power jet that starts from the central engine gets disrupted as soon as it interacts with the dense ISM of these Seyfert galaxies and loses its stability after which the radio plasma simply follows the steepest path in pressure gradient, which is along the minor axis of the host galaxy. This model can also explain the misalignment of the jet with the large scale lobe alignment, typically seen in these galaxies \citep{Schmitt01}. 

In this paper, we try to obtain more insights into the origin, nature, and evolution of the radio emission from Seyfert galaxies using a comparative study between a sample of Seyfert galaxies versus that of starburst galaxies. In the previous papers in this series, we have presented detailed multi-frequency studies of two of the Seyfert galaxies from our sample, viz., NGC\,2639 \citep{sebastian2019b} and NGC\,3079 \citep{sebastian2019}. 
Our primary aim is to try and understand how the linear polarization properties vary between the population of Seyfert galaxies and starburst galaxies. Previous studies of starburst galaxies reveal magnetic fields that follow the spiral arms of the galaxy or the minor axis and is generated as a result of the dynamo process at play in the galaxy \citep{beck2015}. Hence, in the case for a purely starburst origin of radio emission, the magnetic field alignment should be along the galaxy minor axes and the ordering scales similar to sizes comparable to that of the galaxy itself. On the other hand, if the radio emission has a jet origin, the magnetic field orientation need not be along the disk or the galaxy minor axes, and the magnetic field ordering scales can be either in collimated or diffuse outflows. One would also expect to see systematic differences between a Seyfert galaxy sample versus a starburst galaxy sample in terms of fractional polarization and rotation measure features.

This paper is organized in the following manner. In Section~\ref{sampsec}, we have described our sample selection criteria after which we explain the details of our observations and the data analysis procedures in Section~\ref{obssec}. In Section~\ref{resultsec} we elucidate the radio morphology of individual target sources in the context of the existing literature and the polarization properties for the entire sample. We then discuss the implications of our results in Section~\ref{discsec}. We have assumed $H_0$=73~km~s$^{-1}$Mpc$^{-1}$, $\Omega_{m}=0.27$ and $\Omega_{vac}=0.73$ in this paper. Spectral index $\alpha$ is defined such that flux density at frequency $\nu$ is $S_\nu\propto\nu^\alpha$.

\section{The Sample}
\label{sampsec}
We chose a sample of Seyfert galaxies along with a comparison sample of starburst galaxies. The Seyfert galaxies were chosen from the sample of nearby AGNs from the Center for Astrophysics (CfA; \cite{huchra1992}) and extended 12 $\mu$m \citep{rush1993} surveys. Our sample was selected based on the following criteria: (i) all the sources had redshifts below 0.017 so that the angular resolutions correspond to similar physical extents for various sources, (ii) sources were restricted to declinations above $-20^{\degr}$ for EVLA visibility, and (iii) had total lobe to lobe extents derived from \cite{gallimore06} greater than 20$\arcmin$ to be able to choose sources which showed reliable KSR emission spanning several resolution elements and also to be able to undertake a detailed analysis of the morphology of these KSRs.

The starburst galaxies were selected from the samples of \cite{dahlem1998} and \cite{colbert1996b}. We chose five of the seven ``edge-on'' starburst galaxies from \cite{dahlem1998} that showed clear radio emission above and below their galactic disks in arcsecond-scale radio images. These are NGC\,253, NGC\,3079, NGC\,3628, NGC\,4631, and NGC\,4666{\footnote{NGC\,4666 and NGC\,4594  were finally not observed by the EVLA because they were low in the scheduling queue. However, these sources have recently been observed by us at 10~GHz. These results will be presented in a forthcoming paper.}}. Interestingly, three of these, viz., NGC\,253, NGC\,3628 and NGC\,4666, have been classified as LINERs

\begin{landscape}
\begin{table}
\begin{center}
\caption{Observation log and observed parameters.}
\begin{tabular}{ccccccccccc} \hline \hline
%
Source & R.A.&Dec. &Observation &Redshift& Frequency& VLA Array & Beam, PA &Major axis of  & on-source & r.m.s.\\
&  &&Date& &(GHz) & Configuration& (arcsec$^2$, $\degr$)& the beam in kpc & time (in min)&(in $\mu$Jy/beam) \\
\hline
{\bf Seyfert galaxies} & & & \\ \hline

NGC2639    & 08h43m38.10s & $+$50d12m19.99s & 04$-$Nov$-$2017 & 0.0111& 5.5    &B      &     1.1$\times$1.0,    -5.73  &0.24  & 26.75 & 8.829\\
NGC2992    & 09h45m42.00s & $-$14d19m34.99s & 30$-$Jan$-$2018 & 0.0077& 5.5     &BnA   &     1.5$\times$0.9,    -17.01 &0.228 & 29.1  & 10.59\\
NGC3079    & 10h01m57.79s & $+$55d40m47.00s & 16$-$Feb$-$2018 & 0.0037& 5.5     &BnA   &     1.1$\times$0.5,    -87.08 &0.081 & 28.8  & 25.57\\    
NGC3516    & 11h06m47.49s & $+$72d34m07.00s & 05$-$Feb$-$2018 & 0.0088& 5.5     &BnA   &     1.2$\times$0.5,    -73.74 &0.208 & 26.7  & 13.09\\
NGC4051    & 12h03m09.60s & $+$44d31m53.00s & 24$-$Feb$-$2018 & 0.0023& 5.5     &BnA->A&     1.7$\times$0.9,    -78.42 &0.078 & 25.55 & 14.36\\
NGC4235    & 12h17m09.90s & $+$07d11m29.99s & 27$-$Feb$-$2018 & 0.0080& 5.5     &BnA->A&     0.7$\times$0.3,    -59.1  &0.11  & 13.8  & 21.35\\
NGC4388    & 12h25m46.70s & $+$12d39m44.00s & 31$-$Jan$-$2018 & 0.0084& 5.5     &B->BnA&     2.2$\times$1.5,    39.17  &0.248 & 26.6  & 45.08\\
NGC4593    & 12h39m39.40s & $-$05d20m39.00s & 13$-$Feb$-$2018 & 0.0090& 5.5     &BnA   &     2.7$\times$0.5,    -60.27 &0.479 & 26.7  & 33.42\\
NGC5506    & 14h13m14.90s & $-$03d12m27.00s & 27$-$Feb$-$2018 & 0.0062& 5.5     &BnA->A&     1.2$\times$0.6,    -58.8  &0.147 & 26.55 & 134.43\\    
\hline
{\bf starburst galaxies} & & & \\ \hline
NGC1134     & 02h53m41.29s & $+$13d00m50.99s & 03$-$Nov$-$2017 & 0.0121& 1.5 &B&           6.2$\times$4.1,    -58.3  &1.472 & 25.55 & 24.61\\
NGC253     & 00h47m33.10s & $-$25d17m18.00s & 31$-$Oct$-$2017 & 0.0008& 1.5 &B&           9.6$\times$3.2,    -19.36 &0.153  & 29.2  & 205.06\\
NGC3044     & 09h53m40.90s & $+$01d34m46.99s & 26$-$Nov$-$2017 & 0.0043& 5.5 &B&           3.9$\times$1.6,    -55.67 &0.341 & 29.5  & 30.076\\
NGC3628     & 11h20m17.00s & $+$13d35m23.00s & 27$-$Feb$-$2018 & 0.0028& 5.5 &BnA->A&      0.5$\times$0.3,    -68.7  &0.028 & 14.25 & 9.601\\
NGC4631     & 12h42m08.00s & $+$32d32m29.00s & 12$-$Jan$-$2018 & 0.0020& 5.5 &B&           1.5$\times$1.1,    -80.36 &0.06  & 25.55 & 9.659\\
NGC7541     & 23h14m43.90s & $+$04d32m04.00s & 25$-$Sep$-$2017 & 0.0090& 1.5 &B&           4.4$\times$3.9,    -57.33 &0.78  & 26.65 & 39.68\\
UGC903     & 01h21m47.80s & $+$17d35m33.00s & 04$-$Nov$-$2017 & 0.0084& 1.5 &B&           8.3$\times$4.0,     63.79 &1.374  & 24.6  & 22.2\\
\hline
\label{tab1}
\end{tabular}
\end{center}
\end{table}
\end{landscape}

\begin{figure*}
\includegraphics[height=11.0cm,trim = 30 100 20 60  ]{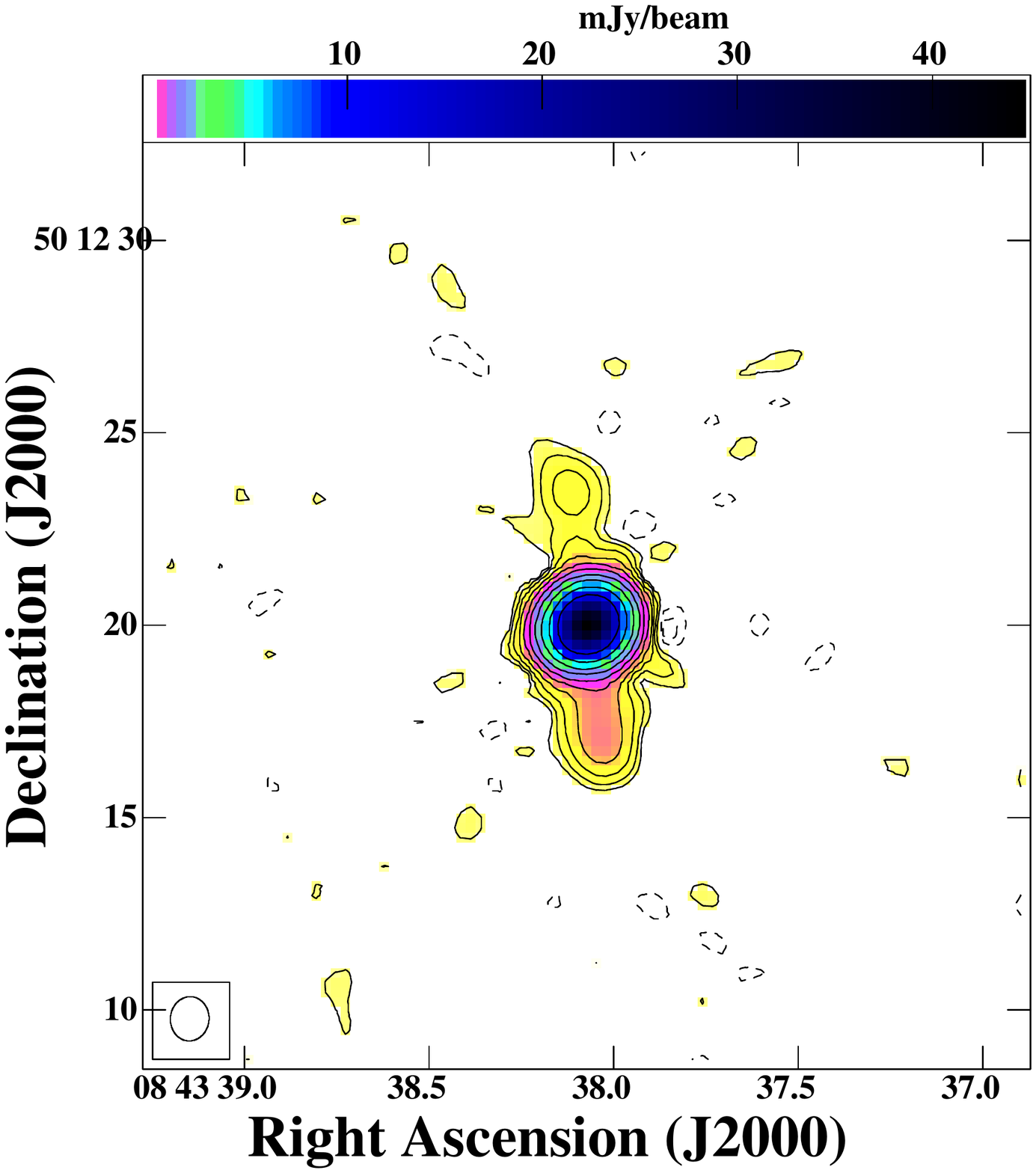}
\includegraphics[height=10.5cm,trim = 40 110 100 100 ]{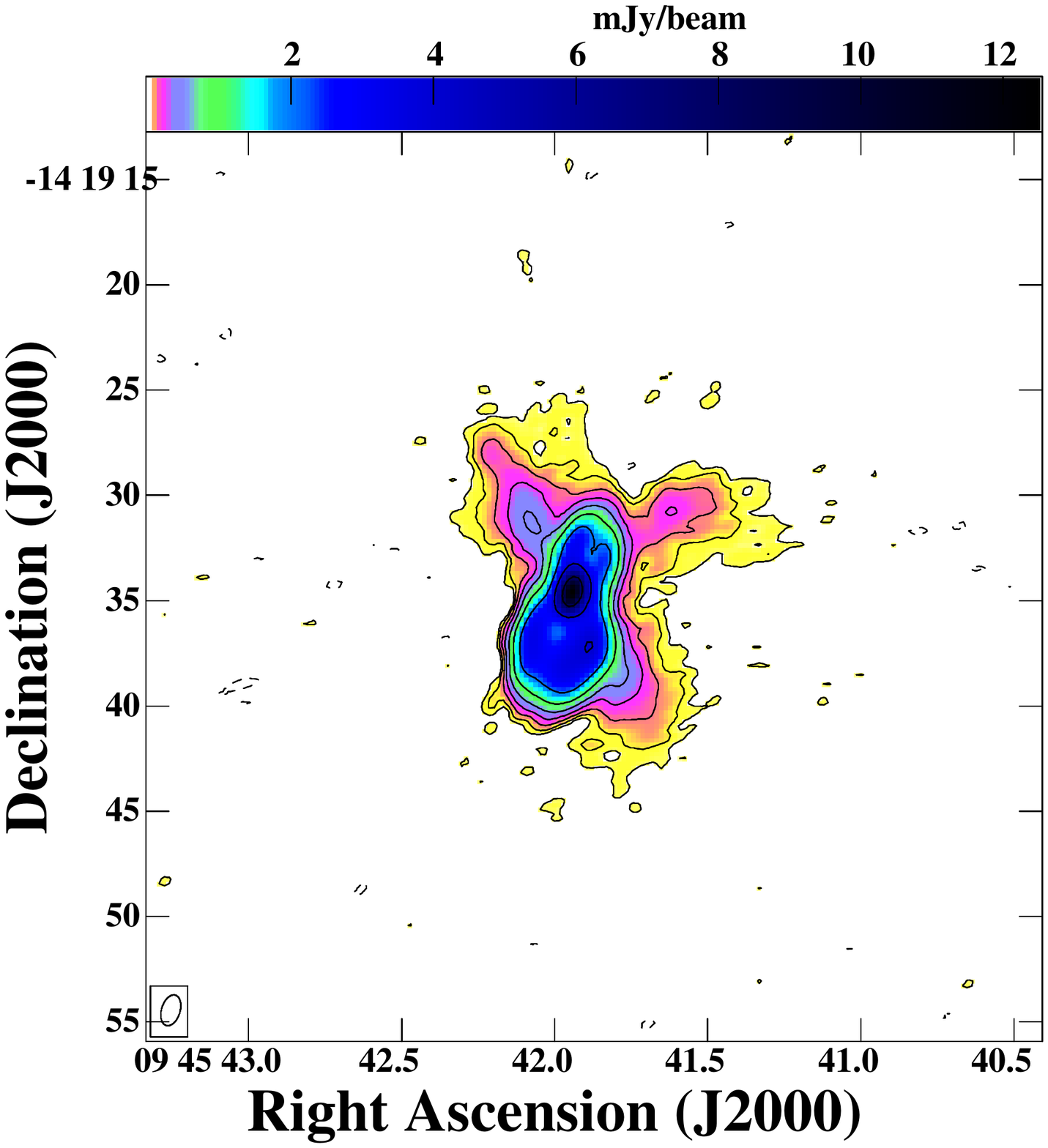}
\includegraphics[width=8.50cm,trim = 20 100 30 100]{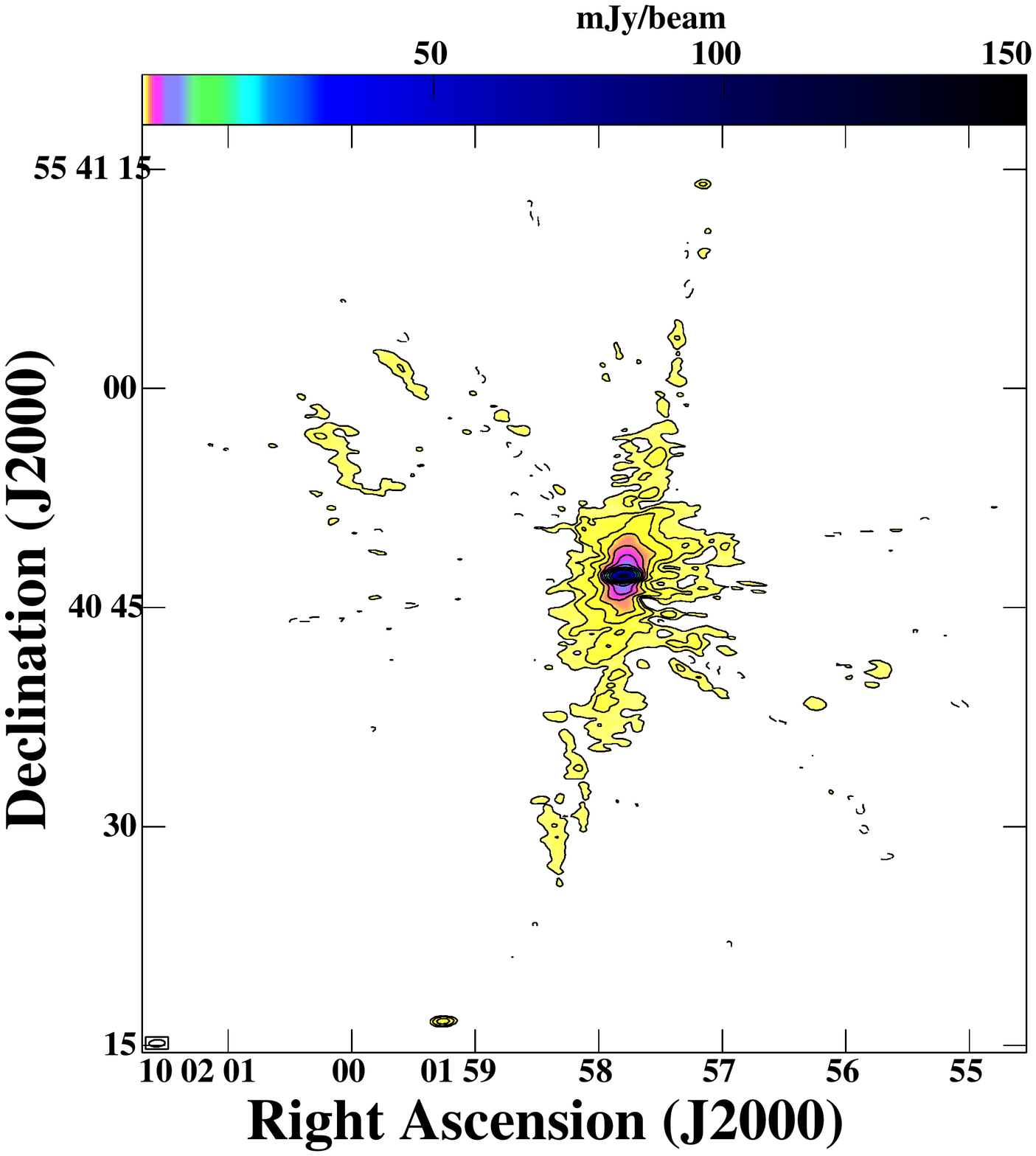}
\includegraphics[width=8.50cm,trim = 20 100 30 100]{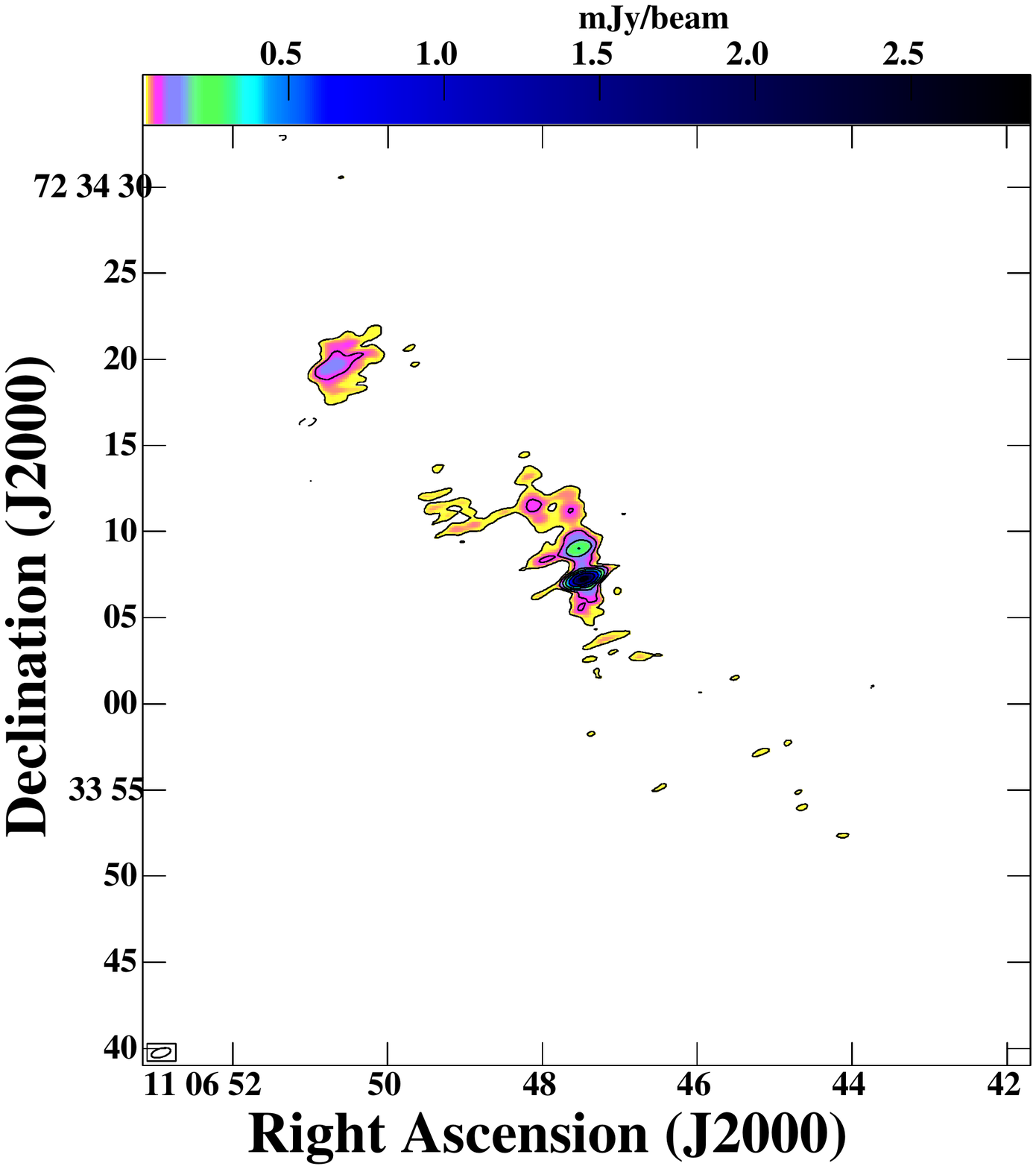}
\caption{Total intensity images of Seyfert galaxies at 5.5~GHz. The contour levels used are 3$\sigma \times $(-2,  -1, 1, 2, 4, 8, 16, 32, ...), where $\sigma$ = 8.83, 10.59, 25.57, 13.085 $\mu $Jy beam$^{-1}$ for NGC\,2639 (top left), NGC\,2992 (top right), NGC\,3079 (bottom left), NGC\,3516 (bottom right) respectively. }
\label{seytotint}
\end{figure*}

\begin{figure*}
\ContinuedFloat
\includegraphics[width=8.5cm,trim = 20 100 0 100 ]{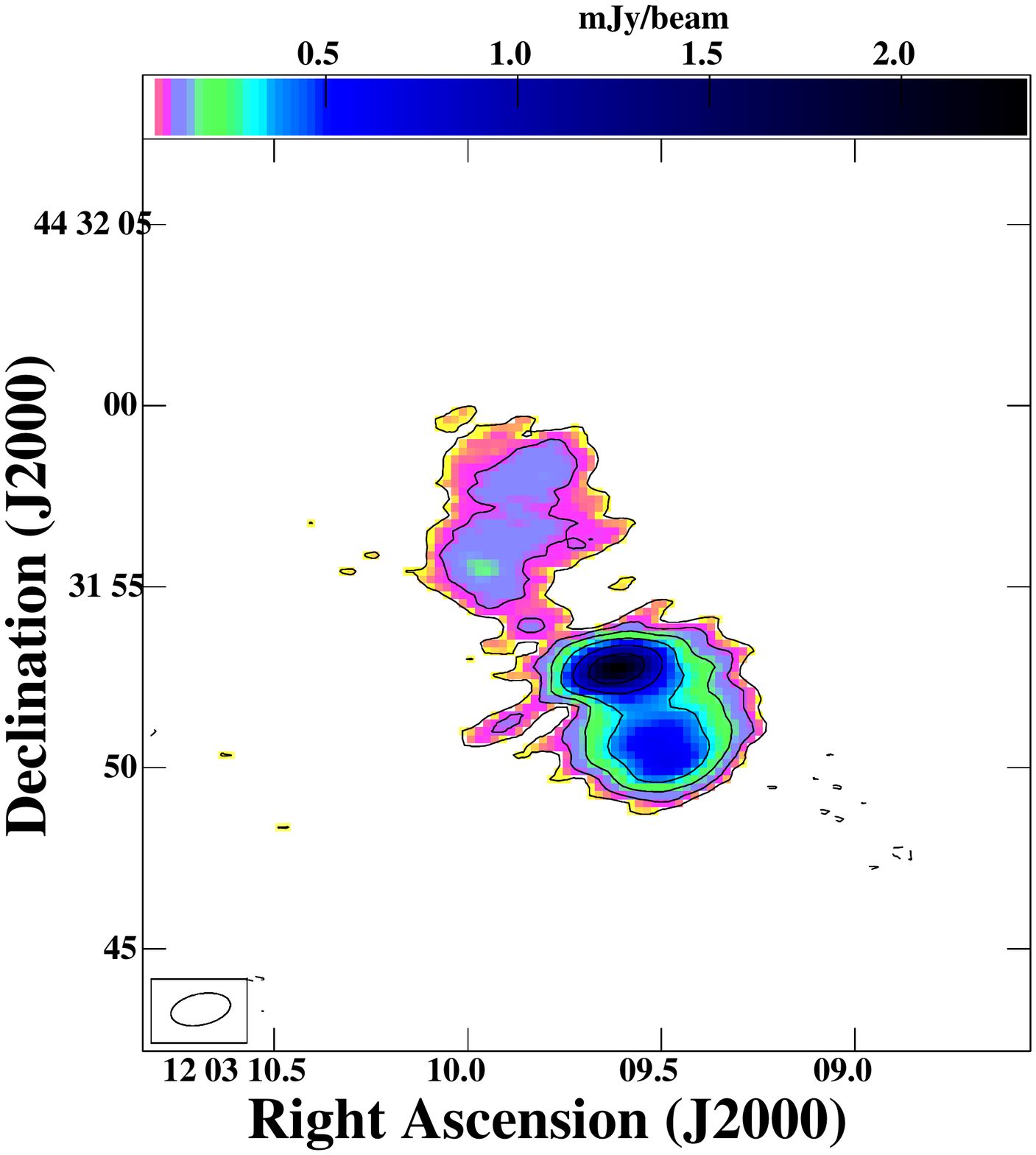}
\includegraphics[width=8.5cm,trim = 20 100 0 100 ]{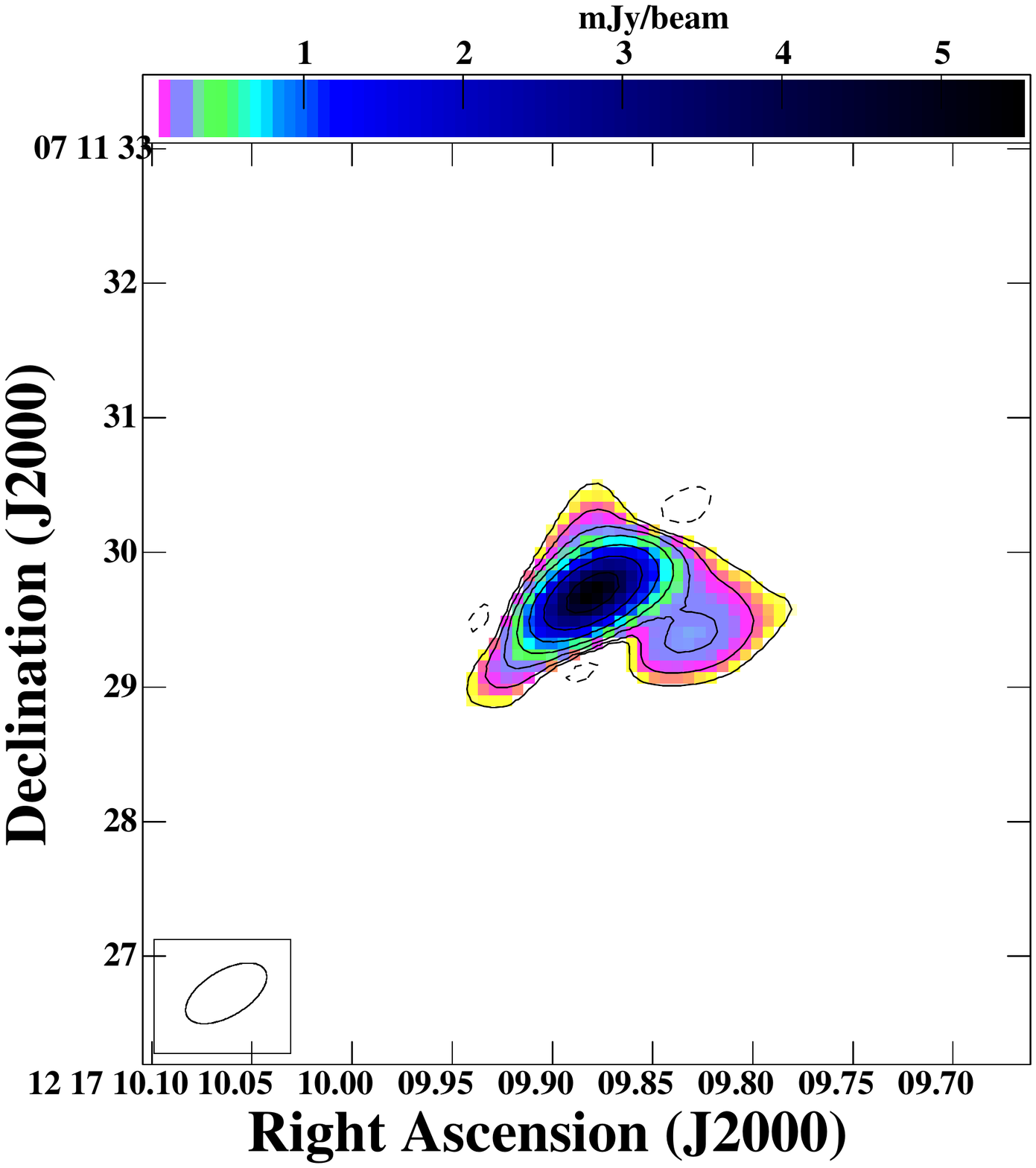}
\includegraphics[width=8.5cm,trim = 20 100 0 100 ]{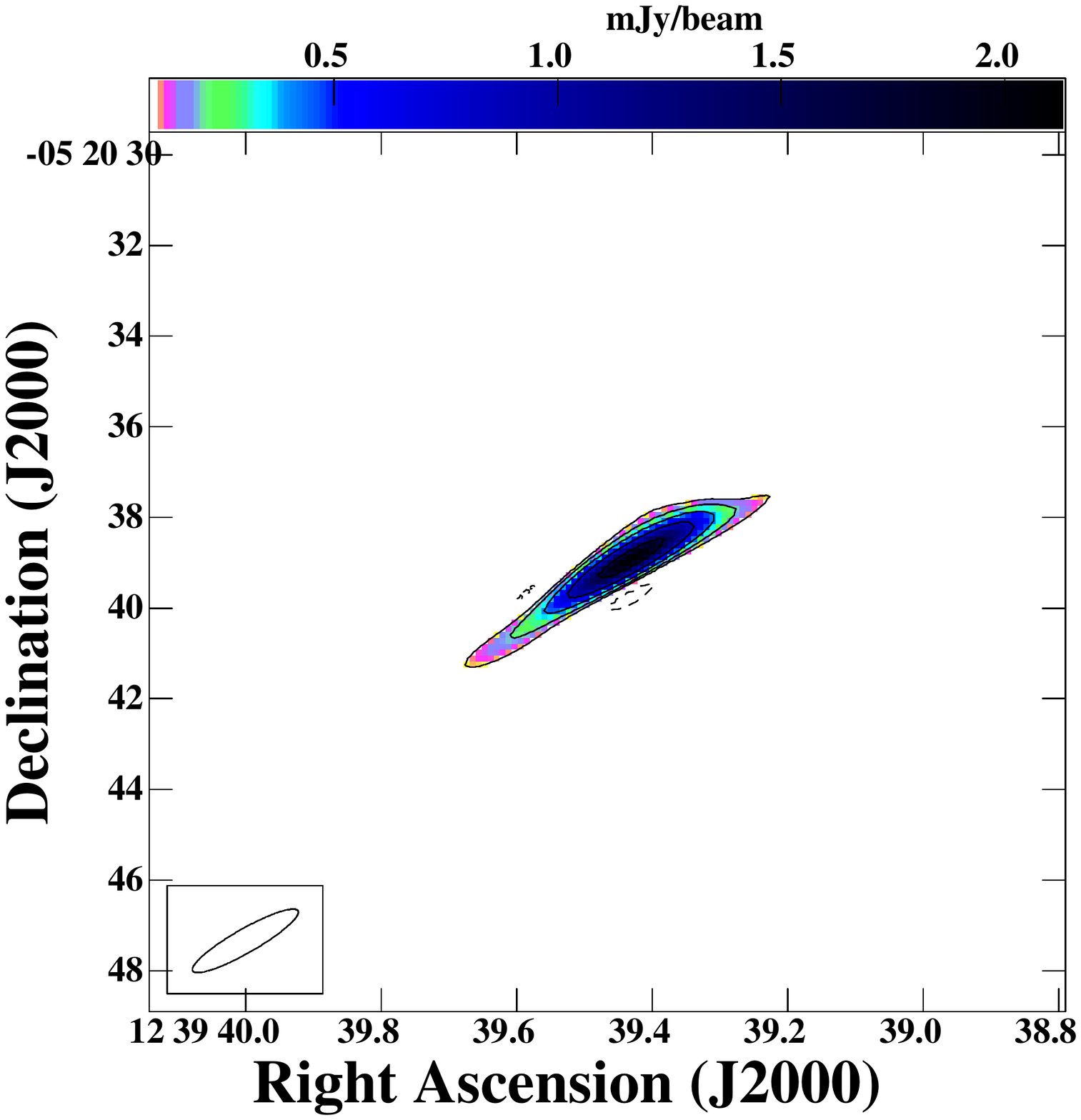}
\includegraphics[width=8.5cm,trim = 20 100 0 100 ]{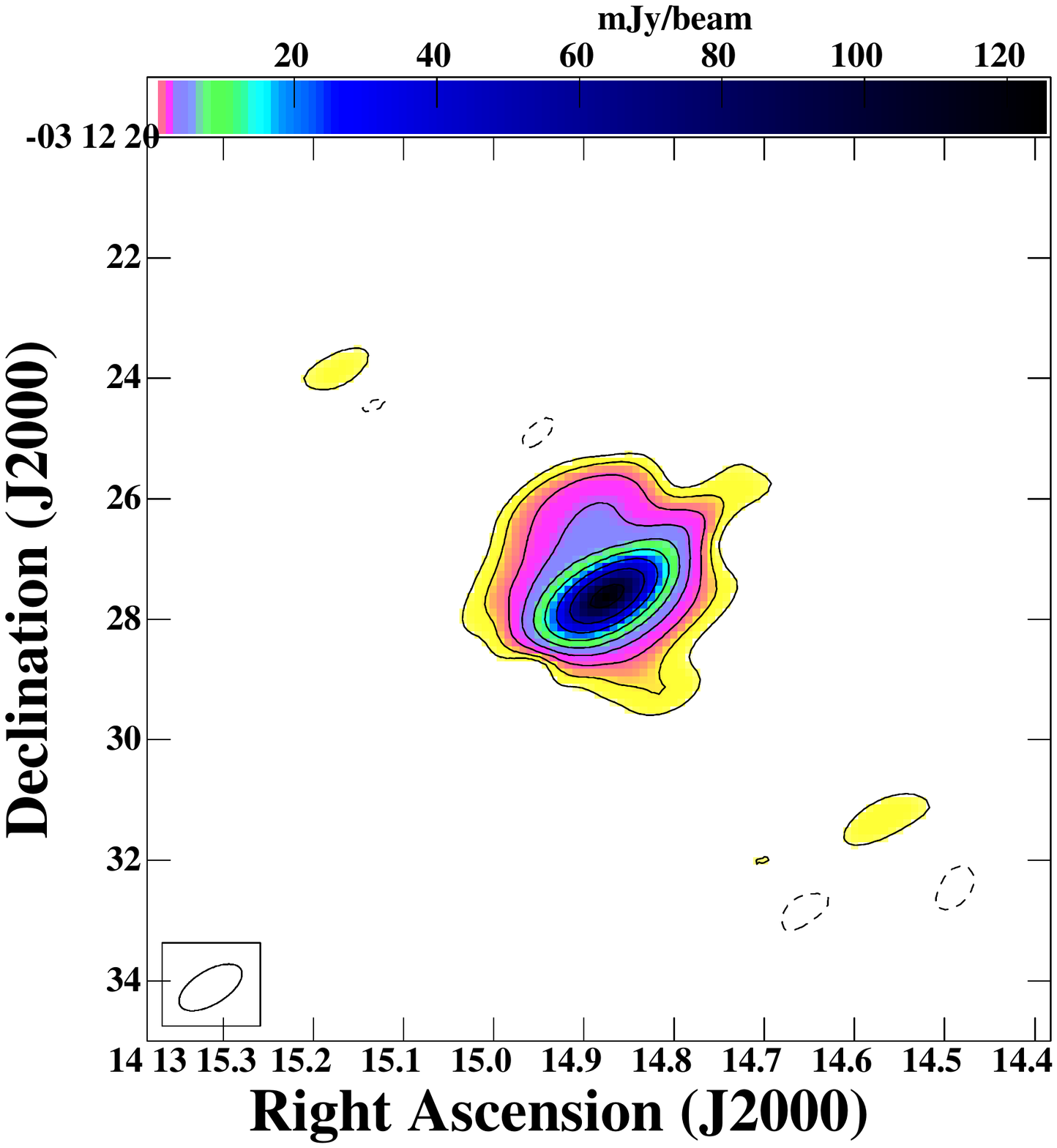}
\caption{Continued --Total intensity images of Seyfert galaxies at 5.5~GHz. The contour levels used are 3$\sigma \times $(-2,  -1, 1 2 4 8 16 32 ...), where $\sigma$ = 14.36, 21.35, 33.42, 134.43 $\mu $Jy beam$^{-1}$ for NGC\,4051 (top left), NGC\,4235 (top right), NGC\,4593 (bottom left), NGC\,5506 (bottom right) respectively.}
\end{figure*}

\begin{figure*}
\ContinuedFloat
\includegraphics[width=10cm,trim = 20 160 0 200 ]{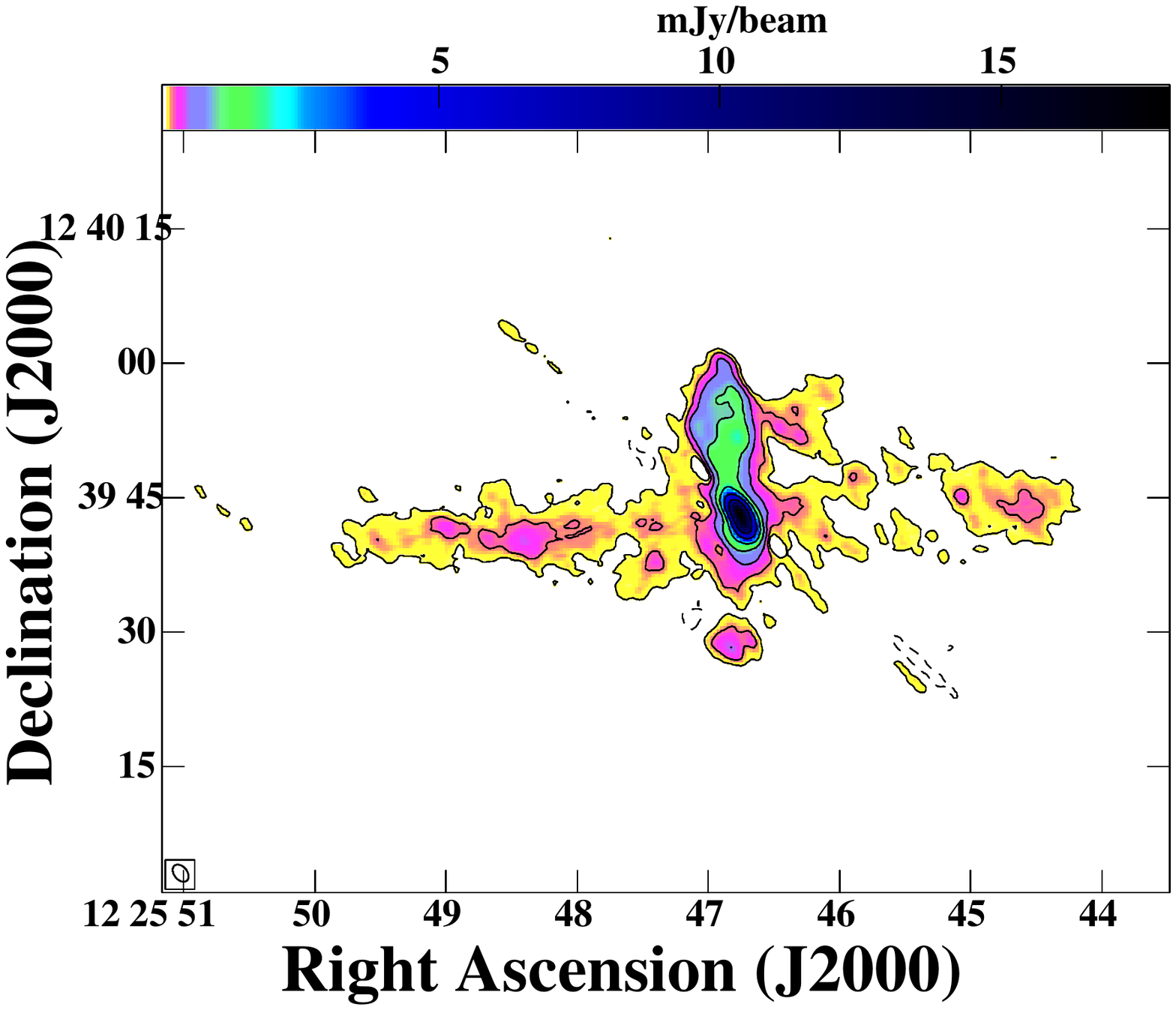}
\caption{Continued --Total intensity images of Seyfert galaxies at 5.5~GHz. The contour levels used are 3$\sigma \times $(-2,  -1, 1 2 4 8 16 32 ...), where $\sigma$ = 45.08 $\mu $Jy beam$^{-1}$ for NGC\,4388.}
\end{figure*}

\begin{figure*}
\includegraphics[width=8.5cm,trim = 20 100 0 100  ]{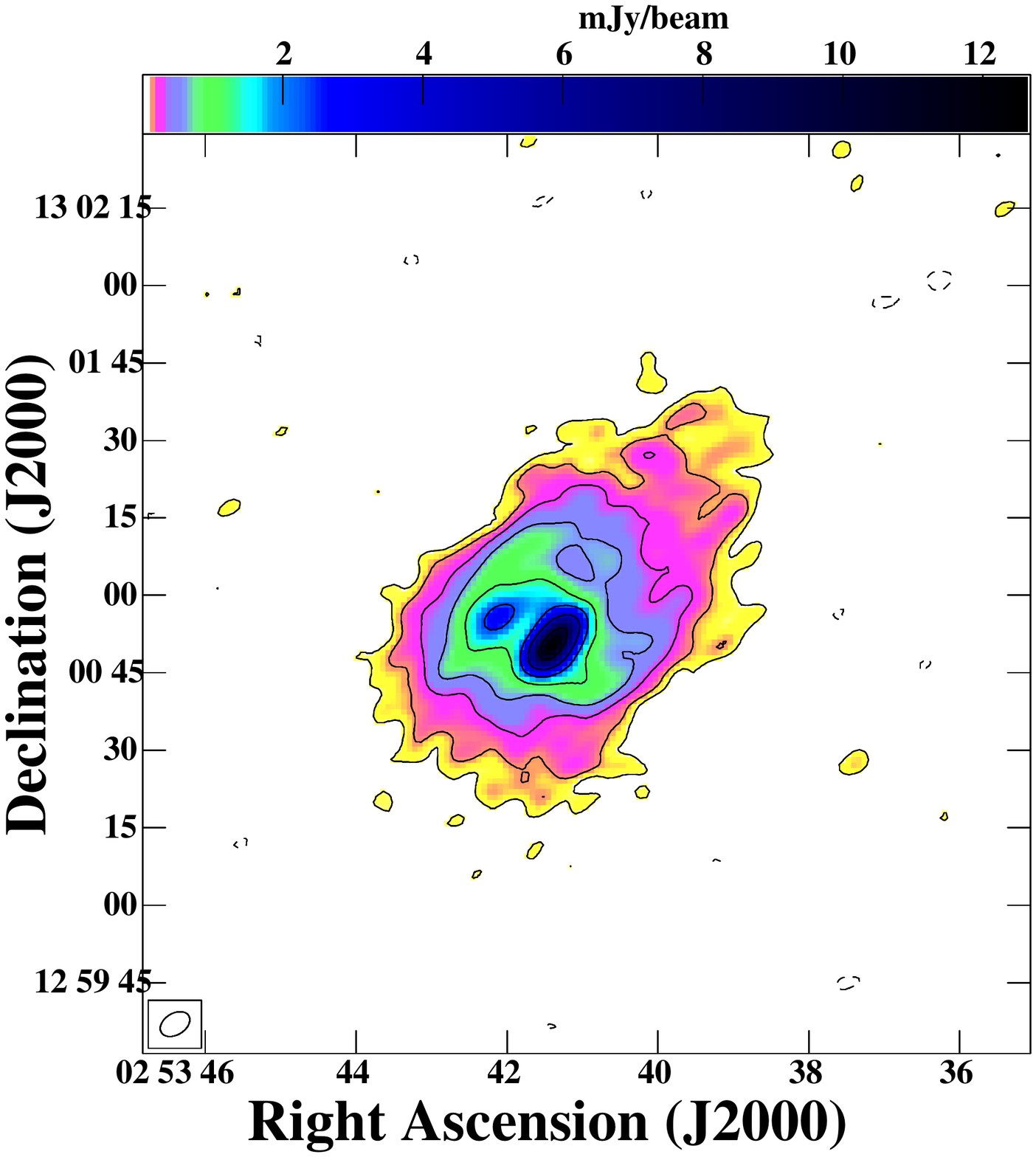}
\includegraphics[width=8.5cm,trim = 20 100 0 100  ]{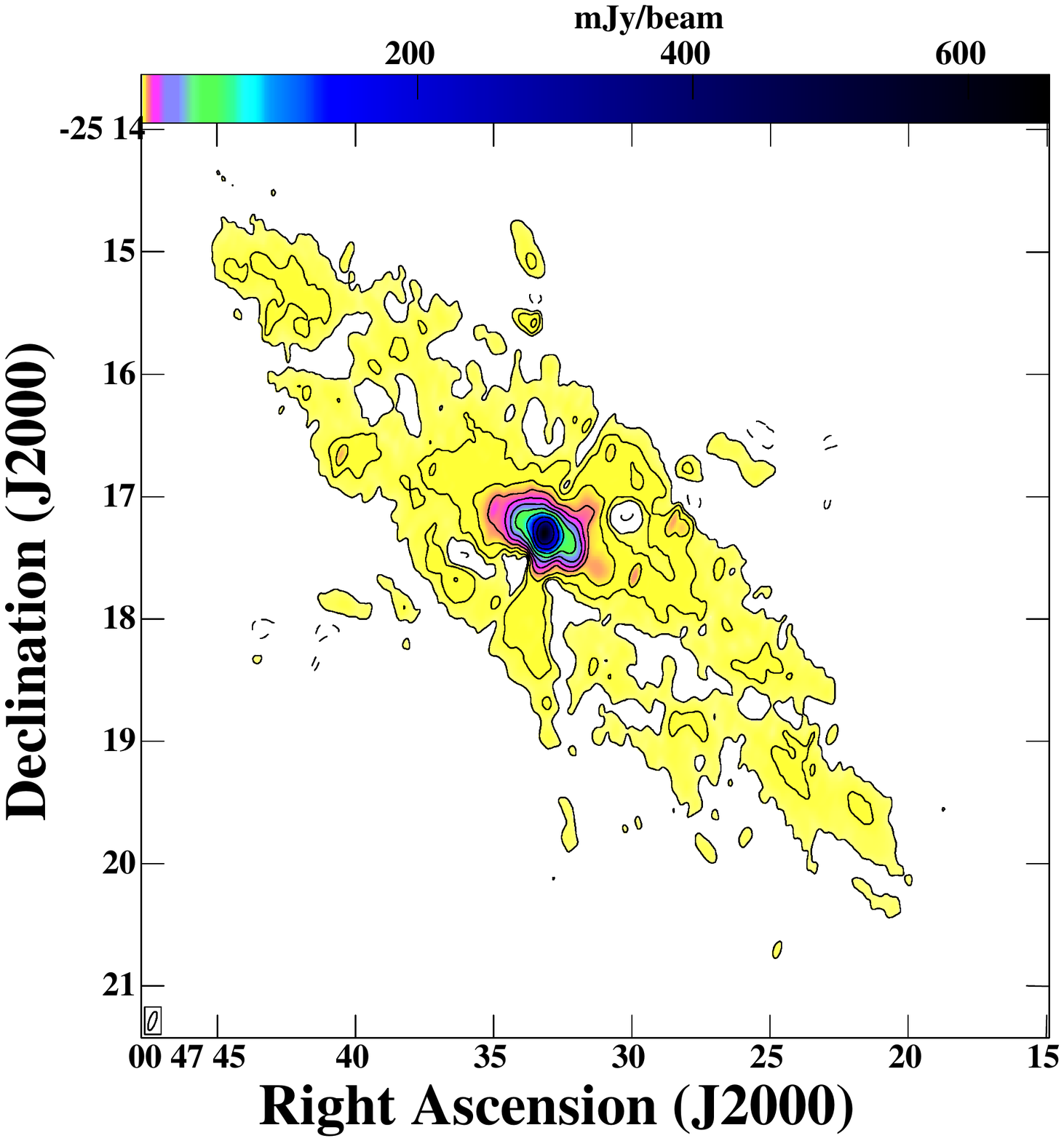}
\includegraphics[width=8.5cm,trim = 20 100 0 100  ]{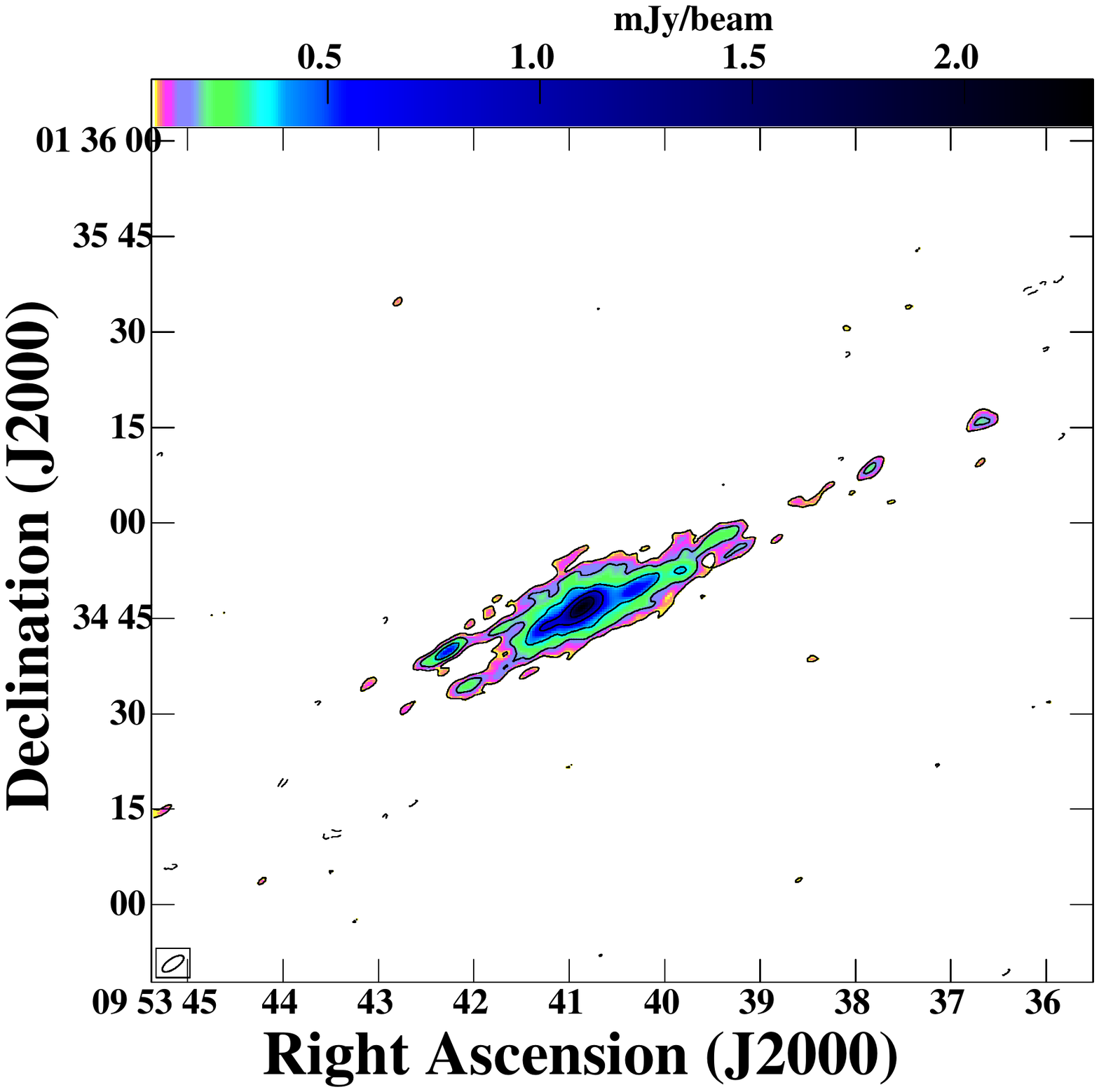}
\includegraphics[width=8.5cm,trim = 20 100 0 100]{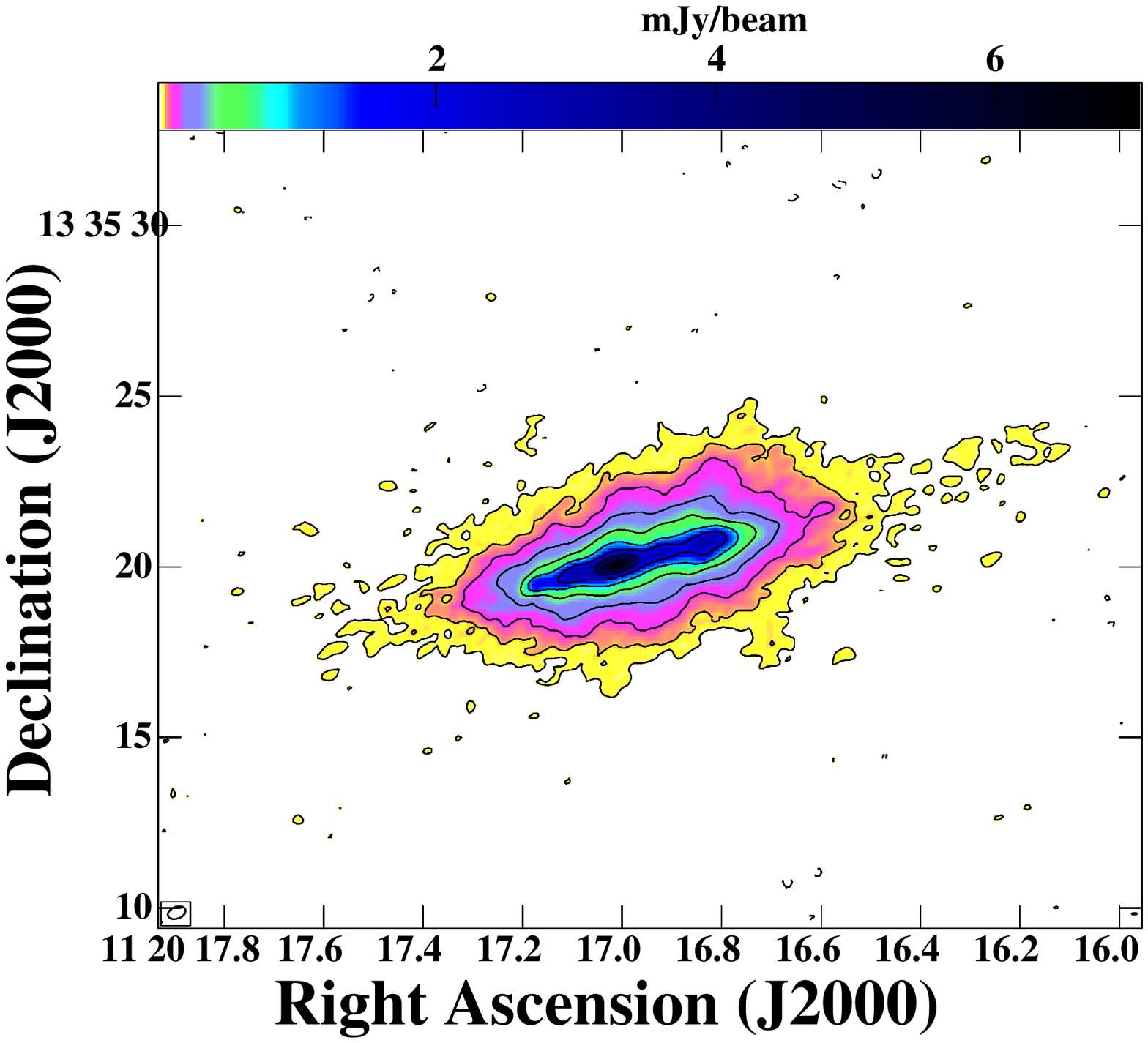}
\caption{Total intensity images of starburst galaxies. The contour levels used are 3$\sigma \times $(-2,  -1, 1, 2, 4, 8, 16, 32, ...), where $\sigma$ =24.61, 205.06, 30.076, 9.60 $\mu $Jy beam$^{-1}$ for NGC\,1134 at 1.5~GHz (top left),  NGC\,253 at 1.5~GHz (top right), NGC\,3044 at 5.5~GHz (bottom left), NGC\,3628 at 5.5~GHz (bottom right) respectively.}
\end{figure*}

\begin{figure*}
\ContinuedFloat
\includegraphics[width=9.0cm,trim = 20 100 0 100]{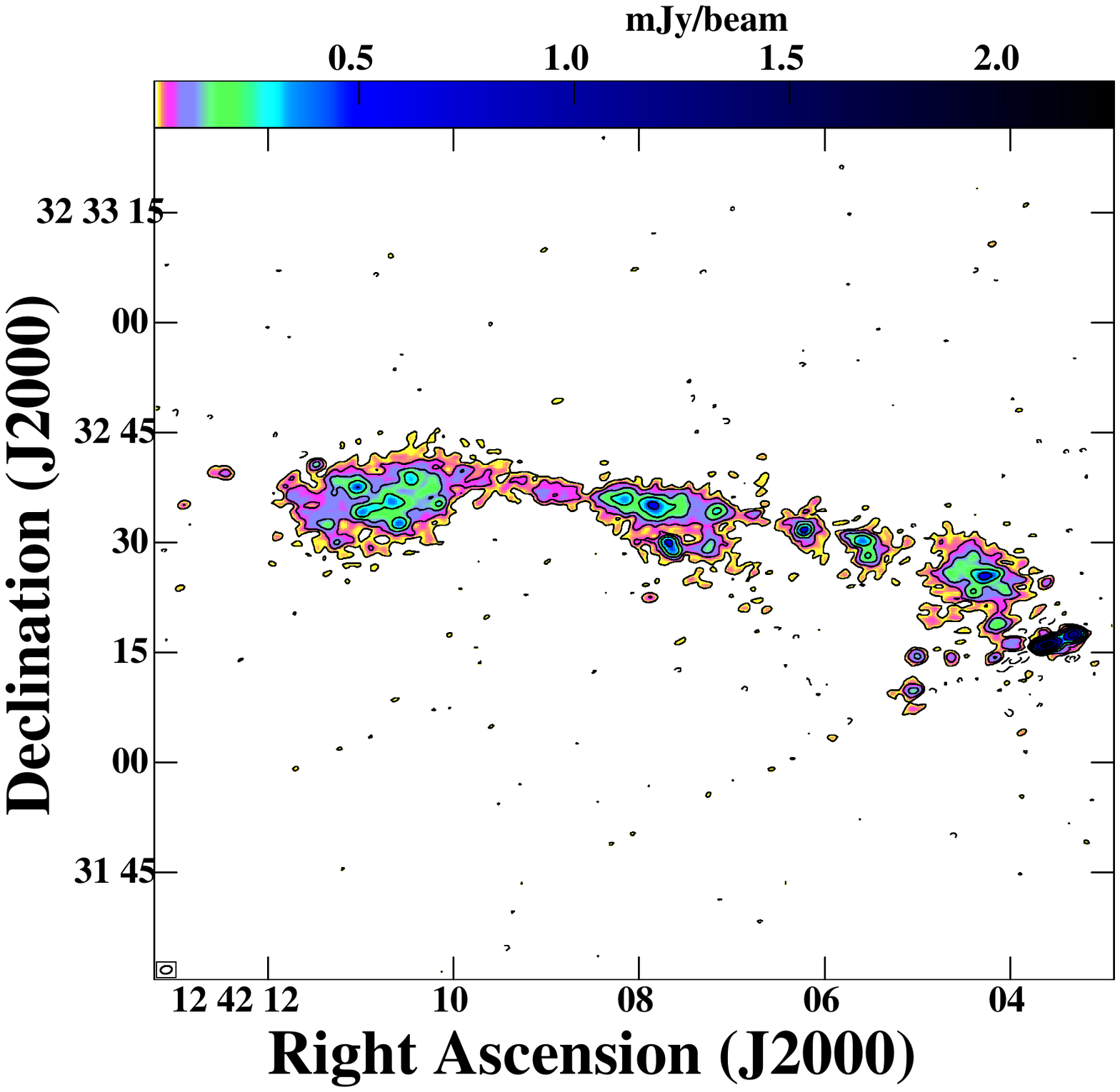}
\includegraphics[width=8.5cm,trim = 20 100 0 100]{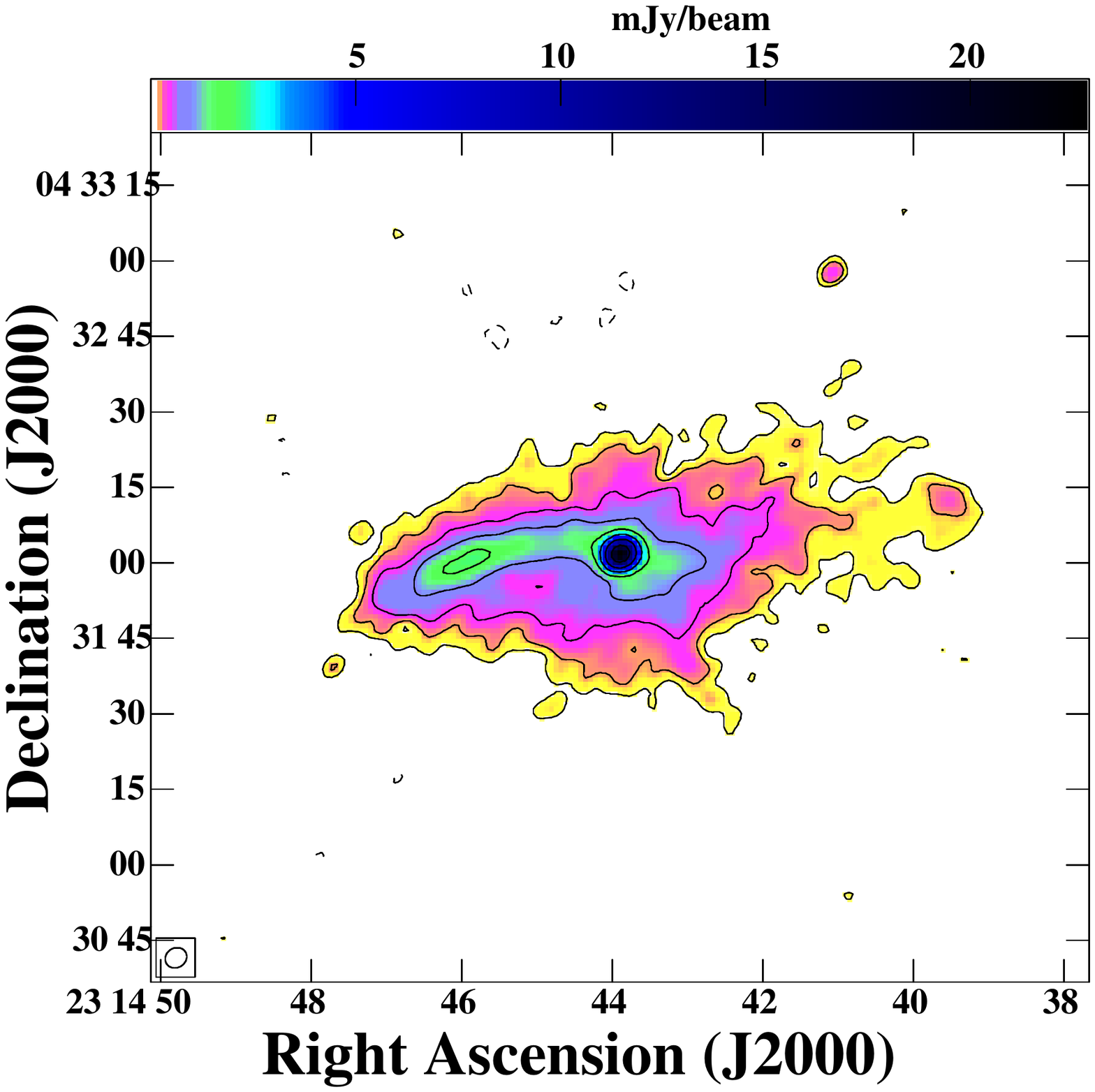}
\includegraphics[width=8.5cm,trim = 20 100 0 100  ]{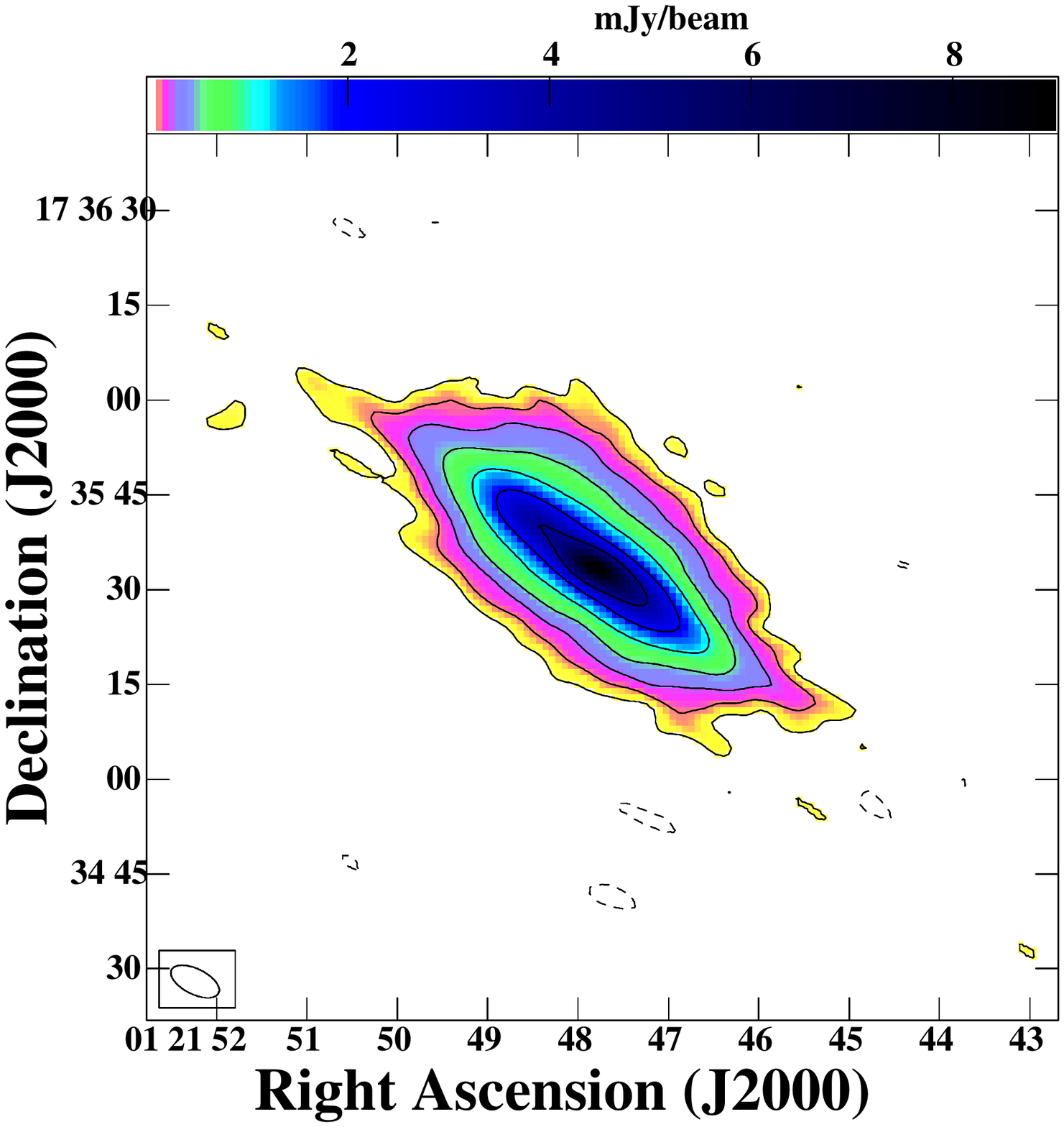}
\caption{Continued--Total intensity images of starburst galaxies. The contour levels used are 3$\sigma \times $(-2,  -1, 1, 2, 4, 8, 16, 32, ...), where $\sigma$ =  9.66, 39.68, 22.2  $\mu $Jy beam$^{-1}$ for NGC\,4631 at 5.5~GHz (top left), NGC\,7541 at 1.5~GHz (top right), UGC\,903 at 1.5~GHz (bottom) respectively.}
\label{sbtotint}
\end{figure*}

%
 and are likely to be hosting AGN nuclei at their cores. To increase the sample size, we also selected ``edge-on'' starburst galaxies from \cite{colbert1996b}, which are matched with the Seyfert galaxies in terms of their radio luminosities, axial ratios, and redshifts. This additional sub-sample includes NGC\,1134, NGC\,3044, NGC\,7541 and UGC\,903. All the nine chosen starburst galaxies have IRAS 60$\mu$m luminosities, L$_{60 \mu m}  \> 2\times10^{43}$~ergs~s$^{-1}$.


\section{Observations and Data analysis}
\label{obssec}
The sample was observed using the EVLA in B-array configuration under the project ID: 17B-074 during 2017-2018. 
All the Seyfert galaxies and three of the starburst galaxies were observed at 5.5 GHz and four of the starburst galaxies were observed at 1.4 GHz.
 From past studies, it is well known that Seyfert galaxies show radio emission from parsec to kpc scales whereas most of the radio continuum studies of starburst galaxies show emission majorly from the large scales comparable to the galaxy itself.
The largest angular scale that can be measured at C-band using B-array configuration is about 29$\arcsec$, so any larger spatial scales of emission will be resolved out in C-band observations. This study serves as a pilot study to understand the ideal frequencies and resolution for carrying out a comparative study between starburst galaxies and Seyfert galaxies. Hence, some of the starburst galaxies were observed at 1.5~GHz and some at 5.5~GHz.

The details of the observations are provided in Table~\ref{tab1}. 3C\,138, 3C\,286, and 3C\,48 were observed to calibrate the flux densities and polarization angles. Suitable phase calibrators were also observed to correct for the ionospheric phase errors. OQ\,208, and 3C\,84 were the unpolarized calibrators observed to correct for the instrumental leakage from these antennas.

{\tt CASA} calibration pipeline for EVLA data was used to carry out the basic editing and calibration of the `Stokes I' data after which polarization calibration was carried out separately. We corrected the cross-hand delays between the two polarizations. The model of the polarization calibrator was fed manually using the {\tt CASA} task `{\tt SETJY}' as elaborated in \cite{sebastian2019}. 

After calibrating for the D-terms using the unpolarized calibrator with the {\tt CASA} task `{\tt polcal}', we corrected for the polarization angle (aka electric vector position angle, EVPA) with reference to the known polarization angle of the polarized calibrator.
Either 3C\,138, or 3C\,286 was used as the polarization angle calibrator for the observations carried out at 5.5~GHz, whereas 3C\,48 was used at 1.5~GHz. 
Since the fractional polarization of 3C\,48 is too low ($\sim$0.5\% at 1.5 GHz) to provide reliable solutions, we combined all spectral windows while determining the cross-hand delays so that these delays are relatively accurate. 3C\,48 was also used to calibrate the polarization angles. However, the polarization angle for 3C\,48 can vary by 30$^{\degr}$ or more at $\sim$1~GHz, resulting in a similar uncertainty in the EVPAs of the sources, which in our sample turned out to be a single galaxy, NGC\,253.

The calibrated target source was then extracted out of the multi-source file using the task `{\tt SPLIT}' in {\tt CASA}.
A few loops of imaging followed by phase-only self-calibration of the target source were carried out before one final round of amplitude and phase self-calibration. We used the {\tt MT-MFS} algorithm while imaging, to correct for errors introduced due to the varied spectral indices of the sources in the field. We finally created the Stokes `Q' and `U' images, and the polarized intensity and polarization angle images from the former.

We then created polarization intensity images that were corrected for Ricean bias, polarization angle images, and fractional polarization images for each of our targets using the task `{\tt COMB}' in ${\tt AIPS}$. While making the polarization angle images we blanked those pixels which had errors greater than 10$\degr$. 
Similarly while making the fractional polarization images, all the pixels which had errors greater than 10$\%$ of the fractional polarization values were blanked.

For the sources observed at 1.5~GHz, we also carried out rotation measure (RM) synthesis \footnote{https://github.com/mrbell/pyrmsynth} to account for bandwidth depolarization. Data were acquired in 10 spectral windows (spws) for sources observed at 1.5~GHz. Separate full Stokes images were made at every spws on which RM synthesis was carried out. The results of our analysis are presented in Section~\ref{subsecpl}.

\section{Results}
\label{resultsec}
\subsection{Radio Morphology}
The total intensity images of Seyfert galaxies and starburst galaxies in our sample are presented in Figures~\ref{seytotint} and \ref{sbtotint}, respectively. The radio morphologies of all the individual sources are discussed in the following sub-sections.

\subsubsection{NGC\,2639}
\cite{martini2003} classified NGC\,2639 as a grand-design spiral galaxy hosting a Seyfert 1.9 nucleus. Our EVLA observations revealed an extra pair of lobes in the north-south direction which appeared to be highly polarized and misaligned by almost 90$\degr$ from the already identified pair of lobes \citep{sebastian2019b}. Our observations favor a scenario where the secondary lobes are generated as a result of stopping and restarting of AGN activity. 

\subsubsection{NGC\,2992}
This Seyfert galaxy is a changing-look AGN that varies from type 2 to intermediate type sometimes accompanied by extreme X-ray activity and back over the span of a few years \citep{gilli2000,ramos2009}. By studying the X-ray variability in NGC\,2992, \citet{marinucci2020} have shown that the region showing X-ray flaring is as close as $\sim15-40~r_g$ in two cases and $>50~r_g$ to the supermassive black hole. \cite{allen1999} found that NGC\,2992 shows biconical outflows seen in [O~{\small III}] and H$\alpha$+N[II] on kilo-parsec scales and \cite{veilleux2001} argue that this outflow is powered by the AGN rather than the starburst. Radio continuum emission shows a bright core and a pair of lobes with 8-shaped superbubble-like morphology with linear extents $\sim$ 500 pc \citep{wehrle1988}. \cite{irwin2017} discovered an additional pair of lobes that are aligned along the minor axis of the host galaxy. This emission was disentangled from that of the disk emission using the high fractional polarization seen in the lobes.

\subsubsection{NGC\,3079}
NGC\,3079 is another Seyfert galaxy with a type 2 nucleus \citep{veron2006} that shows loop-like radio lobes. NGC\,3079 also hosts a central starburst rendering the delineation of the role played by starburst or AGN in the observed radio morphology difficult. Our high-resolution image resolves the ``ring''-like lobes into filamentary structures. We carried out a detailed study of NGC\,3079 and we conclude that the lobe morphology is a result of the central jet, the superwinds, and the galactic magnetic fields \citep{sebastian2019}.

\subsubsection{NGC\,3516}
NGC\,3516 is a Seyfert 1.5 galaxy \citep{Knapen2002} hosted by an S0 type galaxy \citep{perezgarcia1998}. H$\alpha$ and [O~{\small III}] emission line imaging \citep{pogge1989,miyaji1992} have shown elongated features which are aligned with the radio structures \citep{wrobel1988,miyaji1992}. Our radio intensity maps also show a bipolar diffuse feature similar to what is reported in the literature. Our image also shows a knot close to the nucleus in the northern direction, which coincides with a knot in the [O~{\small III}] emission.
Although the northernmost feature appears diffuse, it appears brighter compared to the region in between. It resembles a hotspot-like feature often seen in the jets of FR\,II radio galaxies. 


\subsubsection{NGC\,4051}
NGC\,4051 has a Seyfert 1.2 type nuclei \citep{Ho97} and is hosted by a SAB type galaxy \citep{dumas2007}. NGC\,4051 has also been classified as a narrow-line Seyfert 1 (NLS1) galaxy in the literature \citep{denney2009,yang2013}. \cite{christopoulou1997} using their MERLIN images find a triple source aligned at a PA of 73$\degr$. They also suggest the presence of conical outflows due to the presence of blueshifted components in the [O~{\small III}] emission line profile. We find radio emission which is aligned along the NE direction similar to that found by \cite{Kukula95}. High resolution VLA observations of \cite{berton2018} show the S-symmetry of the extended emission clearly. \cite{christopoulou1997} find that the \lbrack OIII \rbrack emission is aligned along the NE direction coinciding with the emission seen at 8.4 GHz.

\subsubsection{NGC\,4235}
NGC\,4235 is a nearly edge-on, SAa type galaxy \citep{perez1998} with a Seyfert type 1 nucleus \citep{weedman1978,jimenez2000} with high [O~{\small III}] to H$\beta$ line ratios at the core. NGC\,4235 hosts a bright compact X-ray source \citep{fabbiano1992}, whereas \cite{rossa2003} find a bright nucleus and an extended emission feature in H$\alpha$ images. \cite{falcon2006} and \cite{pogge1989} also find extended [O~{\small III}] emission along the major axis, whereas no minor axis extension is detected \citep{colbert1996a}. 

However, the imaging study carried out by \cite{colbert1996b} at 6 cm using the C-array configuration of VLA reveals an extraplanar bubble-like structure which extends up to 9 kpcs along with the compact core emission. The radio source appears unresolved in VLBA imaging, while it shows evidence for variability in flux density in the VLA images \citep{anderson2005}. The imaging study carried out by \cite{hummel1991} also does not show any extended emission. The GMRT images at 325 and 610 MHz presented by \cite{kharb2016} suggest the presence of relic lobe emission. 

\subsubsection{NGC\,4593}
NGC\,4593 has a Seyfert type 1 nucleus \citep{george1998} hosted by an SBb galaxy \citep{gonz1997}. The radio emission appears point-like in our image as well as many of the past high angular resolution studies like those by \cite{UlvestadWilson84A,Thean00,Schmitt01}. We do not detect the faint double lobes seen by \cite{gallimore06} in their lower resolution images. It was noted by \cite{gallimore06} that this kpc scaled emission is aligned along the [O~{\small III}] emission \citep{schmitt2003}.

\subsubsection{NGC\,4388}
NGC\,4388 is also a highly inclined galaxy hosting a Seyfert type 2 nucleus \citep{philips1982}. X-ray studies of NGC\,4388 show the presence of an ample amount of extended soft X-ray emission, which is generated as a result of photoionization by the nuclear activity \citep{iwasawa2003,risaliti2002,cappi2006,beckmann2004}.
Ionized gas studies also show similar extended features, for instance, \cite{pogge1988} find a conical shaped ionized region.

NGC\,4388 is well studied in radio wavelengths, all of which show a double source slightly away from the center of the galaxy \citep{stone1988, hummel1991, irwin2000, condon1987}.

\subsubsection{NGC\,5506}
NGC\,5506 is hosted by an irregular edge-on spiral galaxy. Originally classified as a Seyfert type 2 nucleus \citep{Wilson76}, it was later identified by \citet{Nagar02} as a narrow-line Seyfert 1. Based on emission-line ratios, \citet{Maiolino94} and \citet{Davies05} argue that there is active star formation occurring on $250-400$ parsec scales. \citet{Maiolino94} also found evidence for double-peaked emission lines which were interpreted as the result of a biconical outflow. The extended soft X-ray emission as observed by \citep{Colbert98} is aligned in a direction similar to that the emission-line outflow traced and is also coincident with the radio emission. The IFU images produced by the WiFeS instrument on the ANU 2.3~m telescope also shows biconical outflows in the same direction \citep{Thomas17}. 

\subsubsection{NGC\,1134}

NGC\,1134 was classified as a starburst source based on their radio to FIR flux, the flux at 25 $\mu$m to 60 $\mu$m, and also due to the lack of any diffuse extended structures in radio images. 
Our image also shows radio continuum emission superposing the optical emission from the disk.
\subsubsection{NGC\,253}
\label{253not}
It is a nuclear starburst galaxy, which has star formation activity going on in the inner 200 pc radius \citep{engelbracht1998}. NGC\,253 has a bright central point source in radio coinciding with a faint NIR source \citep{sams1994} and a hard X-ray source \citep{weaver2002} and hence many authors classify the source as an LLAGN \citep{mohan2002, lira2007}. 
NGC\,253 was classified as a LINER galaxy, although it is suggested that the optical line characteristics can be explained as a result of starburst \citep{engelbracht1998}.
\cite{mccarthy1987} identified a shocked bubble in emission-line gas, suggesting outflows.
Diffuse X-ray emitting gas studies also present evidence for biconical outflows in this galaxy \citep{forster2004,dahlem1998}. \cite{carilli1992} discovered the radio continuum halo that extended up to 4~kpc in size from the disk. 

\subsubsection{NGC\,3044}
NGC\,3044 is an edge-on galaxy that is classified as an SB (s) type galaxy \citep{miller2003}.
Narrowband image of H$\alpha$+[N II] shows diffuse emission to both sides of the galaxy probably arising from an expanding superbubble \citep{miller2003}. \cite{collins2000} find that the H$\alpha$ and radio continuum emission possess very similar extents and structure. The VLA observations at 20~cm by \cite{irwin2000} show a loop in the radio continuum to the north-west of the nucleus which we do not find in our images likely owing to the steep spectrum of these features.

\subsubsection{NGC\,3628}
\label{3628note}
NGC\,3628 is a Sb spiral galaxy which has a starburst nucleus and a prominent dust lane \citep{liu2005}.
NGC\,3628 hosts a LINER nuclei and \cite{gonz2006} classify it as an AGN from the spectral fitting of the X-ray data, whereas \cite{flohic2006} suggested that the LINER emission in NGC\,3628 is starburst-related rather than AGN related due to the absence of an X-ray point source at the center of the galaxy. 
In our image, we see radio emission overlapping features in the optical image with finger-like protrusions at several locations across the disk.

\subsubsection{NGC\,4631}

NGC\,4631 is an edge-on spiral galaxy with a highly polarized radio continuum halo with a half-intensity height of about 14~kpc \citep{ekers1977}.
The polarized emission which is X-shaped is mostly extraplanar and the disk is depolarized. The magnetic field structure of the polarized emission is directed radially outwards either indicating the role of outflowing material or the large-scale galactic dynamo or both \citep{golla1994}.
Our 5.5~GHz observations did not have the uv-coverage to recover the diffuse extraplanar emission and the image of NGC\,4631 mostly shows the detailed structure of the radio emission in the disk.
\cite{dahlem1995} show that the extent of the radio emission is similar to the diffuse X-ray emission seen in NGC\,4631.
\cite{strickland2004} argue that the absence of a complete spatial correlation between the X-ray and the radio emission implies that the relativistic electron population, the magnetic fields, and the hot plasma are decoupled from each other.

\subsubsection{NGC\,7541}
NGC\,7541 is an SB type nearly edge-on spiral galaxy. Our radio image at 1.5~GHz shows the spiral arms to the east very clearly. While we do not detect any continuum halo-like emission, we detect diffuse radio emission overlapping with the star-forming disk. \cite{colbert1996b} detect finger-like structures protruding from the disk, which is seen in our images as well.
\subsubsection{UGC\,903}
\cite{colbert1996b} in their radio study of UGC\,903 at 5 GHz find that most of the radio emission have their origin from the disk. They also find that faint finger-like structures that extend out of the disk. 


\subsection{Polarization Properties}
\label{subsecpl}
\begin{table}
\begin{center}
\caption{Upper limit on fractional polarization}
\begin{tabular}{ccc} \hline \hline
Source & Limit on brightest feature (\%) & Limit on diffuse feature (\%) \\
&3$\times \frac{('Q' rms)}{('I' peak)}$& 3$\times \frac{('Q' rms)}{(typical 'I' )}$\\
\hline 


NGC2992 & 0.24 & 8.67 \\ 
NGC3044 & 3.46 & 10.13 \\ 
NGC3516 & 0.82 & 38.77 \\ 
NGC3628 & 0.38 & 3.61 \\ 
NGC4051 & 1.7 & 14.31 \\ 
NGC4235 & 0.51 & 4.73 \\ 
NGC4593 & 12.78 & -- \\ 
NGC4631 & 4.48 & 25.04 \\ 
NGC7541 & 0.31 & 3.28 \\ 
UGC903 & 0.59 & 3.55 \\ 
\hline
\label{tab:fracpollim}
\end{tabular}
\end{center}
{\small Column~1: Target names. Column 2: the upper limit on the fractional polarization calculated using the peak from the total intensity image. Column 3: the upper limit on the fractional polarization calculated using the mean value of  typical low surface brightness diffuse region from the total intensity image. NGC\,4593 is a point source and hence column 3 is left blank.} 
\end{table}

\begin{figure*}
\includegraphics[width=15.0cm,trim =0 275 0 120]{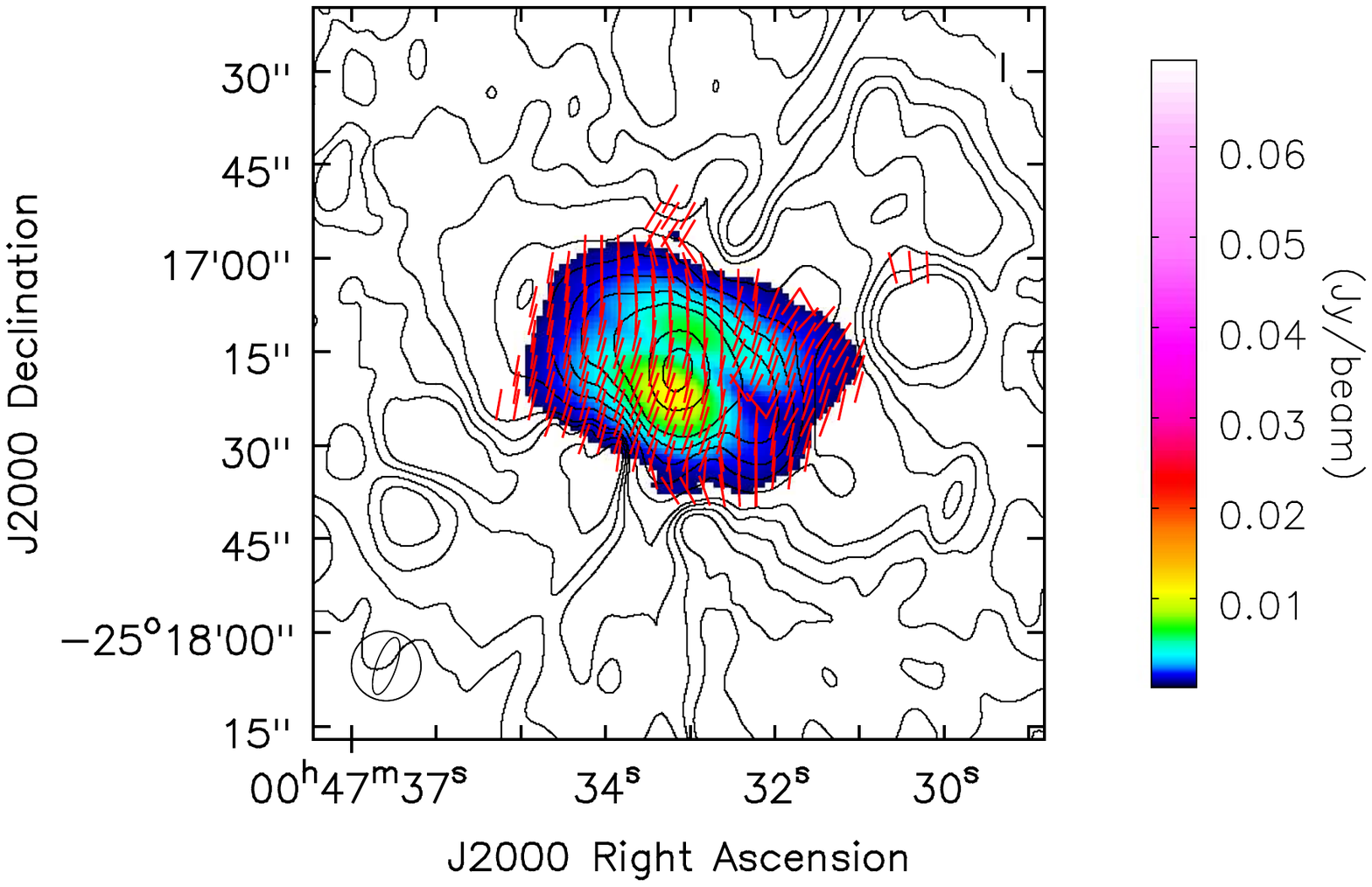}
\caption{Zoomed-in polarization image of NGC\,253: Polarization vectors in red overlaid on color image of the polarized intensity and total intensity contours for NGC\,253 with contour levels = 0.48 $\times$ ( 1, 2, 4, 8, 16, 32, 64, 128, 256, 512) mJy beam$^{-1}$. The polarization intensity and polarization angle images were produced using RM synthesis.}
\label{fig_pol_sb}
\end{figure*}

\begin{figure*}
\includegraphics[height=8.0cm,trim =100 220 50 250]{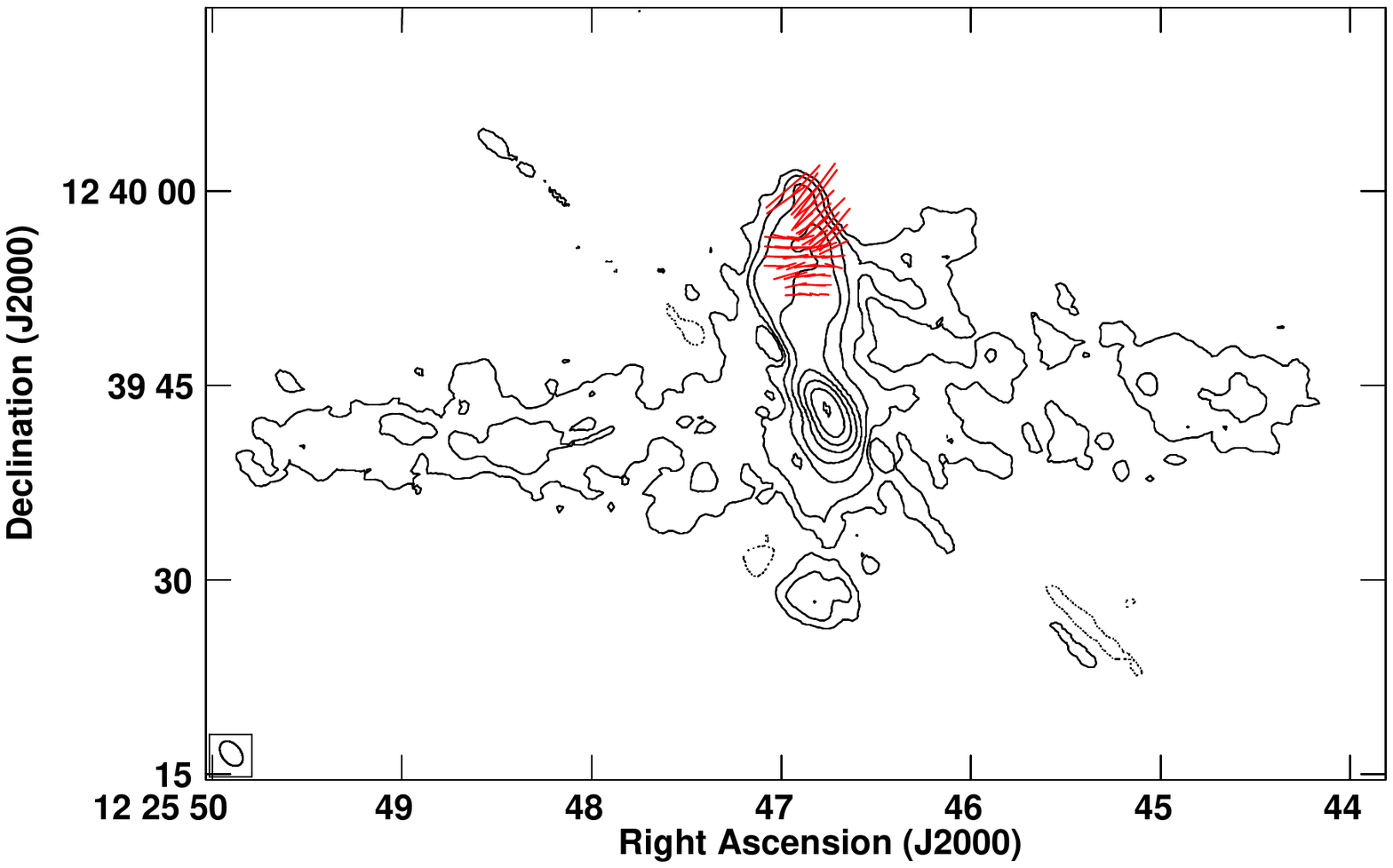}
\includegraphics[height=8.0cm,trim = 10 180 100 160]{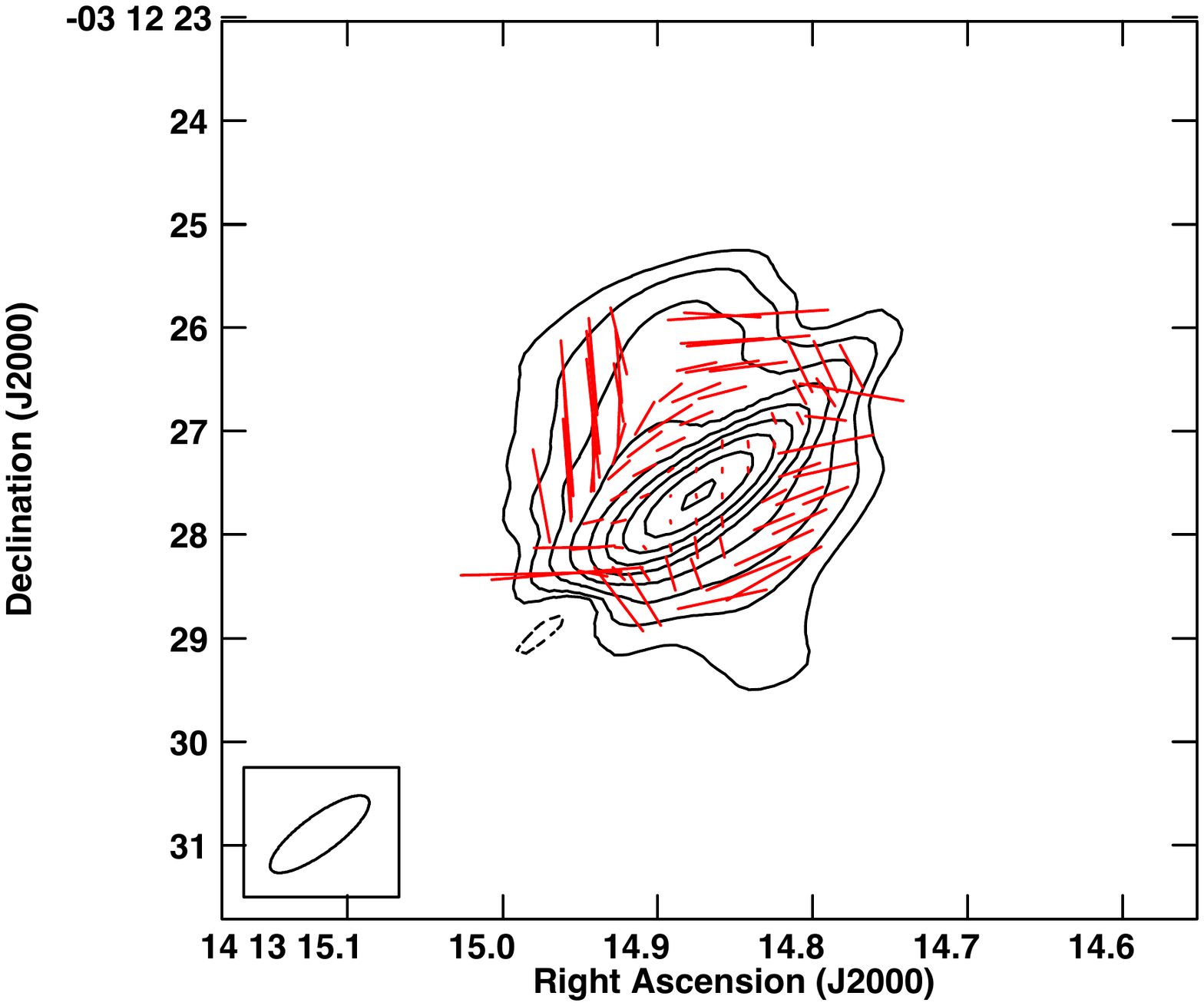}
\caption{Polarization vectors with length proportional to fractional polarization in red overlaid on total intensity contours for (left) NGC\,4388 at 5.5~GHz with contour levels = 0.13 $\times$ (-2, -1, 1, 2, 4, 8, 16, 32, 64, 128, 256, 512) mJy beam$^{-1}$, with one arcsec length of the polarization vector corresponding to 6.8\% fractional polarization and (right) NGC\,5506 at 5.5~GHz with contour levels = 1.17 $\times$ (-0.350, 0.350, 0.700, 1.400, 2.800, 5.600, 11.30, 22.50, 45, 90) mJybeam$^{-1}$ with one arcsec length of the polarization vector corresponding to 6.7\% fractional polarization.}
\label{fig_pol_sey}
\end{figure*}

\begin{figure*}
\centering{
\includegraphics[width=12cm,trim=20 160 0 200]{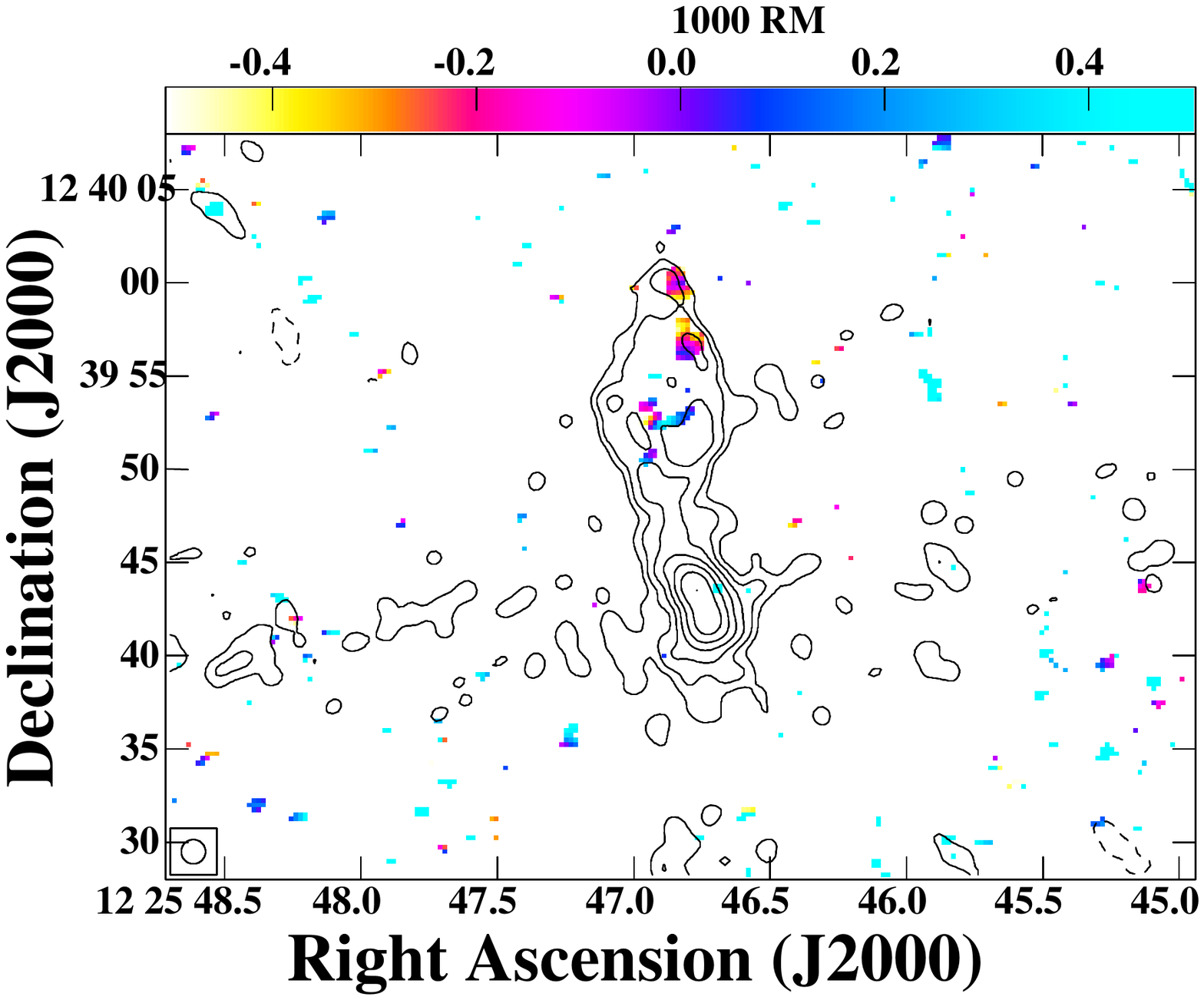}}
\caption{The 5~GHz total intensity radio contours superimposed over the in-band rotation measure image of NGC\,4388 in color. Contour levels: 213$\times$(-2, -1, 1, 2, 4, 8, 16, 32, 64, 128, 256, 512) $\mu$Jy beam$^{-1}$. The beam shown in the lower left corner is of size $1.3\arcsec\times1.30\arcsec$. The RM values are in units 1000 rad~m$^{-2}$.}
\label{fig_rm}
\end{figure*}

\begin{figure}
\centering{
\includegraphics[width=8cm,trim=10 100 0 130]{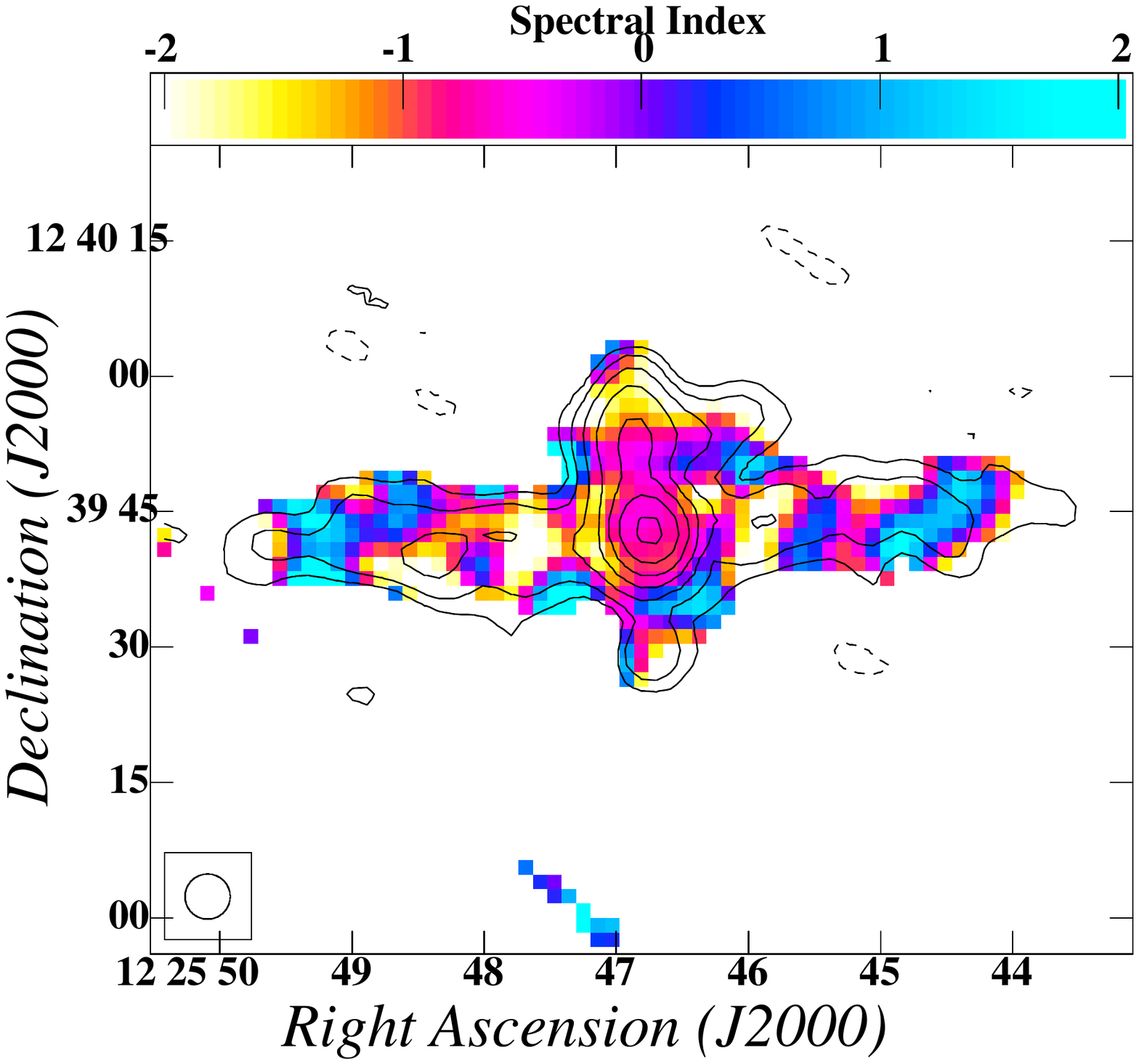}}
\caption{The in-band spectral index image of NGC\,4388 in color with 5 GHz radio contours superimposed. Contour levels: $5\times10^{-4} \times(1, 2, 4, 8, 16, 32, 64, 128, 256, 512)~$Jy beam$^{-1}$. The circular beam shown in the lower left corner is of size $\sim5\arcsec$.}
\label{fignew4388}
\end{figure}

We detected polarization in four Seyfert galaxies (NGC\,2639, NGC\,3079, NGC\,4388 and NGC\,5506) at 5.5 GHz and one starburst (NGC\,253) at 1.5~GHz.
We did not detect polarization in many of our sources, probably due to a variety of reasons including depolarization, sensitivity limitations, and inadequacy of short baselines to map the diffuse polarized emission. Especially at 1.4~GHz, the wavelength dependant depolarization effects may be playing a significant role. 
 
To recover polarized emission which may have been lost due to bandwidth depolarization in the process of imaging, we attempted using rotation measure (RM) synthesis \citep{Brentjens2005} for our starburst galaxies observed at 1.4~GHz. We detected marginal levels of polarization in NGC\,1134, and UGC\,903 which were undetected previously. The mean fractional polarization estimated from regions that showed polarization signal above three sigma threshold turned out to be in the range $\sim 1-4\%$ for NGC\,1134, NGC\,253 and UGC\,903. 
Because of such low levels of polarization along with the lack of organization in the polarization angles, the fidelity of the polarized emission recovered using RM synthesis needs to be confirmed using future observations.
The polarization image of NGC\,253 is presented in Figure~\ref{fig_pol_sb}. The polarization vectors were obtained using the relation, $ \frac{1}{2}$ tan$^{-1}( \frac{U}{Q})$. We note that the polarization angle calibrator 3C48 used for this dataset is only 0.5\% polarised at 1.5 GHz \footnote{https://science.nrao.edu/facilities/vla/docs/manuals/obsguide/modes/pol}. Therefore the polarization angles could in principle have large errors.
In our study, the starburst galaxy NGC\,253 shows polarized emission near the core with typical mean fractional polarization $\sim$ 4\%. It is worth noting here that NGC\,253 has been suggested to possess a weak LINER-like AGN in its center.

Therefore, we detect polarized emission in 4/9 ($\sim$44$\%$) of Seyfert galaxies at 5.5~GHz, 3/4 (75\%) of starburst galaxies observed at 1.4~GHz and 0/3 starburst galaxies at 5.5~GHz. 
A direct one-to-one comparison between the fractional polarization in the starburst galaxies at 1.4~GHz and Seyfert galaxies is limited because of the difference in frequencies and resolution. The starburst galaxies might be facing more bandwidth and beam depolarization with the additional complication of picking up more diffuse emission.

Despite the lower frequency of observation, the detection rate is higher for starburst galaxies at 1.4~GHz, which, if real, would imply that the ordering scales of polarized emission in starburst galaxies is larger and is preferably picked up by the lower resolution observations at 1.4~GHz. Such a situation is expected when the magnetic fields are ordered due to the large-scale mean-field dynamo mechanism operating on the scales similar to the structures of the galaxy itself. On the other hand, the radio emission which owes its origin to the jets at resolutions higher than the galaxy size scales is easily explicable. 
Owing to the small sample size and the fact that only four starburst galaxies were observed at 1.4~GHz and three at 5.5~GHz, our results must be treated as indicative pending additional data.

The upper limits on the fractional polarization of our sources calculated using the rms noise of the Stokes 'Q' images are tabulated in Table~\ref{tab:fracpollim}.
The lobes of NGC\,2639 and NGC\,3079 show polarization fractions as high as 30\% at 5.5~GHz \citep{sebastian2019b,sebastian2019}. 
While it is true that our observations suggest that Seyfert galaxies are more probable to show polarized emission at the given frequency ($\sim$ 5~GHz) and resolution ($\sim$ 1 $\arcsec$), this is not a universal trend. For example, we did not detect polarization in NGC\,2992 with an upper limit of 8.7\% on the most diffuse feature. \cite{irwin2017} find an average fractional polarization of 6.4\% for NGC\,2992 observed at C-band using the C and D arrays. On the other hand, none of the starburst galaxies have polarization fractions above 25\% even in the most diffuse features at either of the frequencies.


Figure~\ref{fig_pol_sey} shows the polarization images of NGC\,4388 and NGC\,5506.
The northern lobe of NGC\,4388 shows polarized emission towards the northern edge. There appears to be two distinct polarized regions, one within the lobes and one along the north-western edge. The fractional polarization in the northwestern edge and that within the inner regions of the lobes turn out to be 24$\pm$4.5\% and 13$\pm$3\%. The inner regions of the northern lobe show electric vectors aligned perpendicular to the axis implying magnetic field vectors that are aligned along the jet axis. More interesting is the alignment of the electric vectors at the north-western edge of the lobe. The electric field vectors are aligned perpendicular to the edge, implying tangential magnetic fields.

The high levels of fractional polarization along with the tangential alignment of magnetic fields and the tightening of contours are expected if shock compression is leading to the amplification of magnetic fields at the edges. Yet another interesting morphological detail is the diffuse extension towards the west which is present only after the end of the highly polarized feature in the north-west. 
\cite{damas-segovia2016} presented the lower-resolution polarization images of NGC\,4388 from observation carried out using C and D array configuration at 6.0~GHz. They recover much more diffuse polarized emission which extends up to 5~kpc. Their images mostly show poloidal magnetic fields similar to those seen in the inner regions of the lobes in our image. Figure~\ref{fig_rm} shows the RM image of NGC\,4388 with radio total intensity contours overlaid. The RM values show gradients in two regions along the north-western edge of the northern lobe, and even an RM inversion in the region at declination 12$\degr$~39$\arcmin$~55$\arcsec$. Interestingly, such RM gradients have been inferred to suggest the presence of helical magnetic fields \citep[e.g.,][]{Kharb09,Clausen11}. Clearly, the magnetic field structure in the lobes of NGC\,4388 is complex.

Figure~\ref{fignew4388} shows the spectral index image of NGC\,4388 generated by dividing the data into into four sub-bands and using the beginning and end frequency sub-band images. We have uv-tapered the data to achieve a coarse resolution of $\sim$ 5 $\arcsec$ to discern the variation of the spectral index across the lobes .

We do not find an obvious correlation between the implied magnetic field compression at the lobe edge in NGC\,4388 and the in-band spectral index values, in that there is no clear spectral flattening at the lobe edge . Such spectral flattening has been observed in the lobes of the Seyfert 1 galaxy Mrk\,231 by \citet{ulvestad1999} and Seyfert 2 galaxy NGC\,3079 by \citet{sebastian2019} and could be consistent with magnetic field compression as well as re-acceleration of charged particles at the lobe edges, where the lobes interact with the surrounding medium. 

The outflow-like radio emission observed in NGC\,5506 is also polarized with a complex morphology for the polarization electric vectors. The inferred magnetic field vectors appear to follow the outflow. The fractional polarization has an average value of 7.5\% which increases towards the edge of the outflow. The radio emission does not show clear edges like those observed in NGC\,3079 or NGC\,4388. Shock acceleration, therefore, might not be the primary source for the polarized emission in this source.

Amongst several of the studies which explored the radio continuum emission from Seyfert galaxies, a few involved understanding their polarization properties. \cite{ho2001} detected polarized emission from 12 out of the 52 sources that they observed. Several other detailed polarization studies of individual Seyfert sources include \cite{wilson1983,Duric88,crane1992,kaufman1996,elmouttie1998,Kharb06}. 
Several studies have noted a difference in the magnetic field structures inside parsec-scale jets of various AGN sub-classes like low-energy peaked BL Lacs (LBLs) and high-energy peaked BL Lacs (HBLs) with LBLs predominantly showing transverse inferred magnetic fields and HBLs showing longitudinal inferred magnetic fields, similar to those observed in quasars \citep{Kharb08,Gabuzda92,Cawthorne93}.
We inspected the images of the Seyfert galaxies published in the literature in conjunction with ours. In nearly seven out of ten sources with well-resolved polarized radio emission, the inferred magnetic field vectors are aligned with the lobe edges. In four of these sources the inferred magnetic field is largely poloidal along the outflow direction, while in most other sources, it is hard to attribute a uniform trend for the polarization vectors in the lobe. In the remaining sources, the inferred magnetic field orientation is neither parallel nor perpendicular to the outflows. 

The CHANG-ES survey \citep{irwin2012} has obtained polarization data on a large number of nearby spiral galaxies. \cite{krause2020} find that the fractional polarization values observed in spiral galaxies range from 10 to 40\% when imaged at a resolution of 12$\arcsec$. These high degrees of polarization are in contrast to the negligible values seen in our observations at higher resolutions. 
\cite{krause2020} also noted that increased fractional polarization comes from large-scale diffuse emission from galactic haloes on scales larger than 1~kpc or several kpcs. We conclude that the fractional polarization is higher in the lobes of Seyfert galaxies in high-resolution observations compared to starburst galaxies, although more polarization data are needed to confirm this result.

\section{Discussion}
\label{discsec}

\begin{figure*}
\includegraphics[height=5.7cm,trim = 0 0 0 0 ]{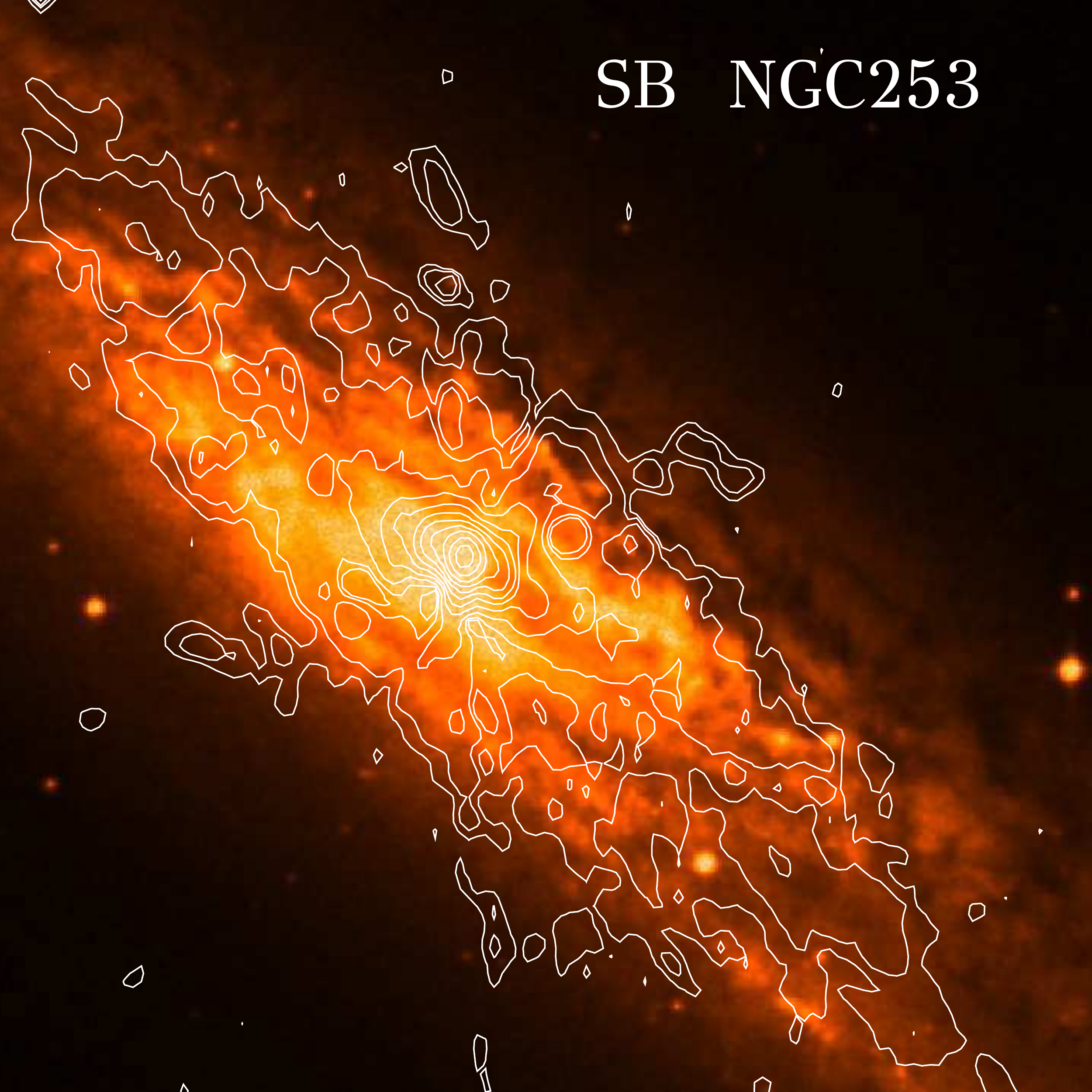}
\includegraphics[width=5.7cm,trim = 0 0 0 0 ]{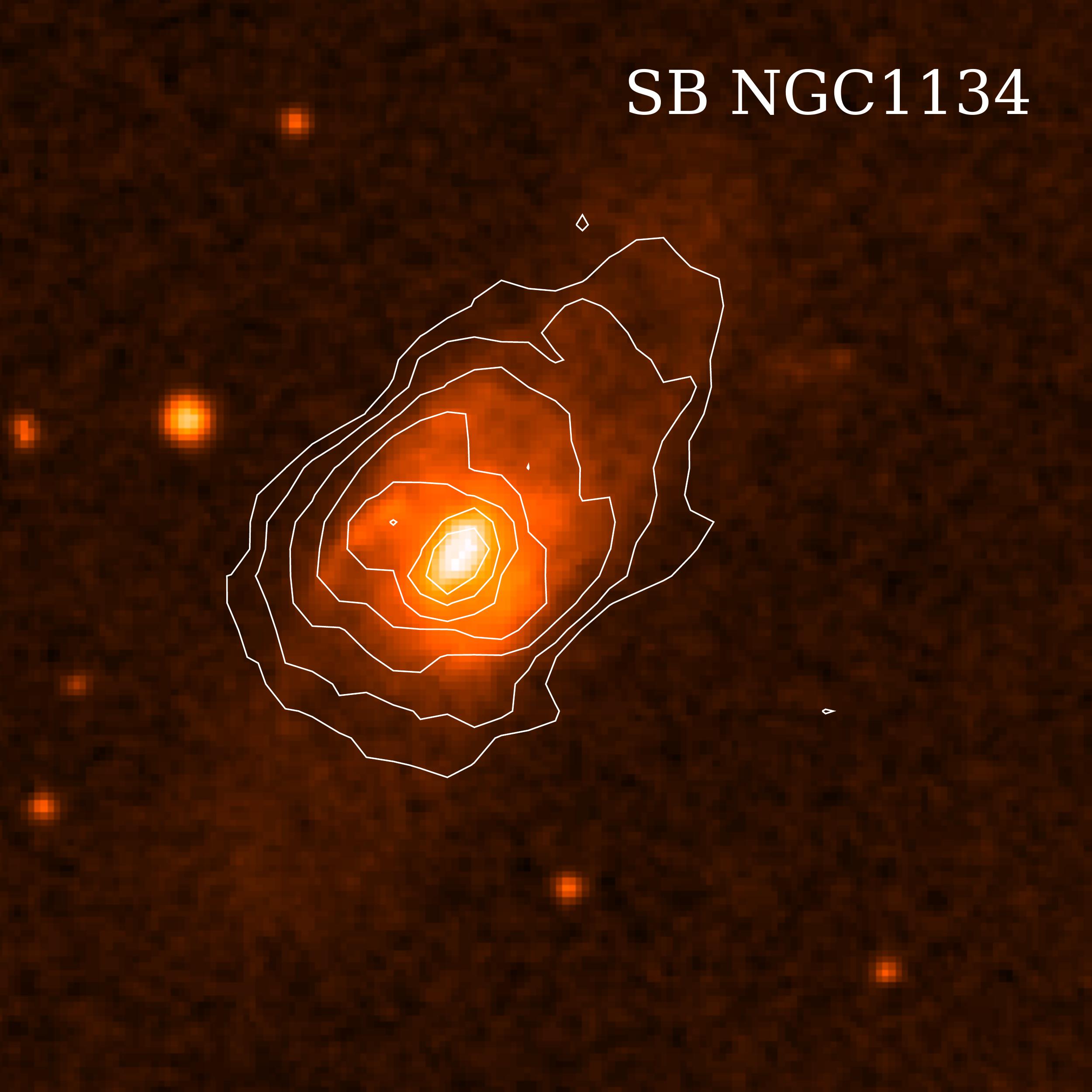}
\includegraphics[width=5.7cm,trim = 0 0 0 0 ]{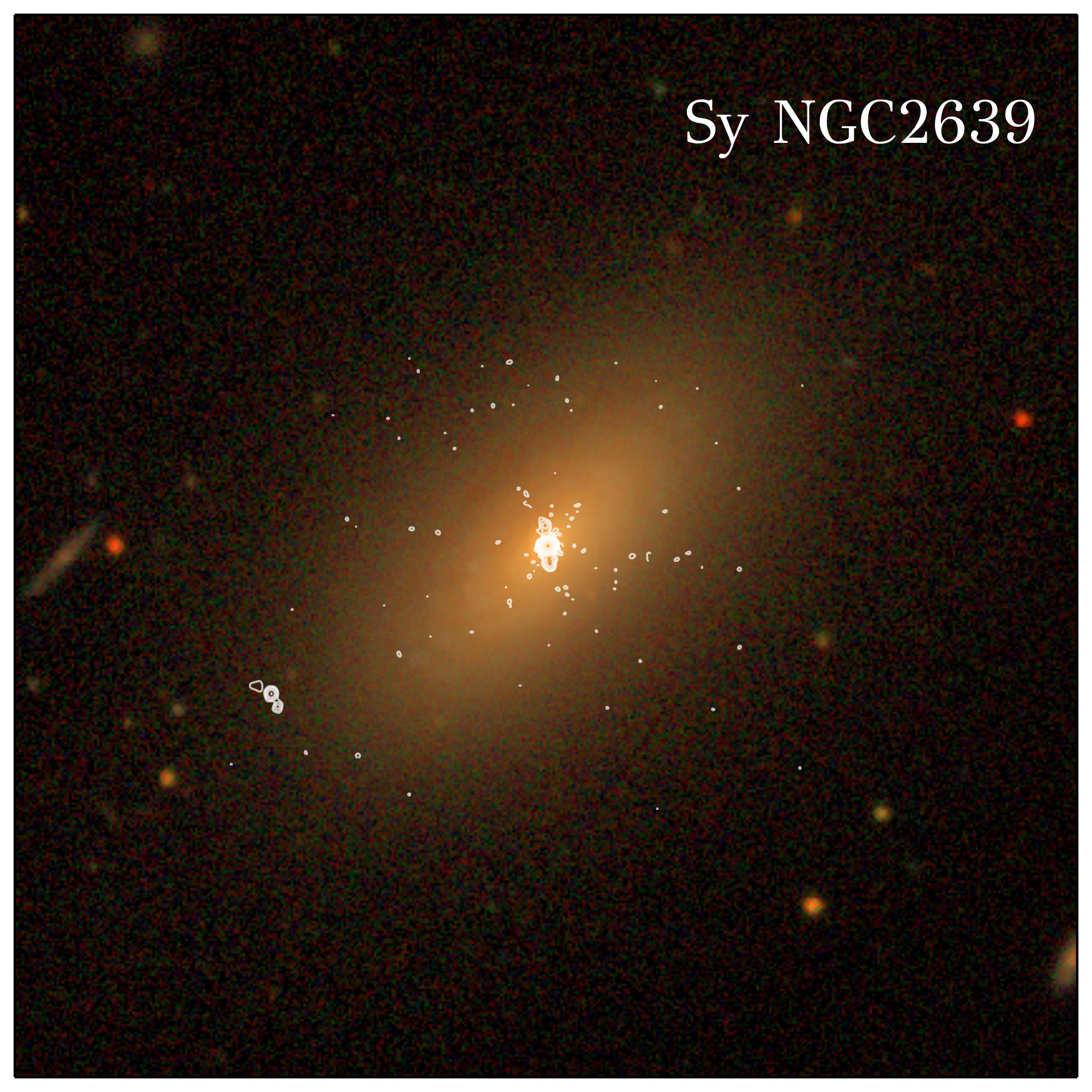}
\includegraphics[width=5.7cm,trim = 0 0 0 0 ]{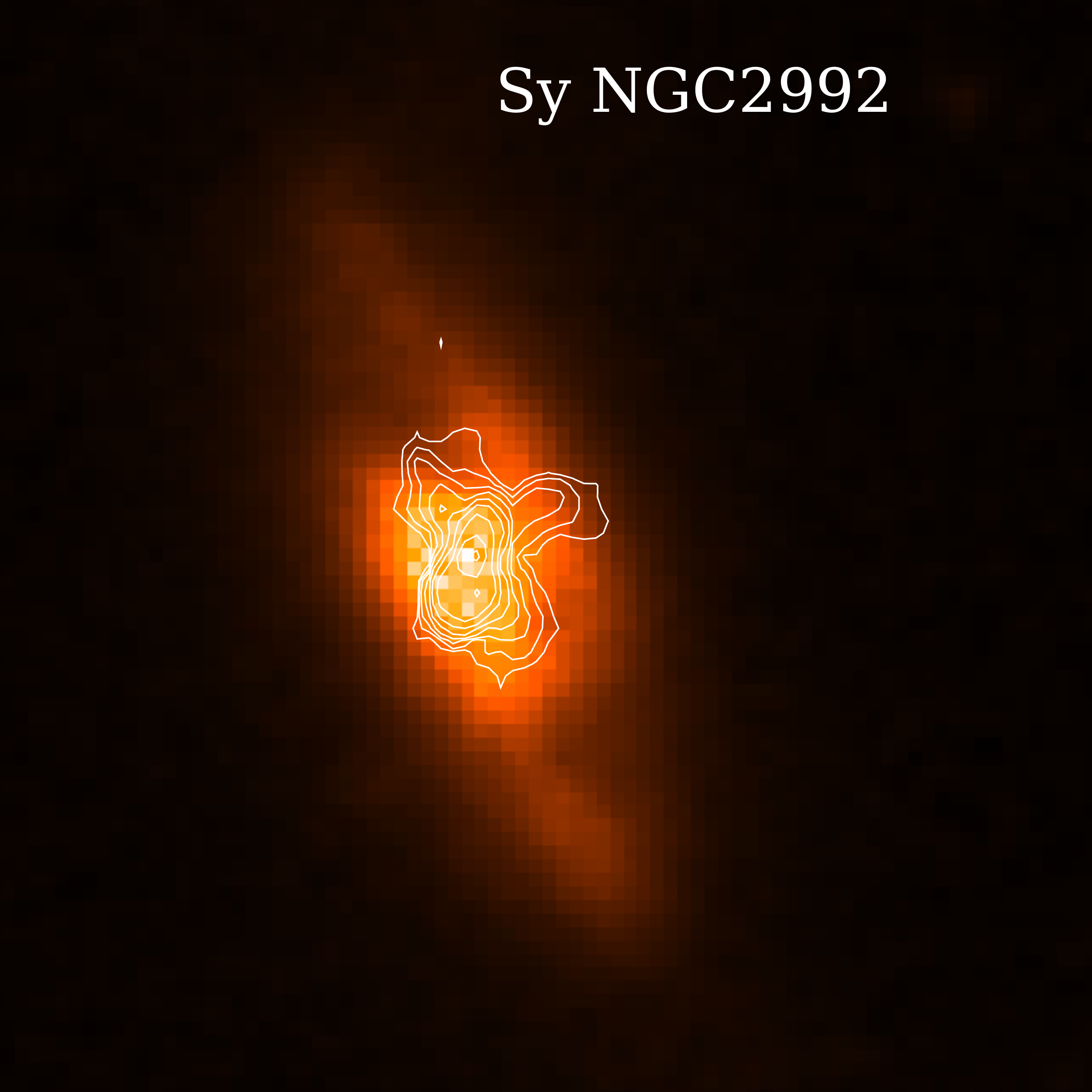}
\includegraphics[width=5.7cm,trim = 0 0 0 0 ]{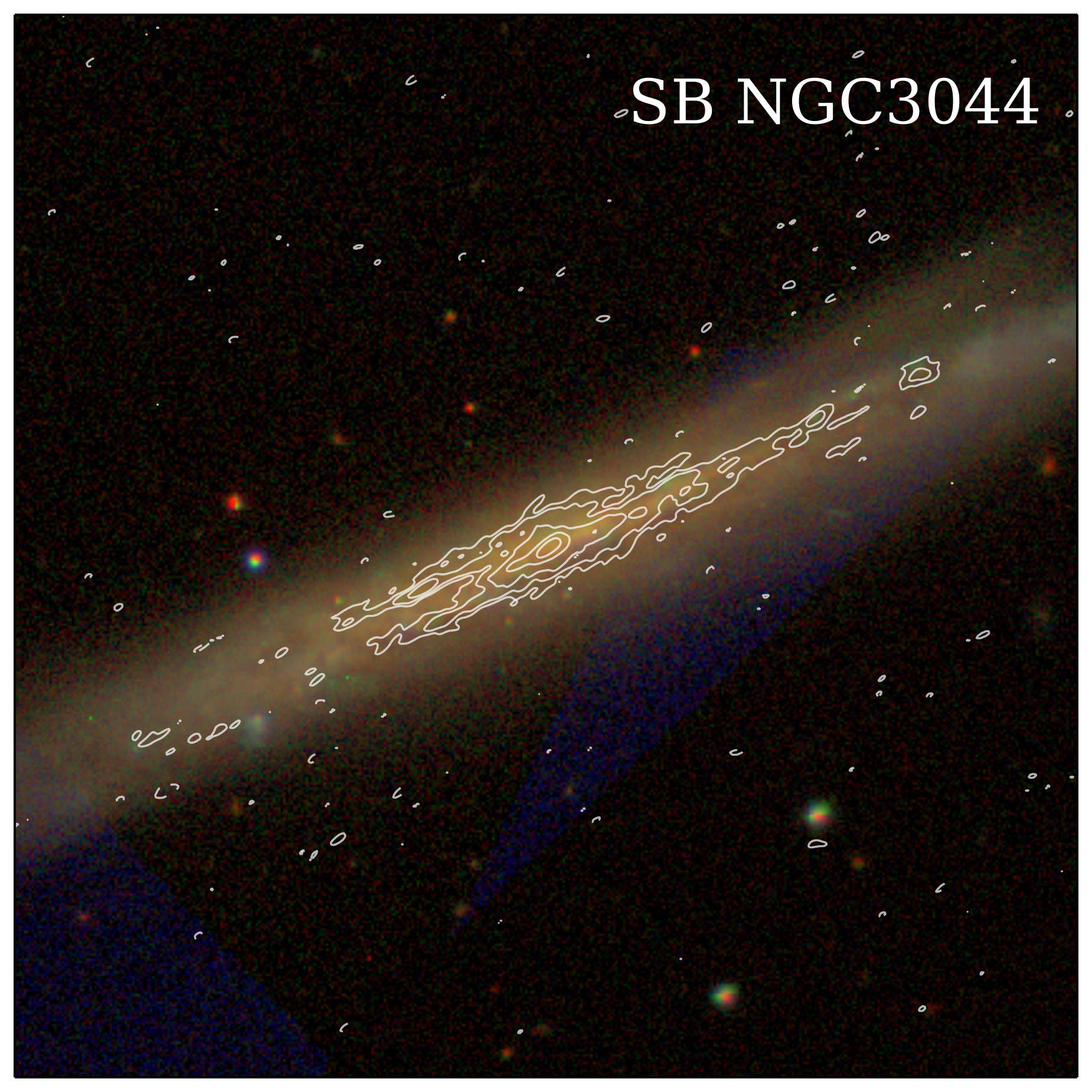}
\includegraphics[width=5.7cm,trim = 0 0 0 0 ]{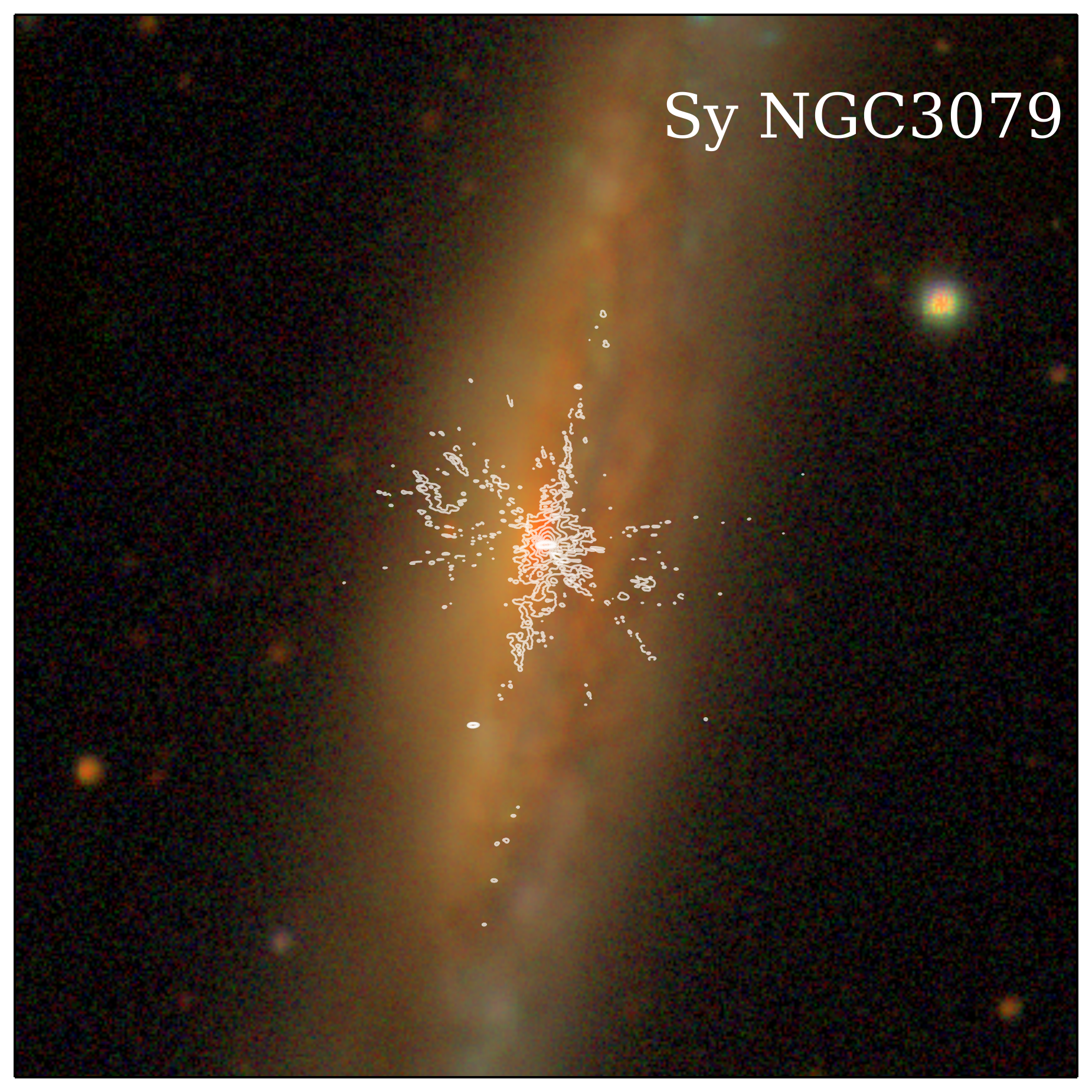}
\includegraphics[width=5.7cm,trim = 0 0 0 0 ]{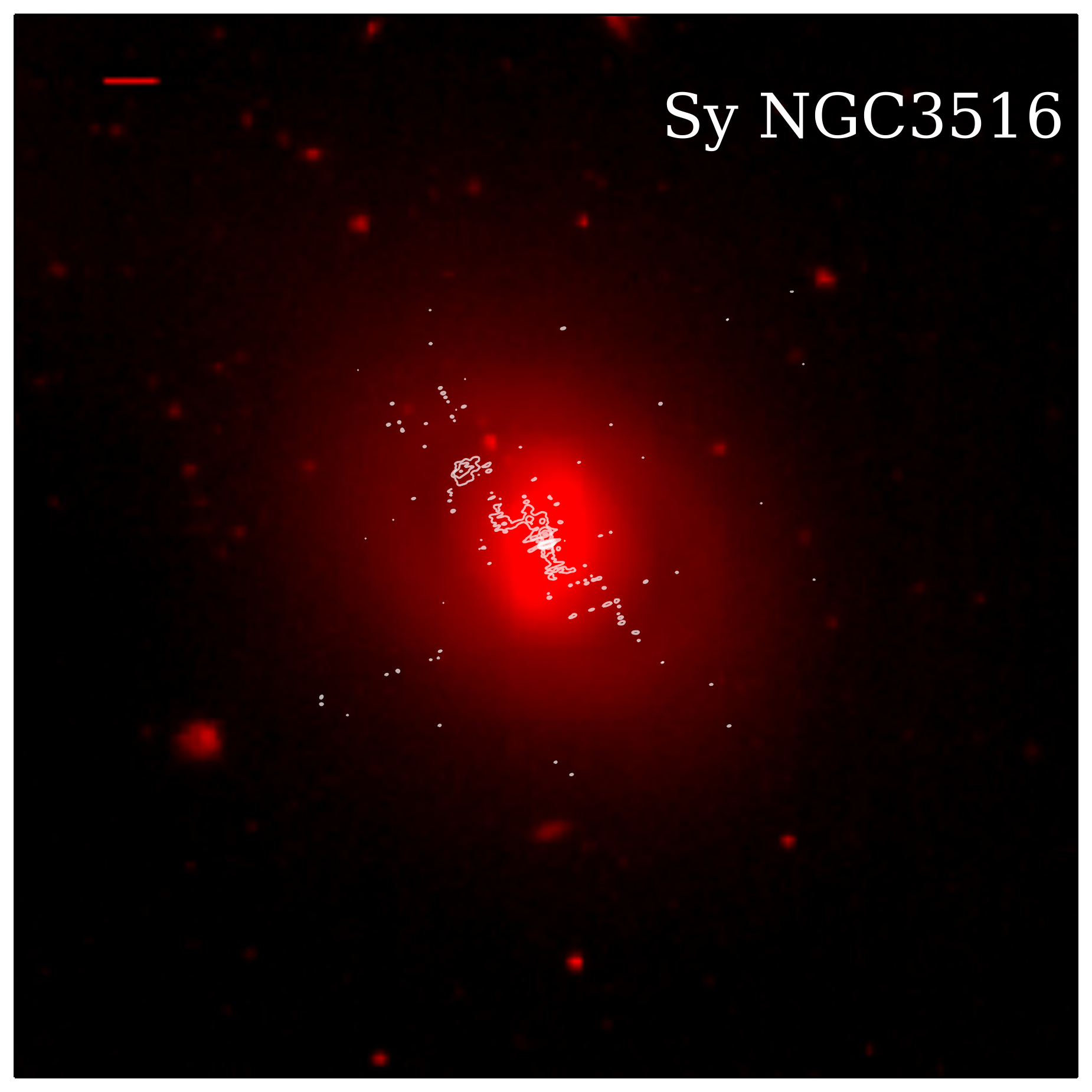}
\includegraphics[width=5.7cm,trim = 0 0 0 0 ]{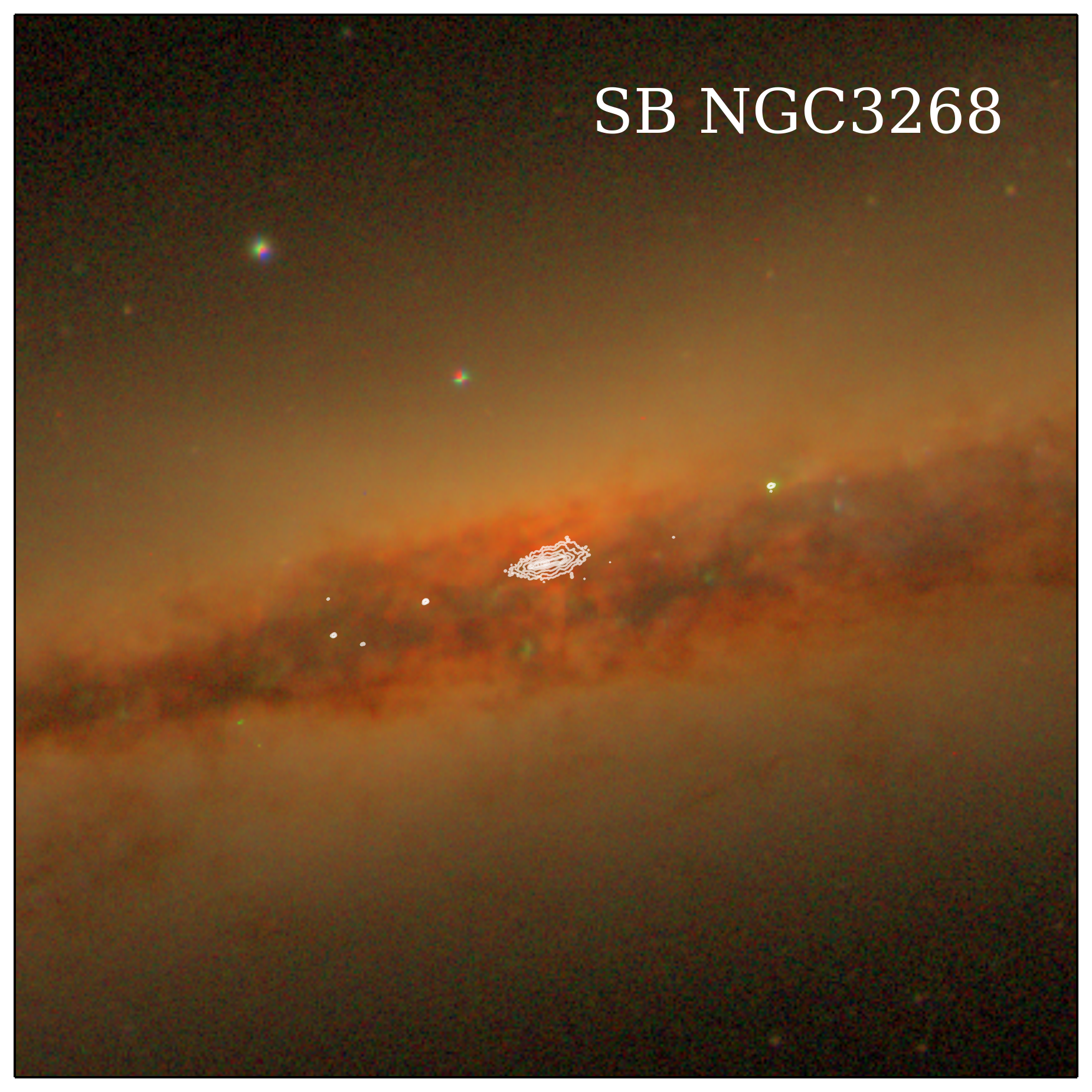}
\includegraphics[width=5.7cm,trim = 0 0 0 0 ]{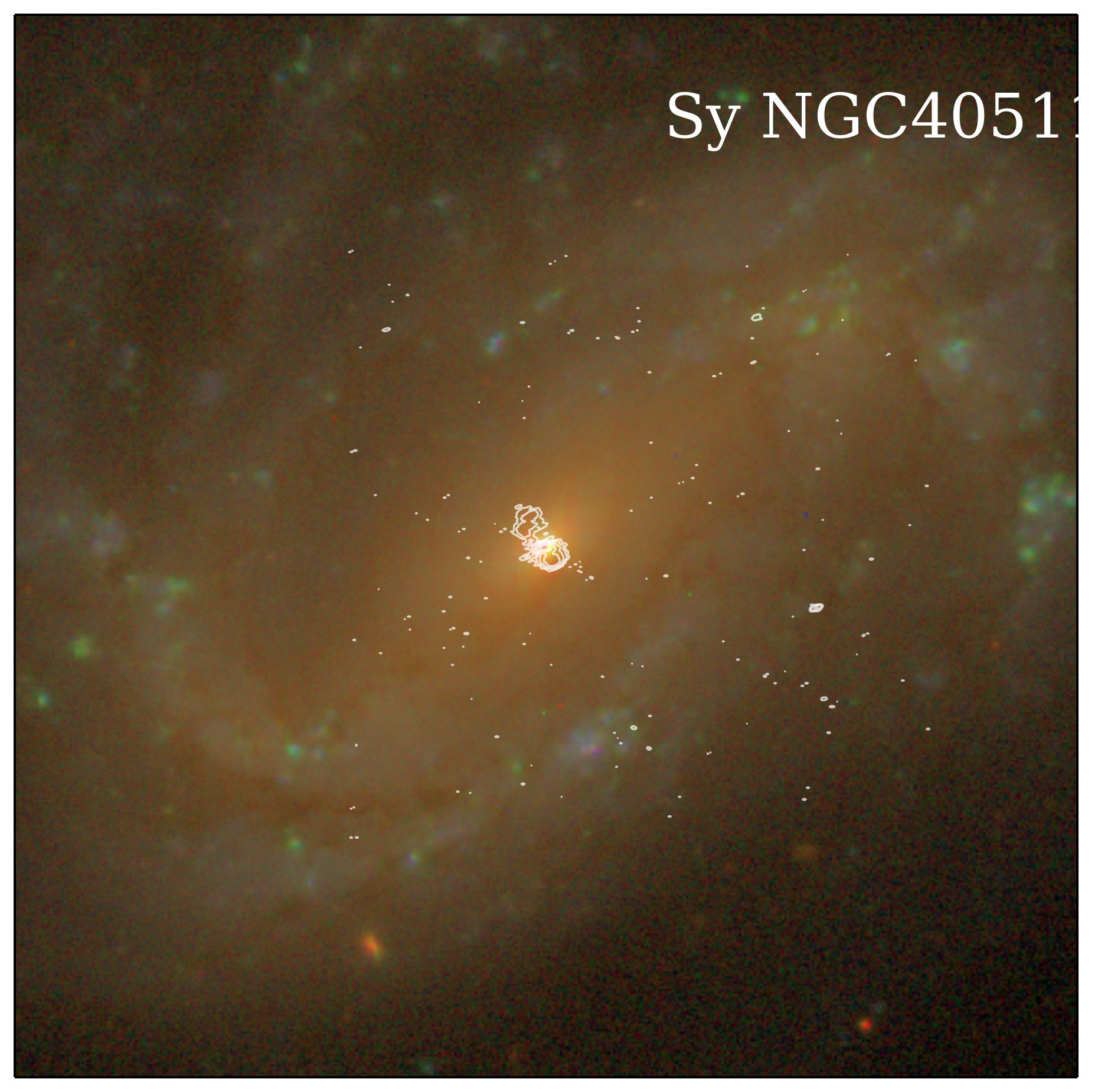}
\includegraphics[width=5.7cm,trim = 0 0 0 0 ]{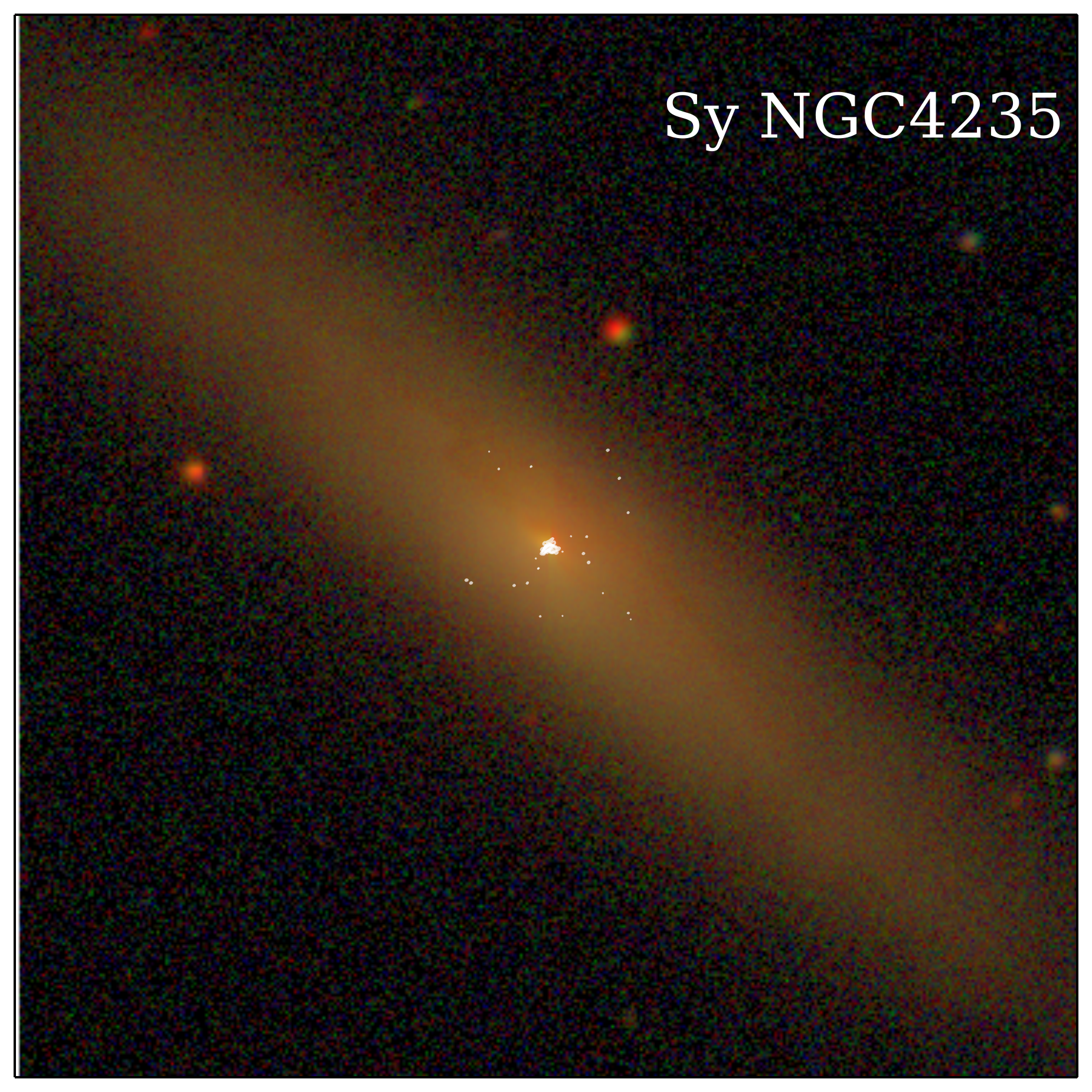}
\includegraphics[width=5.7cm,trim = 0 0 0 0 ]{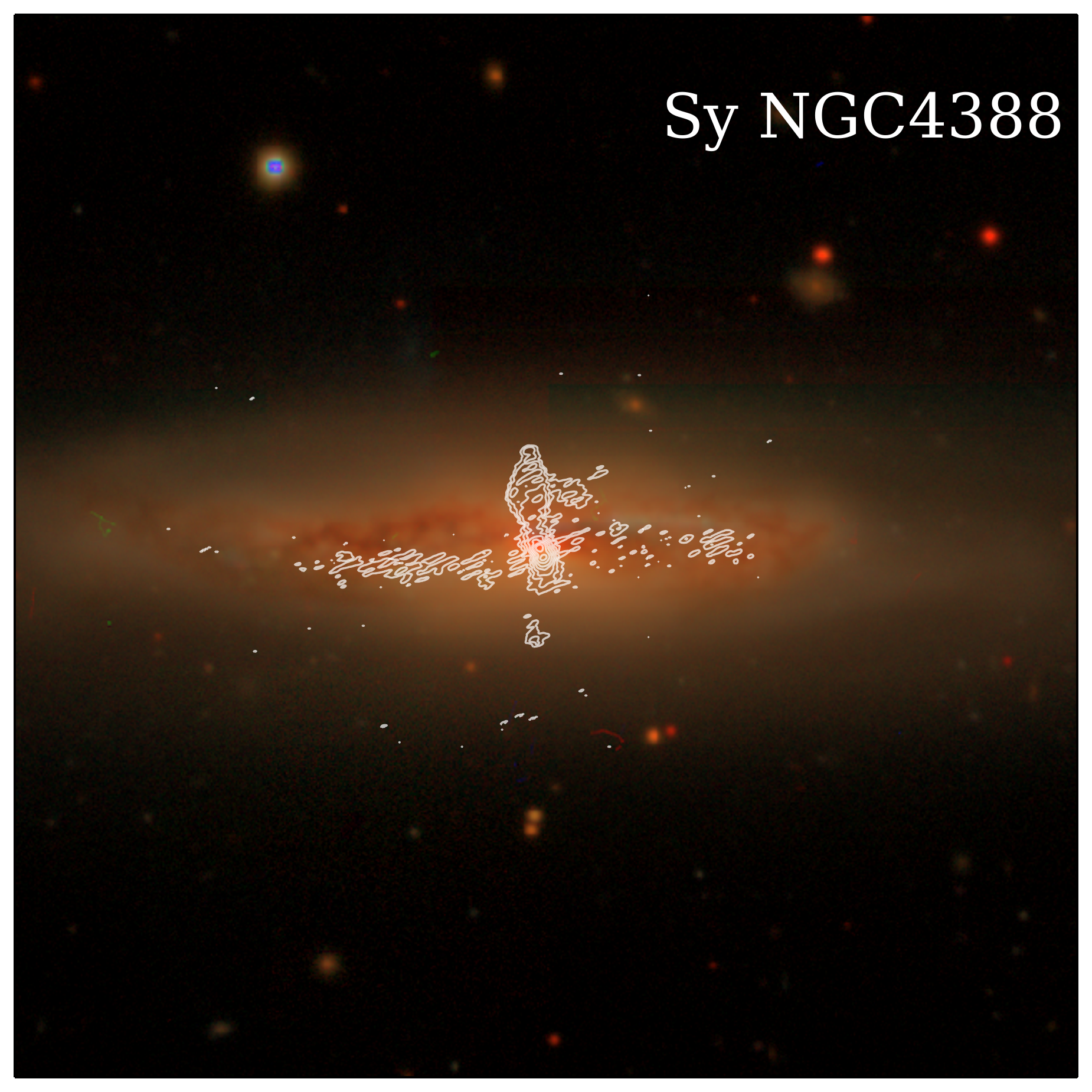}
\includegraphics[width=5.7cm,trim = 0 0 0 0 ]{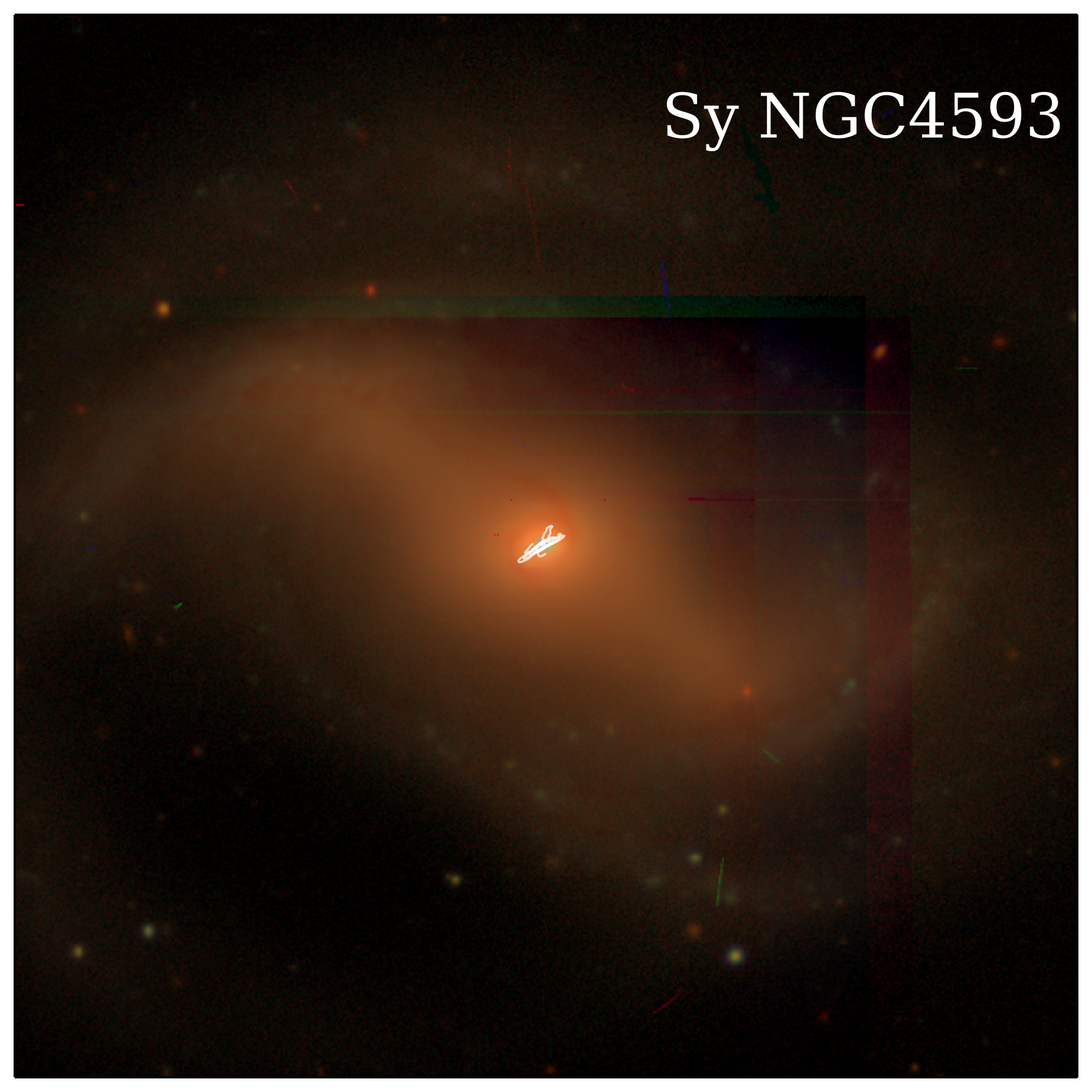}

\caption{Radio total intensity emission overlaid in white contours over the optical color composite images. The optical images are taken from various optical surveys namely, DECaLS, MzLS, SDSS and DSS depending on the availability of data in these surveys. We have used only the DSS r-band image to make the overlays of NGC\,253, NGC\,1134 and NGC\,2992 due to the poor quality of the color composite image.}
\label{optical_overlay}
\end{figure*}
\begin{figure*}
\ContinuedFloat
\includegraphics[width=5.7cm,trim = 0 0 0 0 ]{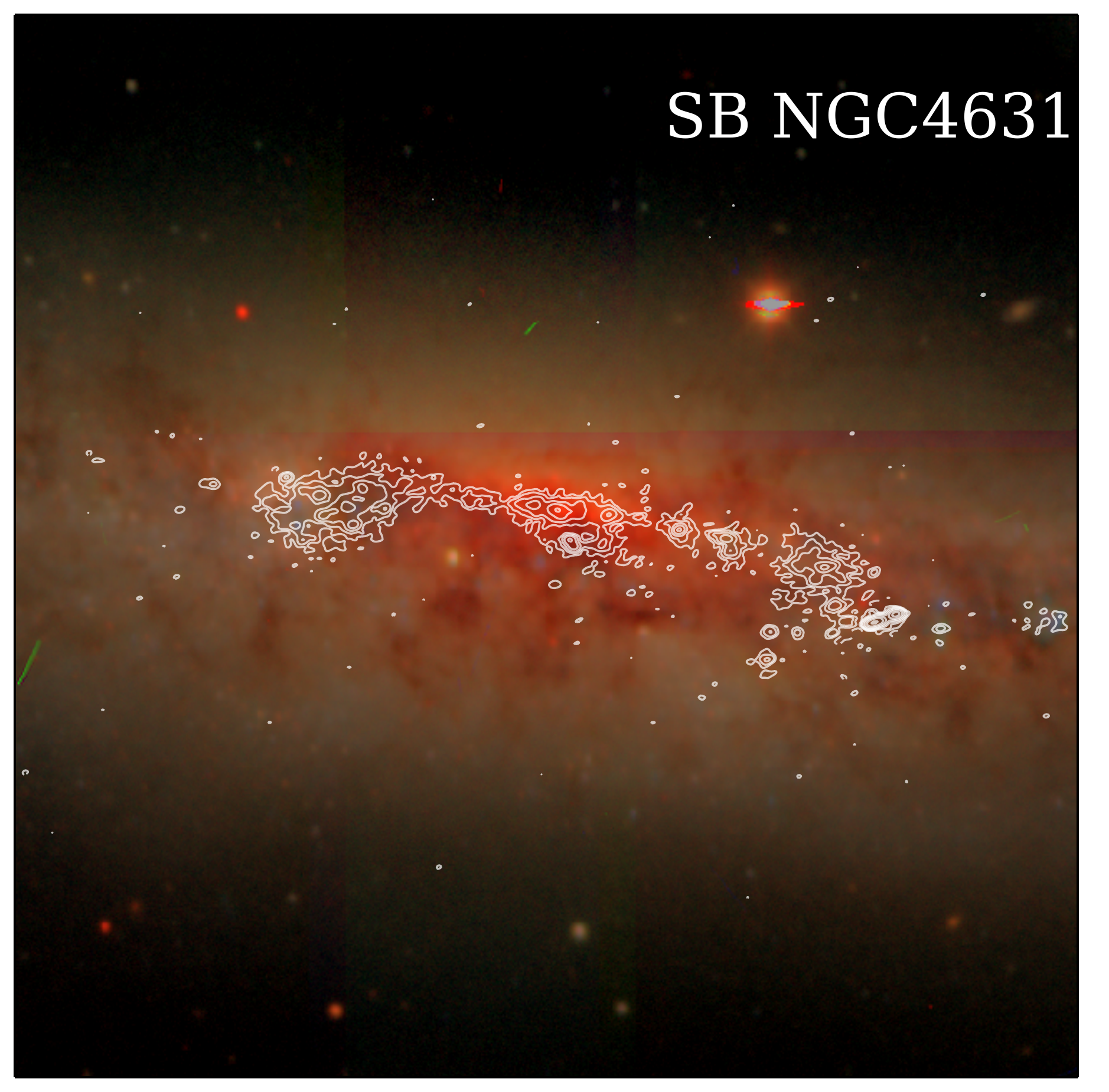}
\includegraphics[width=5.7cm,trim = 0 0 0 0 ]{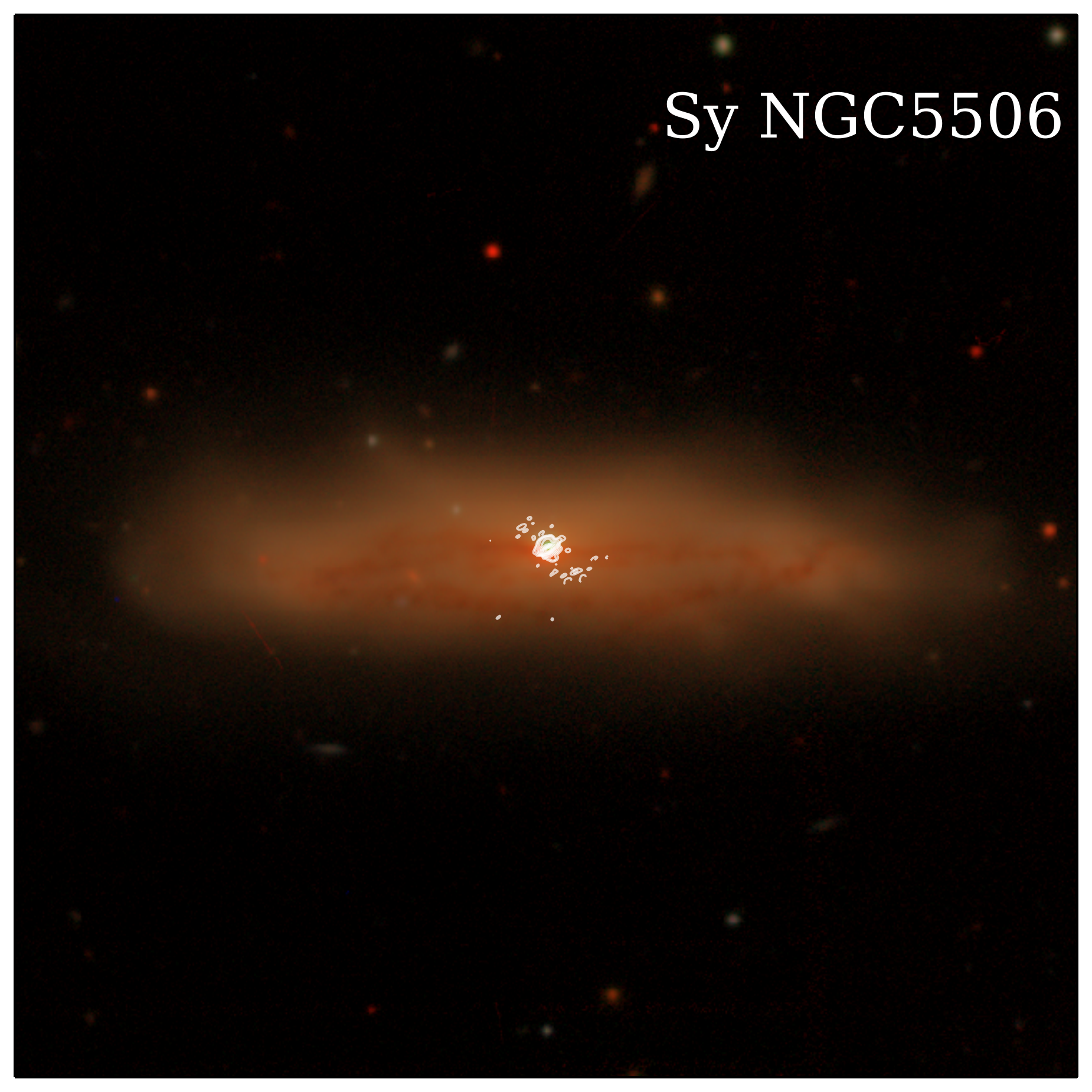}
\includegraphics[width=5.7cm,trim = 0 0 0 0 ]{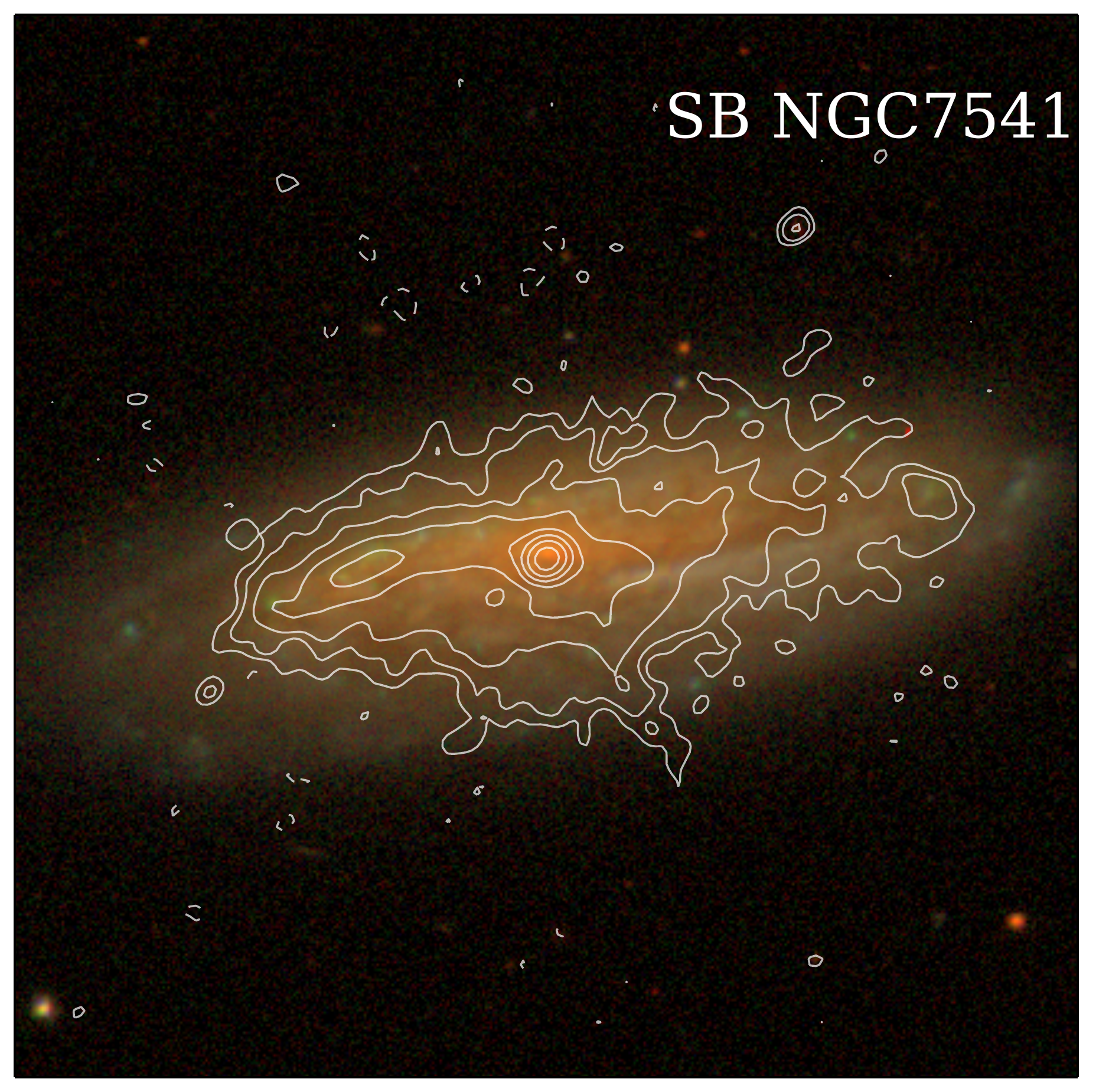}
\includegraphics[width=5.7cm,trim = 0 0 0 0 ]{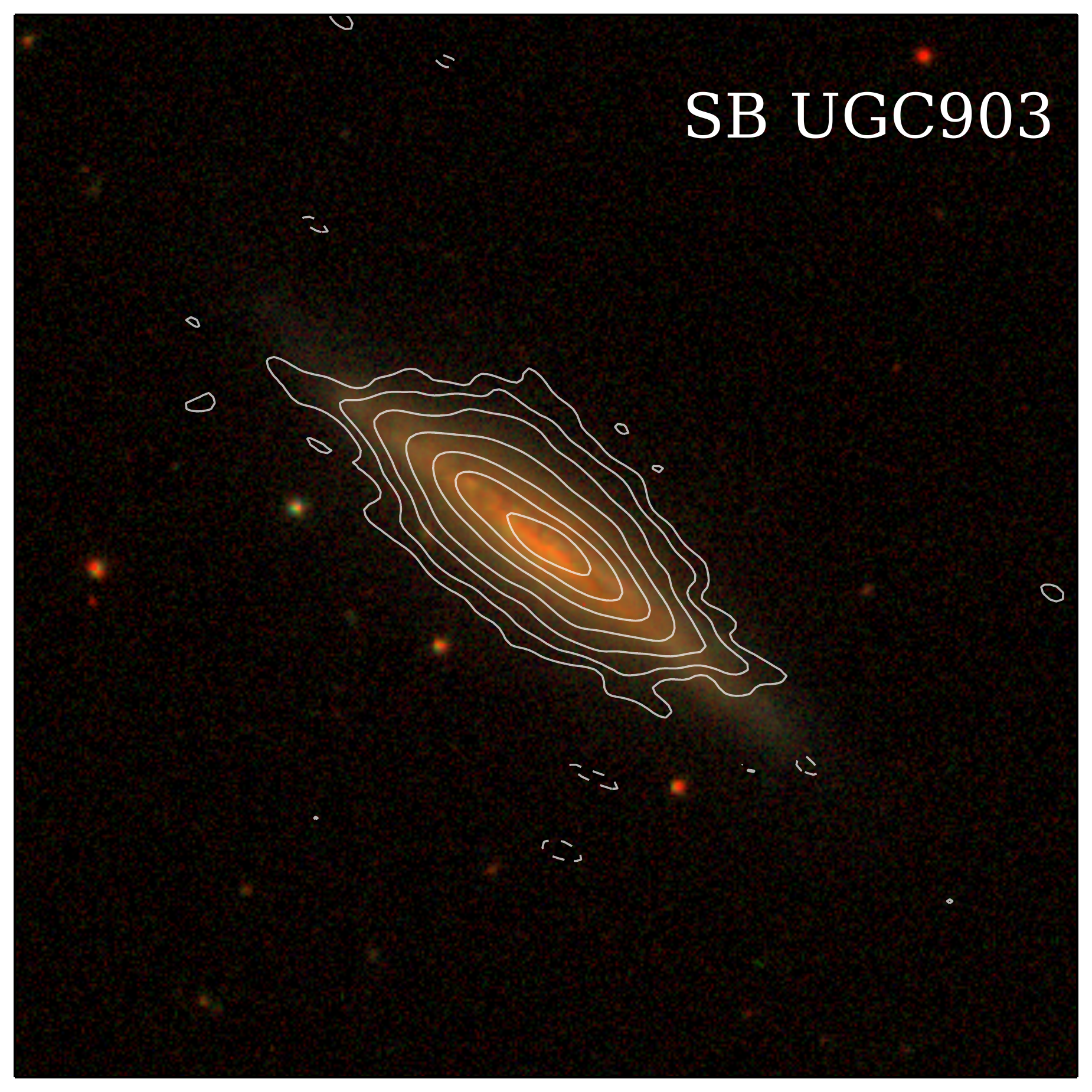}
\caption{Continued--Radio total intensity emission overlaid in white contours over the optical color composite images.}
\label{optical_overlay2}
\end{figure*}

The radio total intensity contours overlaid on the optical images are shown in Figures~\ref{optical_overlay}. It can be noticed that the radio continuum emission in starburst galaxies mostly follows the optical disk star formation regions, whereas the Seyfert galaxies most often show either lobe-like features without any corresponding optical features or nuclear core-dominated emission along with emission from the star-forming disk.
The diffuse lobe-like structures in Seyfert galaxies are very different in morphology from those seen in both FR\,I, and FR\,II radio-galaxies. NGC\,2639, NGC\,2992, NGC\,3079, NGC\,4051, and NGC\,4388 show lobe-like structures with a pinched in morphology near the AGN core. Most of them appear brighter at the farthest edges and sometimes show S-shaped symmetry like in the case of NGC\,4051 and NGC\,3079. NGC\,3516 also shows S-symmetry although, the morphology appears somewhat intermediate between bulbous shaped lobes of a Seyfert galaxy and an FR\,II type radio galaxy showing hotspot and lobes.

On the other hand, most cases of galaxies dominated by star-formation show top open bi-conical outflow like morphology rather than a dome-shaped appearance. For example prototypical starburst galaxies like NGC\,253 \citep{heesen2011} and M\,82 \citep{adebahr2013} show bi-conical radio emission which is spatially aligned and correlated with X-ray and optical line emission. Such correlated emission was explained to be a result of the entrainment of the cosmic ray electron population from the galactic disk and the frozen-in magnetic fields along with the outflows which are traced using X-ray or emission-line gas \citep{suchkov1994,suchkov1996,cecil2001,adebahr2013}.
\cite{krause2020} through a stacking analysis find that the linear polarization in star-forming galaxies is X-shaped in morphology. This X-shaped morphology has been attributed to an interplay of the galactic outflows and the dynamo action of the host galaxy itself.

An attractive possibility to explain the bulbous-shaped morphology is to invoke the dynamical influence that the galaxy may have in the evolution of the Seyfert jets. A simple hydrodynamical model would favor a frustrated jet model, according to which the intrinsically low-power jets gets disrupted by the dense ISM percolating these late-type galaxies, and expands into the ambient medium along the minor axis, which offers the least resistance \citep{gallimore06}. 
However, such a model is rendered insufficient to explain the closed dome-like morphology, for which the large-scale magnetic fields that thread the galaxy was invoked \citep{sebastian2019,henriksen2019}, because some of the solutions of a force-free magnetic field of the galaxy resemble the morphology of NGC\,3079.

But even in this case, one would expect alignment along the minor axis. While this is true for some cases, like NGC\,3079, NGC\,4388, etc there are many where this is not the case, for example, NGC\,2639 and NGC\,3516.   
More importantly, Seyfert galaxies like Mrk\,6 which show this closed dome-type morphology in several directions misaligned with each other \citep{Kharb06} pose a problem for the simple picture of galactic magnetic fields being responsible for the dome-like structure.

Hence, we conclude that the peculiar dome-type morphology of the radio emission seen in Seyfert galaxies is linked to the active nuclei itself although the environmental influences might still be playing a role. The difference in the alignment itself can be produced as a result of multiple epochs of activity \citep{Kharb06,sebastian2019}.
In the following subsections, we try to understand the individual contributions from the AGN accretion, jet, and the star formation to the kpc-scaled radio emission.

\subsection{Correlation between star-formation and radio emission}

\begin{figure}
\includegraphics[height=7.15cm,trim = 0 0 0 0 ]{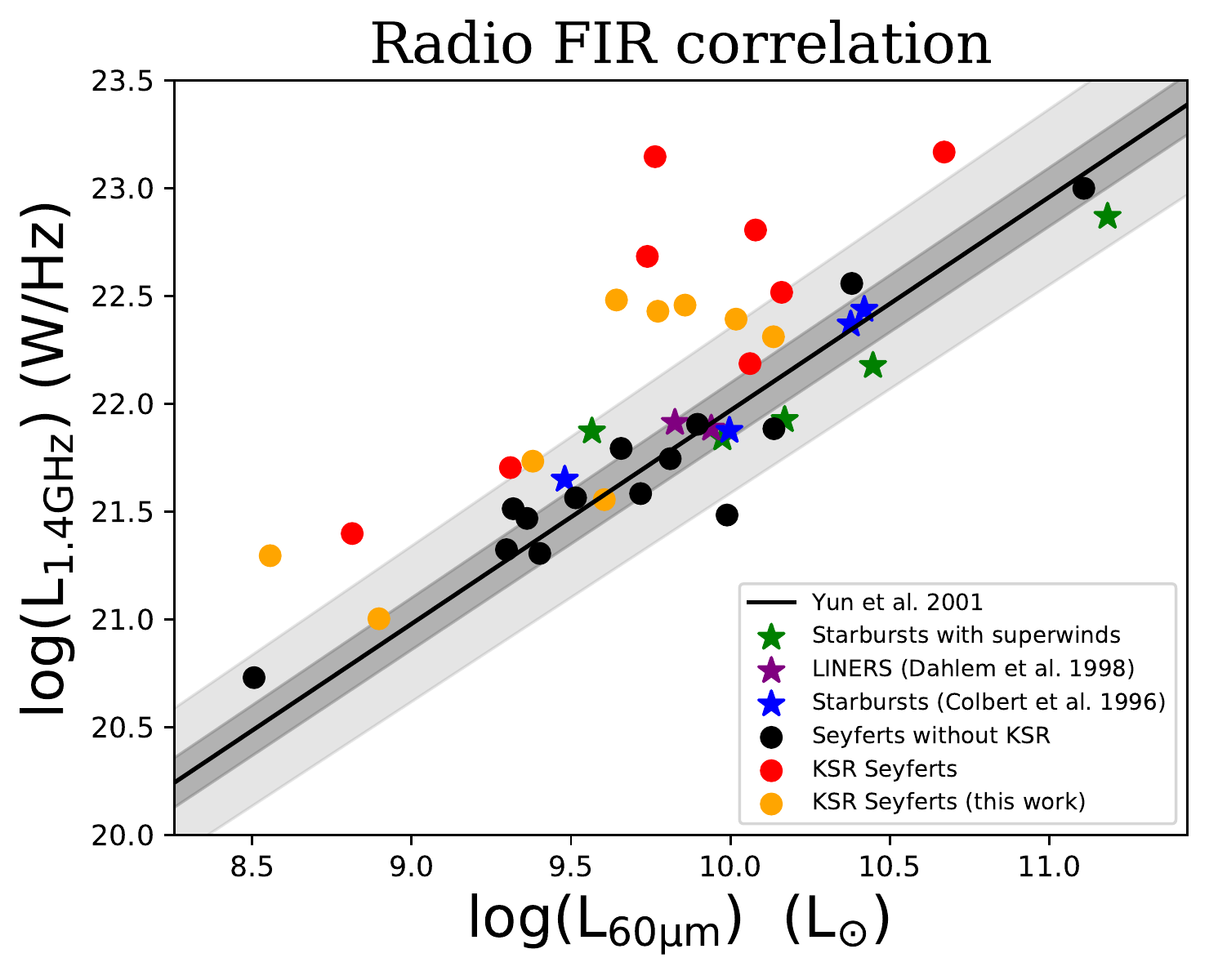}
\caption{Plot of FIR luminosity at 60 $\mu$m vs the 1.4 GHz luminosity in W~Hz$^{-1}$. The black line shows the radio-FIR correlation whereas the region shaded using dark grey and light grey represents the 1$\sigma$ and 3$\sigma$ boundary limits of the radio-FIR correlation from {\protect\cite{yun2001}}.}
\label{radiofir}
\end{figure}

The radio-FIR correlation is one of the tightest astrophysical correlations known \citep{Condon92}. It is proposed that the radio emission and the relativistic plasma seen in Seyfert galaxies arise due to mainly three reasons (i) jet-related, (ii) winds and the resulting in-situ acceleration of the thermal electrons to relativistic energies at shocks via diffusive shock acceleration \citep{blandford1987} and/or (iii) star-formation-related. Any emission related to either of the former two reasons should show deviation in the form of a radio-excess from the radio-FIR correlation. It is however not clear if the shocks created due to winds like those by starburst galaxies/ AGN accretion are powerful enough to generate significant radio emission which is comparable to that generated by supernovae/ H\,II regions which result in the radio-FIR correlation. A simple observational test is to examine whether a radio excess is seen in starburst galaxies which show evidence for superwinds but do not host an AGN such that there is no contamination from jets. 

In figure~\ref{radiofir}, we have plotted the radio luminosity at 1.4~GHz as a function of the far-infrared luminosity at 60$\mu$m. We have also plotted the black line corresponding to the radio-FIR correlation as derived by \cite{yun2001}. We obtained the total radio flux densities from the NVSS images at 1.4~GHz, and the IR flux densities from the IRAS catalog \citep{fullmer1989}. The dark grey and the light grey shaded regions represent the 1$\sigma$ and 3$\sigma$ deviations from the radio-FIR correlation respectively \citep{yun2001}. The dots represent the Seyfert galaxies, whereas the stars represent starburst galaxies. We increased the sample size of Seyfert galaxies by including more sources that showed radio flux densities above 10.0 mJy in NVSS from the sample presented in \cite{gallimore06}. The red and black dots represent Seyfert galaxies that host KSR and that without much evidence for KSR respectively. 

The starburst galaxies chosen from the \cite{colbert1996b} represented by blue stars were chosen such that it matches the radio-luminosity of Seyfert galaxies in our sample. 
The three starburst galaxies from \cite{dahlem1998} showed diffuse X-ray emission which was interpreted as evidence for superwinds. To increase the sample size of superwind-hosting starburst galaxies, we further chose four more superwind hosting galaxies from \cite{heckman1990}, which did not show any evidence for an AGN at its core, namely M\,82, NGC\,4194, NGC\,1222 and NGC\,1614. The purple and green stars represent the starburst galaxies hosting superwinds taken from both \cite{heckman1990} and \cite{dahlem1998}. However, the purple stars are the galaxies from \cite{dahlem1998} which showed LINER nuclei, although the origin of this LINER emission is not indisputably AGN related (see Section~\ref{253not} and \ref{3628note}). Moreover, they follow the radio-FIR correlation within 1$\sigma$ error bar.

It is evident from figure~\ref{radiofir} that most Seyfert galaxies with KSR show excess radio emission in comparison to the radio-FIR correlation, with 13/17 Seyfert galaxies having radio flux densities above the 3$\sigma$ limits. However, most of the Seyfert galaxies without any KSR are consistent with the radio-FIR correlation. This was previously noted by \cite{gallimore06} as well.
The starburst galaxies with superwinds are consistent with the radio-FIR correlation and often lies below the correlation. Hence, it is clear that shocks generated due to the supersonic velocities of the winds typically seen in these systems are not powerful enough to create radio emission comparable with that related to star-formation itself. 
Hence, any excess radio emission seen in Seyfert galaxies can be conclusively attributed to the jets or magnetically launched winds by an AGN present rather than radiatively/thermally driven winds that leads to radio emission via shock acceleration. 

\subsection{Investigating the correlation of radio emission with NLR properties}

The correlation between the narrow line region and radio emission in Seyfert galaxies, in particular, that between the [O~{\small III}] line luminosity, its FWHM, and the radio luminosity has been known for a while \citep{debruyn1978,heckman1981,whittle1985,xu1999}. These emission-line regions were found to be spatially aligned with the extended radio emission and sometimes surrounding the radio emission \citep{haniff1988,capetti1996,falcke1998}.
Extended emission-line regions were also found extending up to tens of kpcs were found to be spatially associated with large scale radio emission in radio galaxies as well \citep{fosbury1986,baum1989}. Moreover, such extended emission-line regions are more frequently found in steep spectrum sources rather than flat-spectrum sources. 
\cite{leipski2006} found that the size of the radio sources in radio-quiet quasars correlates well with the size of the NLR.
All these pieces of evidence are suggestive of a scenario where the material in the NLR is ionized as a result of shocks generated by the relativistic radio plasma interacting with the gas around it.

Several other studies preceded by \cite{veilleux1991a,veilleux1991b} also revealed a correlation between the narrow line region kinematics and the radio luminosity.
In a recent study, \cite{zakamska2014} found that the [O~{\small III}] line-width containing 90\% of the emission line flux correlates well with the radio luminosity in these systems. The radio emission was interpreted to be a result of the shock acceleration that happens due to the supersonic velocities of the outflows driven by the accretion in these systems rather than a jet.

Seyfert galaxies, which are also mostly radio-quiet in nature, prove to be ideal candidates that can be studied in detail to test both these scenarios, namely the jet origin of the radio emission which then shock ionizes the medium or the radiation pressure -driven wind origin which is responsible for both the radio emission via shocks and the NLR emission which constitutes the photo-ionized wind.

\begin{figure*}
\includegraphics[width=8.5cm,trim = 0 0 0 0 ]{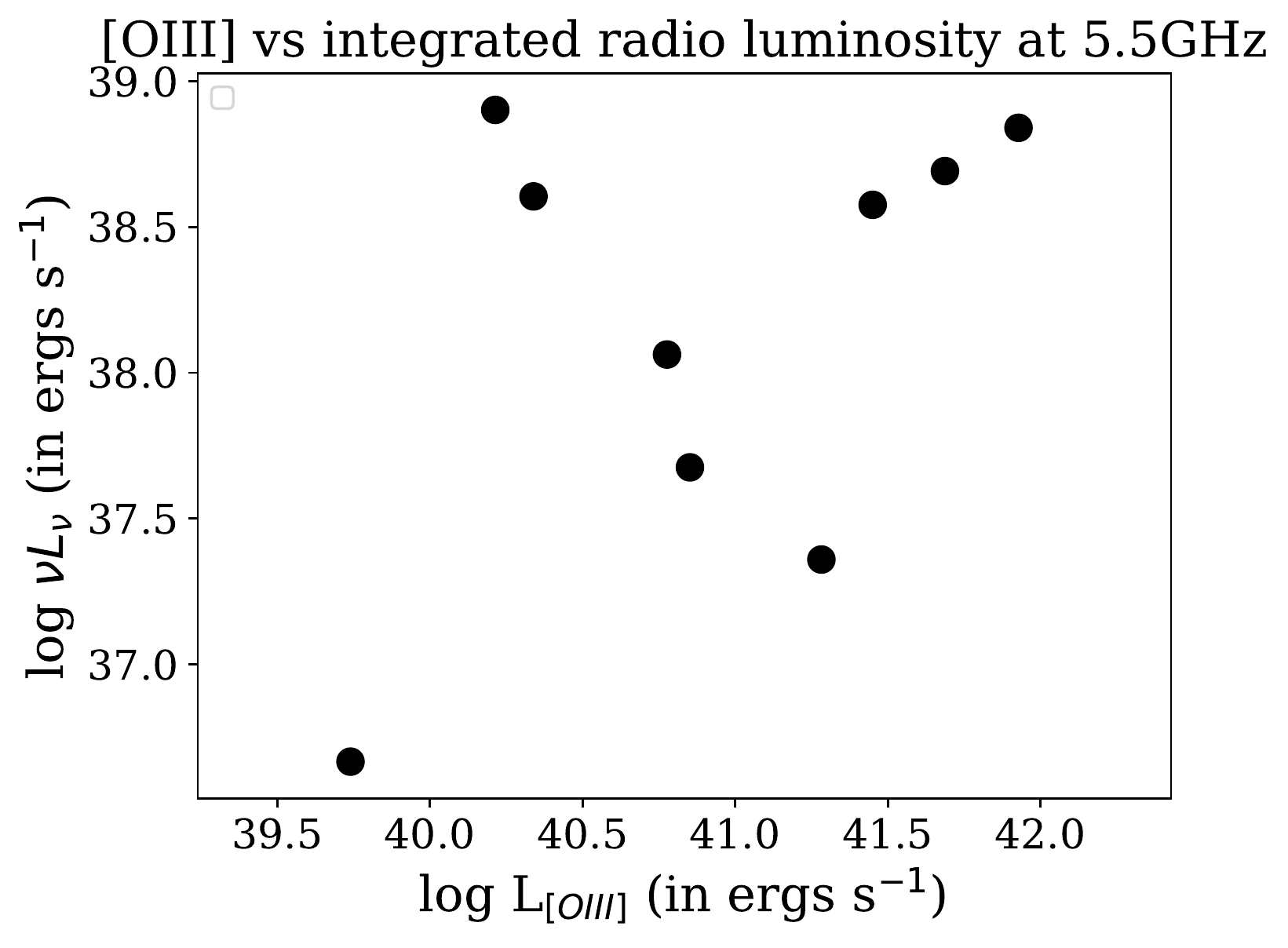}
\includegraphics[width=8.5cm,trim = 0 0 0 0 ]{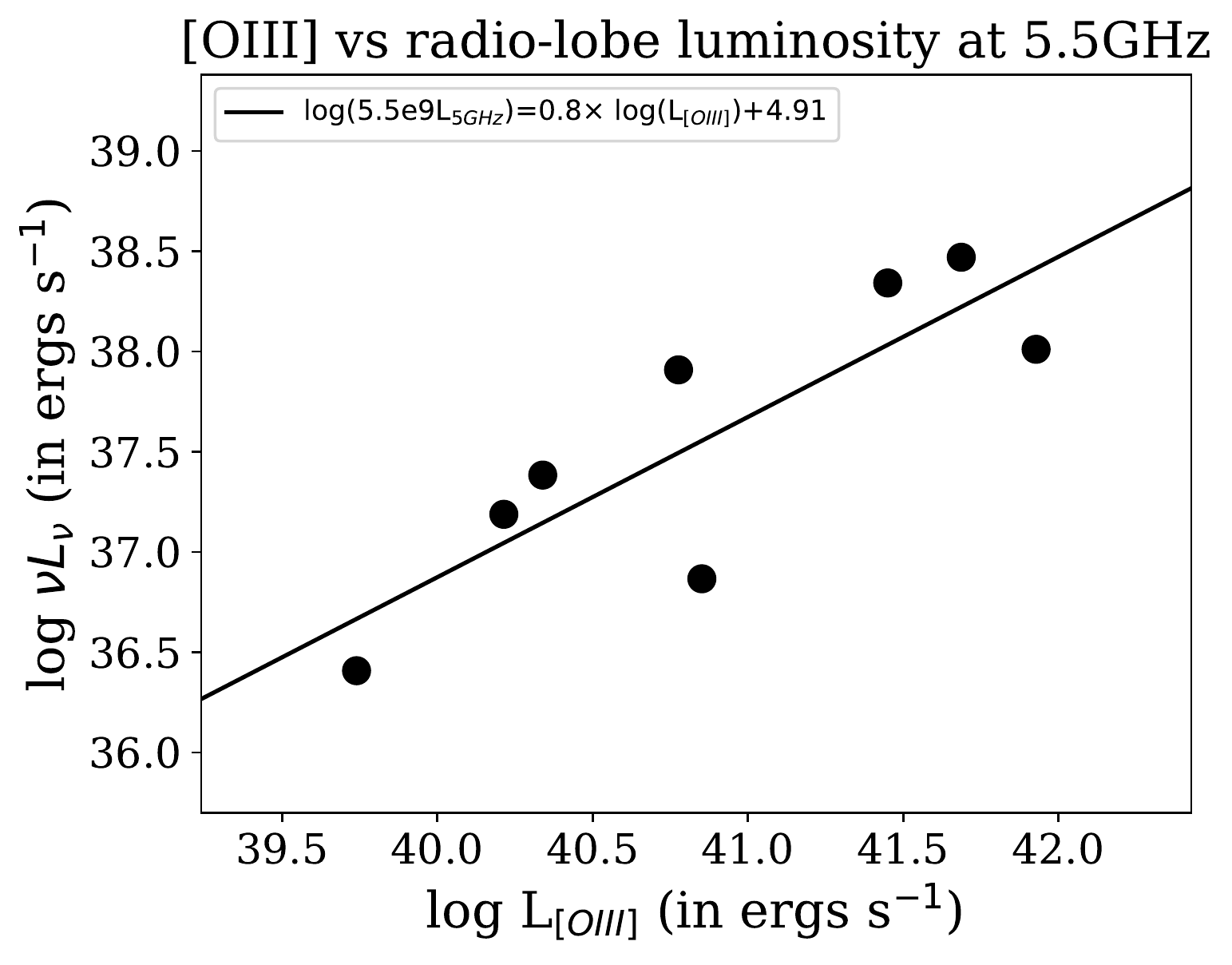}
\caption{Left: [O~{\small III}] line luminosity versus the integrated radio luminosity at 5.5~GHz. Right: [O~{\small III}] line luminosity versus the radio lobe luminosity at 5.5~GHz.}
\label{oiicorrelations}
\end{figure*}

In this paper, we study how the radio luminosity in our sample of Seyfert sources correlate with the [O~{\small III}] line luminosity. We obtained the [O~{\small III}] line luminosities of our Seyfert galaxies from the compilation presented in \cite{malkan2017}. We also obtained the H$\alpha$ and H$\beta$ line luminosities from \cite{malkan2017} to correct the [O~{\small III}] line luminosities for extinction following \cite{bassani1999} and assuming a Balmer decrement of 3.0 \citep{osterbrock2006}. Figure~\ref{oiicorrelations} shows the plot of [O~{\small III}] line luminosity plotted against the integrated radio luminosity ($\nu L_ {\nu}$) at 5.5 GHz including the extended lobes, bright core and galaxy emission in the left panel, whereas in the right panel we have plotted the radio luminosity exclusively coming from the extended radio lobes. We defined regions to avoid the central unresolved core and any emission which is related to the host galaxy disk. In NGC\,4593, we do not detect any extended emission and hence we did not include this source in the plot shown in the right panel. We avoided the radio emission from the galactic disk in the NGC\,3079, and NGC\,4388 while estimating the lobe luminosity.

Interestingly, we do not find a statistically significant correlation between radio luminosity ($\nu L_ {\nu}$) and [O~{\small III}] luminosity ($\mathrm{L_{[O~{\small III}]}}$) when we use the integrated source radio luminosity. The Spearman's rank correlation coefficient is 0.3 with a two-tailed p-value of 0.43. On the other hand, when we use only the extended radio emission and avoid the core contamination, we find that there is a significant correlation between $\nu L_ {\nu}$ and $\mathrm{L_{[O~{\small III}]}}$. The corresponding Spearman's rank correlation coefficient is 0.79 and the two-tailed p-value turns out to be 0.021. We used the least-square fitting to determine the slope of the relation between the extended radio emission and the line luminosity and the relation between the $\nu L_ {\nu}$ and $\mathrm{L_{[O~{\small III}]}}$ is quoted in Figure~\ref{oiicorrelations}.

It must be noted that there are several uncertainties in the extended flux density estimation. Our observations were carried out using the VLA B-array configuration as a result of which we might be missing flux from very diffuse structures. Also, the contamination from the star-formation related radio emission can not be completely accounted for, although efforts to avoid radio emission from location coinciding with the disk were taken. Despite these caveats, there appears to be a strong correlation between the radio luminosities from lobe-like features and the $\mathrm{L_{[O~{\small III}]}}$.

It was previously noted by \cite{rawlings1989} and \cite{baum1989} that there is a correlation between 178 MHz radio luminosity (dominated by extended radio emission) and the line luminosity whereas there is not much evidence for an intrinsic correlation between 5~GHz radio emission (often core dominated) and the NLR emission.

The [O~{\small III}] line being a forbidden line can only be found in the low-density regions like the NLR. Hence, the radio emission coming very close to the core, will not lead to [O~{\small III}] line emission and could be a plausible reason for the lack of correlation, when we take the core flux densities into account. 
However, the correlation of the [O~{\small III}] emission with the radio emission is suggestive of a common origin as was suggested by previous papers as well.

\begin{figure*}
\includegraphics[height=7cm,trim = 0 0 0 0 ]{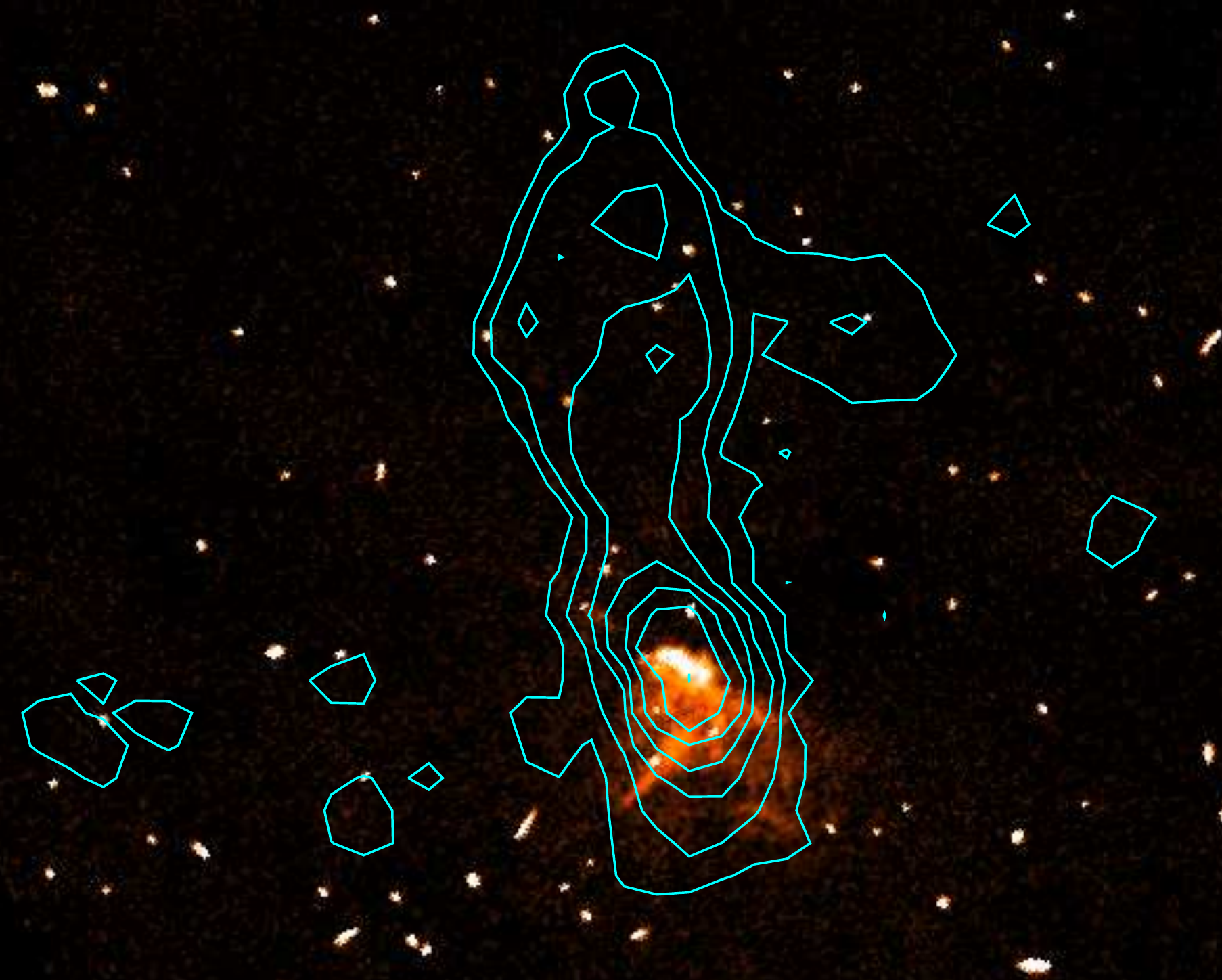}
\includegraphics[height=7cm,trim = 0 0 0 0 ]{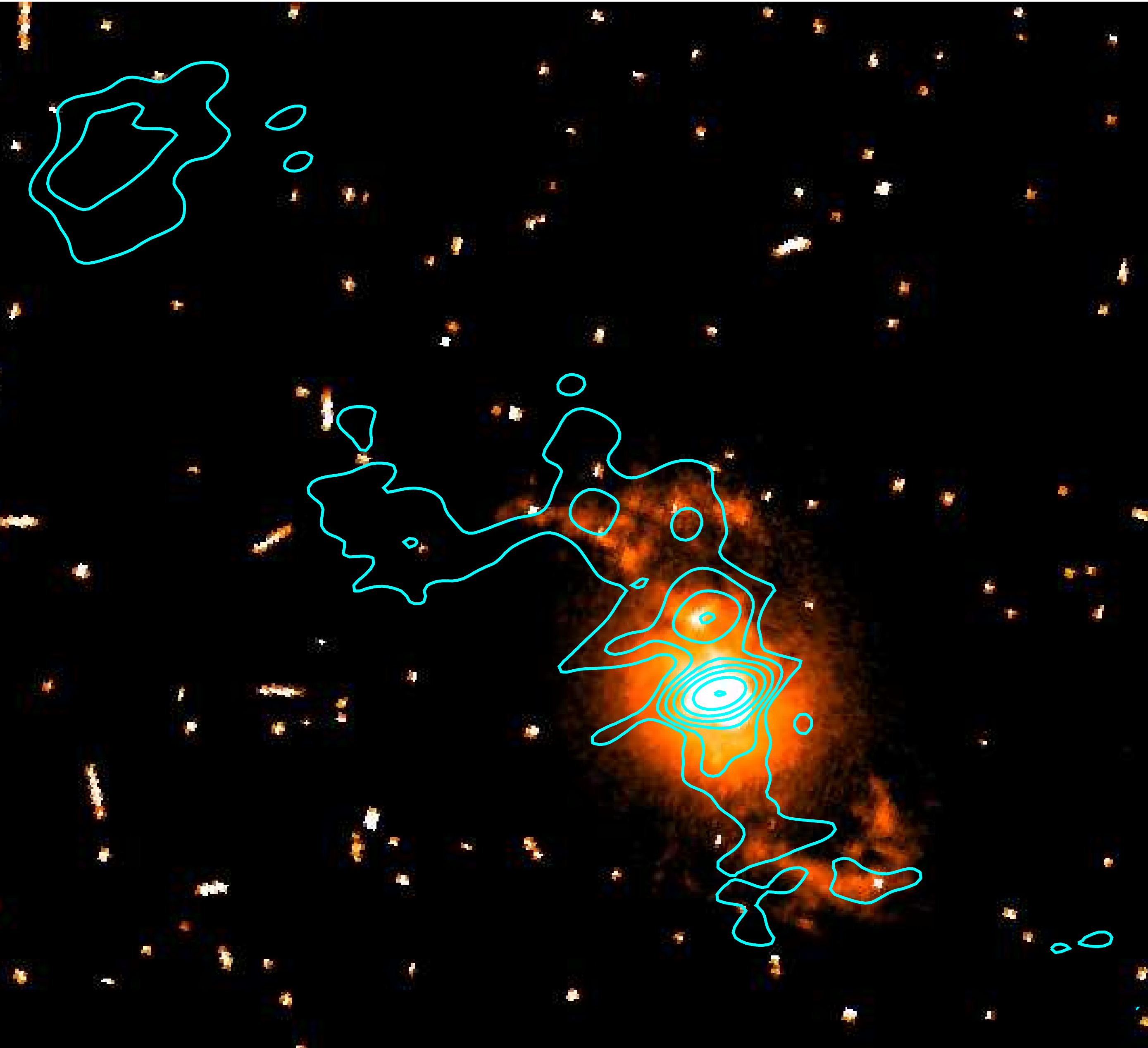}
\caption{Overlay of radio continuum intensity contours in cyan on [O~{\small III}] emission line images of NGC\,4388 (Left) NGC\,3516 (Right)  }
\label{oiii_overlay2}
\end{figure*}

To further investigate the origin of the extended radio emission itself, we created overlays of the radio total intensity emission over the HST narrow-band images of [O~{\small III}] emission for NGC\,4388 and NGC\,3516 (see Figure~\ref{oiii_overlay2}). These were the only two Seyfert galaxies in our sample which showed extended radio emission and also had HST [O~{\small III}] narrow-band images.
There appears to be a spatial overlap in the radio emission and the [O~{\small III}] emission in both these objects. The underlying [O~{\small III}] emission is rather filamentary in nature in both these systems. The [O~{\small III}] emission is only seen towards the south of the core in NGC\,4388 probably due to greater dust extinction towards the north. It is interesting to note that the [O~{\small III}] emission has a conical morphology aligned along the minor axis, which can be easily explained to be due to either wind shock-ionization or photoionization by the accretion disk emission.

On the other hand, NGC\,3516 is another example where the [O~{\small III}] emission and the radio emission are aligned almost along the major axis. Moreover, both the [O~{\small III}] and radio emission posses an S-shaped symmetry rather than a biconical shape.
Such symmetry can be introduced if a precessing jet leads to the [O~{\small III}] emission rather than a conical outflow and can not be explained using simple photoionization by the central engine.

\cite{falcke1998} studied a sample of seven Seyfert 2 galaxies using VLA radio images and HST narrow-band images. 
Majority of them show a spatial overlap between the emission line gas and the off-nuclear radio emission. More interestingly, four of their sources show evidence for an S-shaped symmetry in their emission line images. Clearly defined ionization cones are seen only in NGC\,4388 and MrK\,573 although the morphology of Mrk\,573 in their images resembles the bulbous kpc-scaled radio emission in these Seyfert galaxies.

Also, in the previous section, we have argued from the study of radio-FIR correlation in these systems that the strengths of the shocks generated at galaxy-wide scales is inadequate to generate the amount of radio emission seen in these systems via shock acceleration. 
This along with the S-shaped symmetry seen in the emission-line images for several of these galaxies strongly suggest that jet-driven shocks are a major contributor, or are driving the photo-ionized outflows at least in some of these sources, if not all.



\subsection{Relative energy input from various mechanisms}

\begin{table}
\begin{center}
\caption{Kinetic energy injection rates due to jets, AGN radiation and SFR}
\begin{tabular}{cccccccc} \hline \hline
Source & S$_{jet}$ & dE/dt$_{tot}$  & f$_{jet}$ & f$_{AGN}$ & f$_{SFR}$ \\
&(mJy) & (erg s$^{-1}$)& & & \\
\hline

NGC2639     & 44.53     & 43.36     & 0.906     & 0.001     & 0.094 \\ 
NGC2992     & 12.48     & 43.21     & 0.251     & 0.583     & 0.166 \\ 
NGC3079     & 151.8     & 43.12     & 0.704     & 0.007     & 0.289 \\ 
NGC3516     & 2.88     & 42.7     & 0.31     & 0.544     & 0.147 \\ 
NGC4051     & 2.31     & 41.79     & 0.24     & 0.242     & 0.518 \\ 
NGC4235     & 5.49     & 42.56     & 0.611     & 0.353     & 0.036 \\ 
NGC4388     & 4.47     & 42.99     & 0.206     & 0.331     & 0.463 \\ 
NGC4593     & 2.12     & 42.89     & 0.161     & 0.644     & 0.195 \\ 
NGC5506     & 125.05     & 44.09     & 0.15     & 0.836     & 0.014 \\ 

\hline
\label{tab2}
\end{tabular}
\end{center}
{\small Column 1: The target source. Column 2: The core flux density. Column 3: The total mechanical energy injection rates from the three mechanisms. Columns 4,5,6: Fractional contribution to the total injection rates from the jet, AGN accretion and star formation respectively.} 
\end{table}

\begin{figure*}
\includegraphics[height=6cm,trim = 0 0 0 0 ]{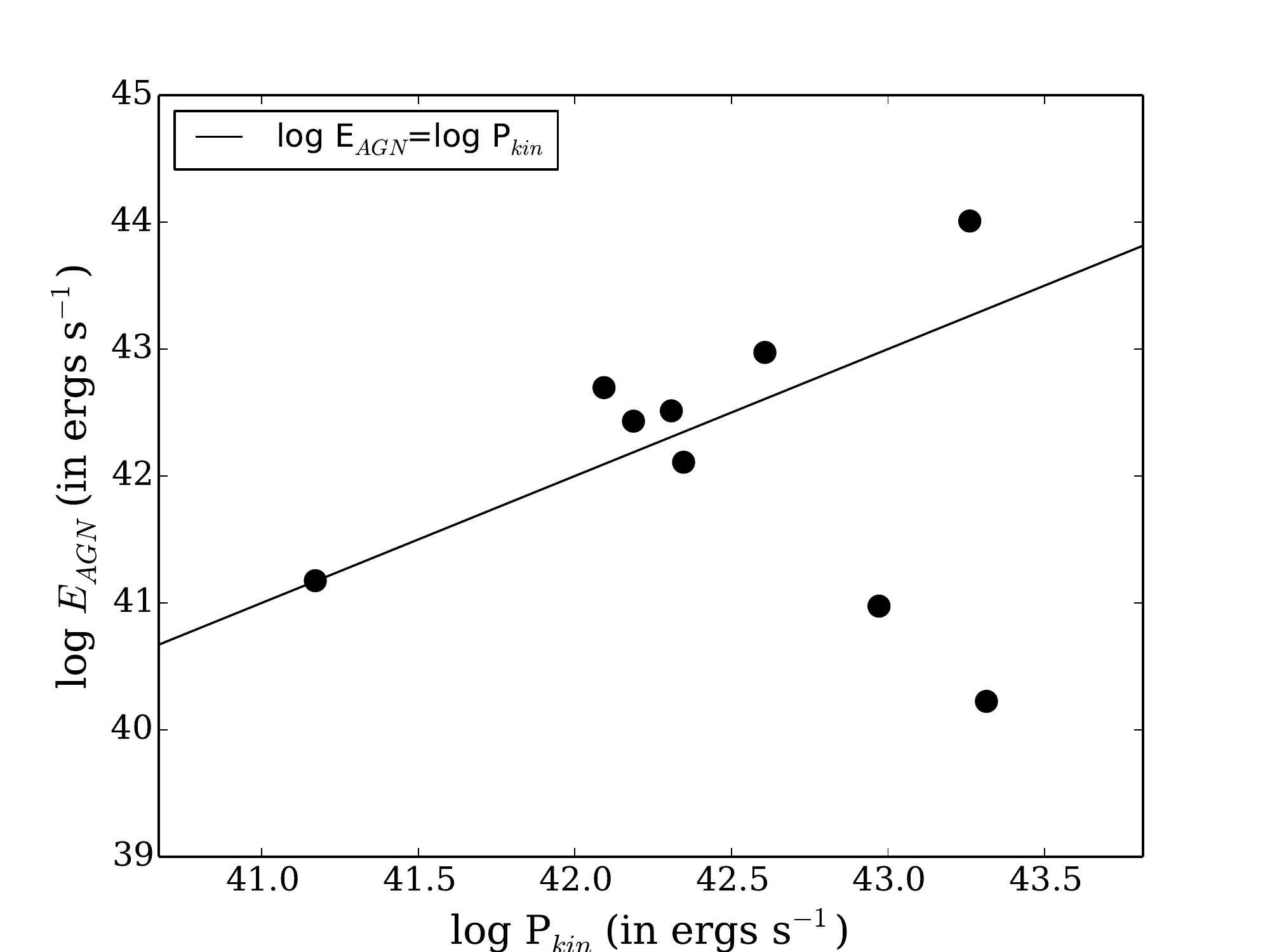}
\includegraphics[height=6cm,trim = 0 0 0 0 ]{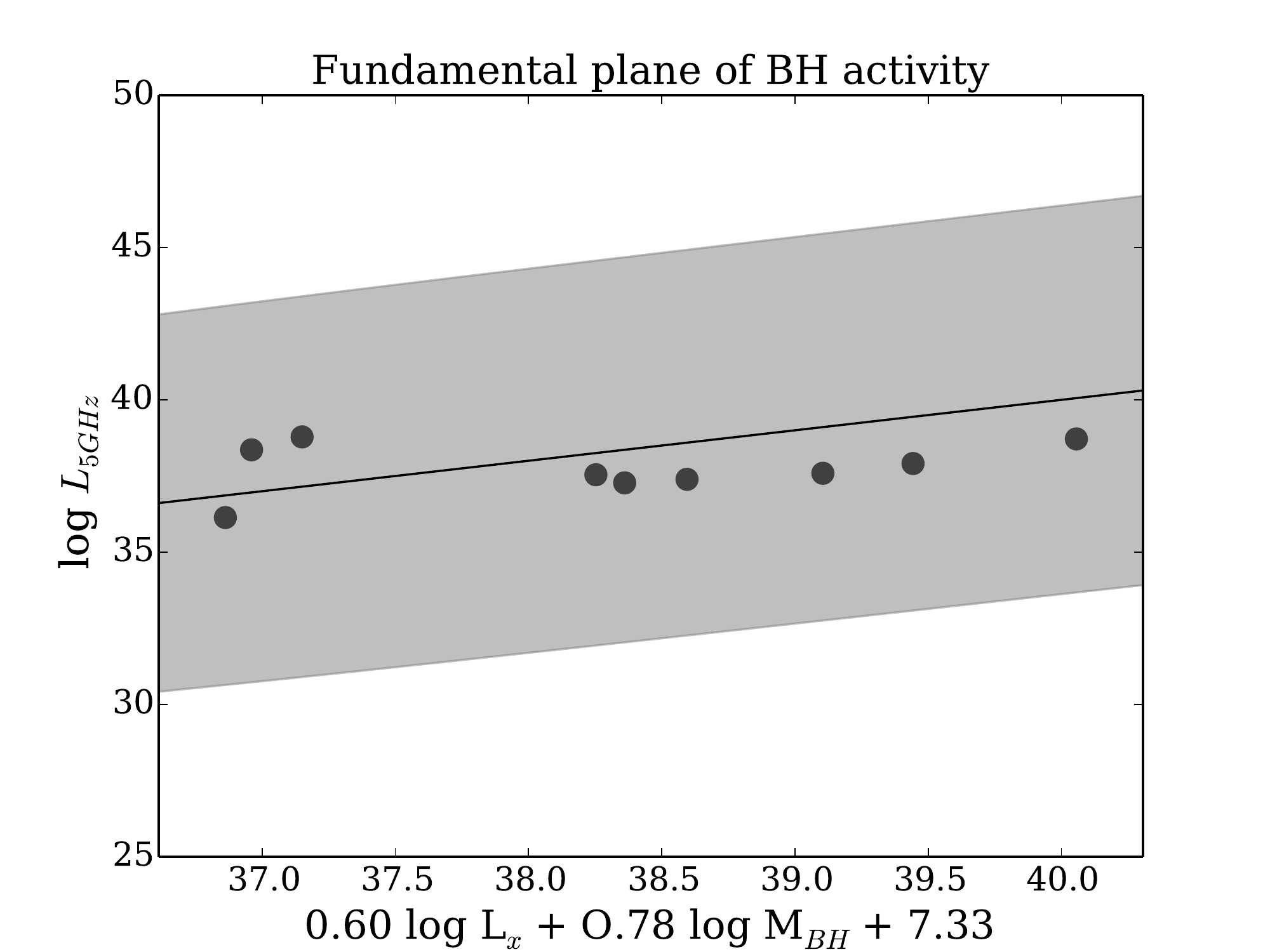}
\includegraphics[height=6cm,trim = 0 0 0 0 ]{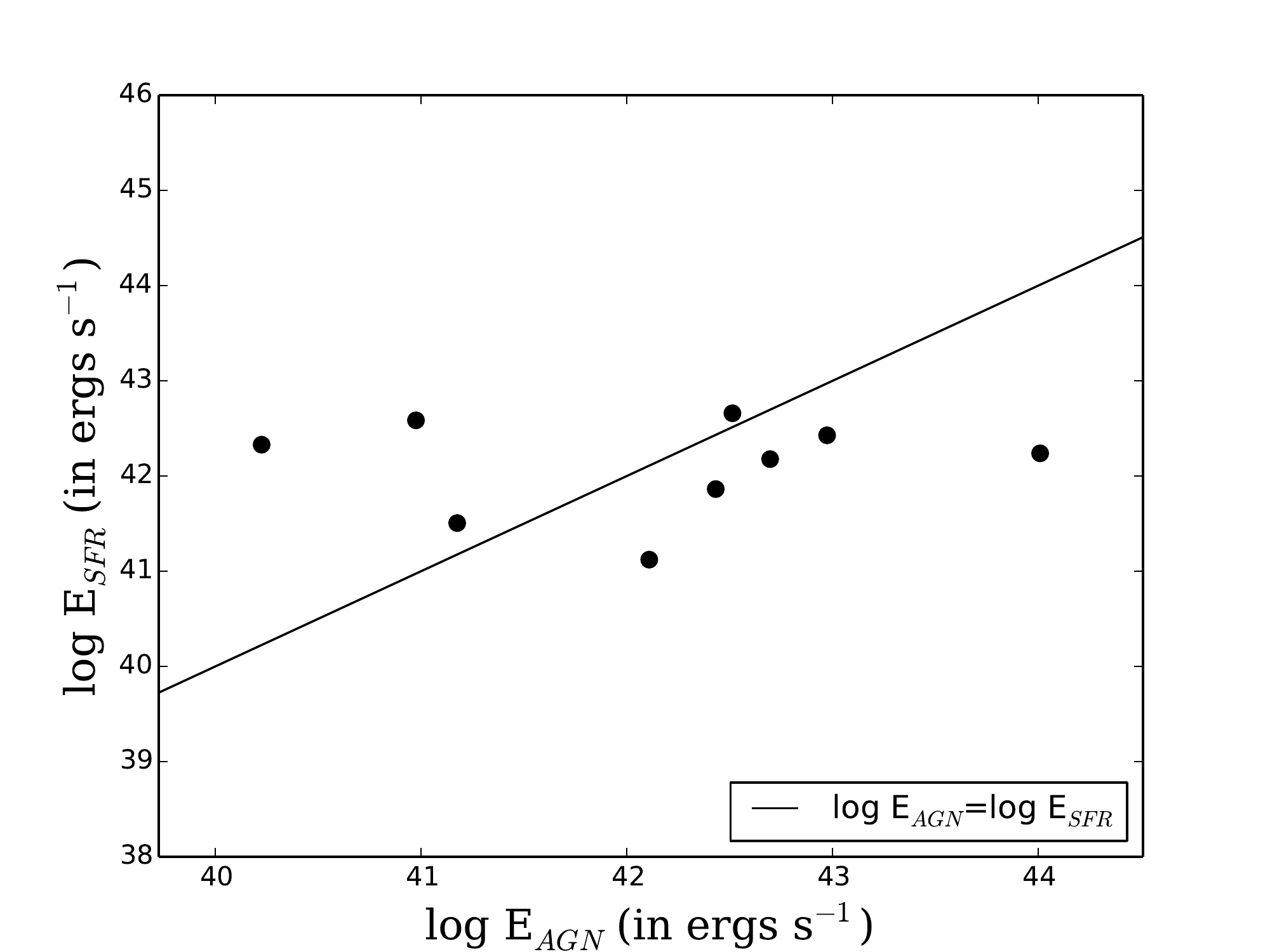}
\includegraphics[height=6cm,trim = 0 0 0 0 ]{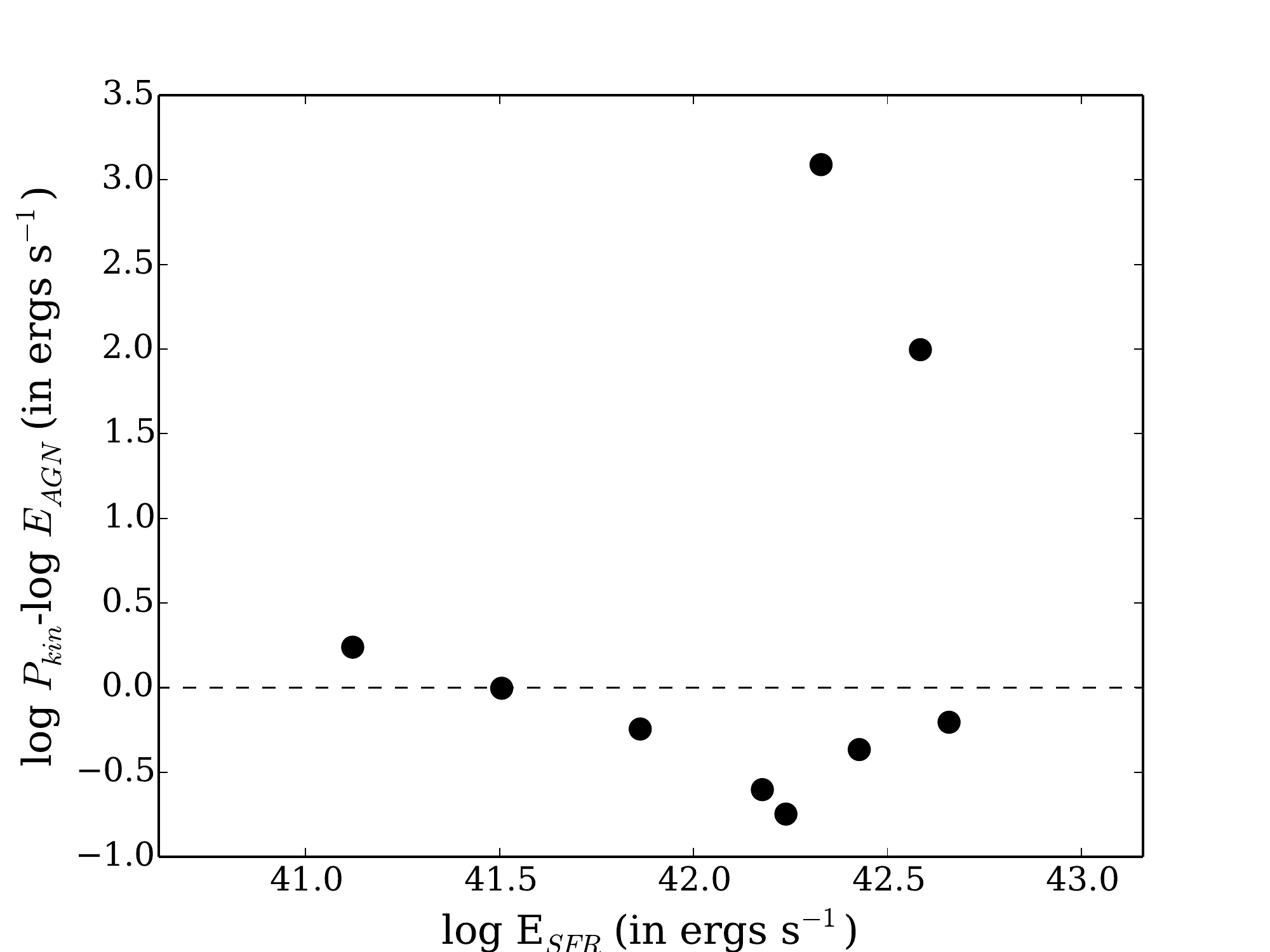}
\caption{Top left: jet power versus AGN energy input rate from bolometric luminosities for the Seyfert sample. The identity line is indicated. Top right: The Seyfert sample in the fundamental plane of BH activity, along with its well-known scatter from observations in the literature in grey. Bottom left: AGN energy input rate from bolometric luminosities versus energy input from SFR. Bottom right: Energy injected from SFR versus the difference in AGN and jet power.}
\label{energyop}
\end{figure*}
 
In this section, we compare the relative contribution of each of the sources of energy namely AGN bolometric luminosity, star formation, and the jet.  

\subsubsection{Jet power}
Several studies like \cite{birzan2004} and \cite{cavagnolo2010} derive empirical relations between the jet power and the radio luminosities by estimating the mechanical work done by the radio jets in inflating the X-ray cavities.  
These studies which use the total flux density of the radio features, mostly utilize lower frequency ($\nu \leq$ 1.4~GHz) data. It was noted by \cite{birzan2008} that the scatter for the relation derived using flux densities at lower radio frequency, namely, 327~MHz was lower compared to 1.4~GHz. Our observations were carried out using the VLA at 5~GHz in the B-array configuration, which is not the ideal setup to measure the total extended emission from lobes of the radio sources due to the complications like steep spectrum lobes and missing flux due to inadequate short baselines. 
\cite{merloni2007} derive a similar relation between 5~GHz radio luminosity and jet power, however only using the nuclear radio luminosities rather than the integrated radio emission. We rely on the following equation from \cite{merloni2007} to estimate the jet power. 
\begin{equation}
\mathrm{log (P_{kin}) = (0.81\pm 0.11)~log (L_{5GHz})+11.9^{+ 4.1}_{-4.4}}
\end{equation}


We estimated the core flux densities from our EVLA images at 5.5 GHz by fitting a Gaussian model with an underlying constant background to the core in the Seyfert galaxies. We used task `JMFIT' in AIPS to do the fitting and the peak value is listed in the Table~\ref{tab2}.

\subsubsection{Kinetic energy from AGN bolometric radiation}
 We assume that a fraction ($\sim$5\%) of the total AGN bolometric luminosity gets injected into the galaxy as kinetic energy \citep{dimatteo2005,nesvadba2017}.
 We derive the bolometric luminosities using hard X-ray (2-10 kev) luminosities taken from \cite{malkan2017}.
 \cite{duras2020} computed a universal bolometric correction, K$_X$ for a sample of both type-1 and type-2 AGN as a function of both X-ray luminosity and bolometric luminosity. They show that while K$_X$ remains nearly unchanged for sources with log(L$_{bol}$/L$_{\odot}$) < 11, it starts to increase beyond it. We used K$_X$=  a{\Large(}1+{\large(}$\frac{log(L_x/L_{\odot}\large{)}}{b})^c$ {\Large)}  
to obtain the bolometric AGN luminosities for our sources, where a=15.33, b=11.48 and c=16.20.

\subsubsection{Energy injection rate due to star formation}
Star formation accompanied by supernovae and stellar winds leads to the effective thermalization of the kinetic energy, which in turn leads to the formation of hot bubbles filled with tenuous plasma. This hot bubble expands into the lower pressure medium giving rise to a galaxy-wide outflow \citep{veilleux2005}. 
For kinetic energy injected per solar mass of stars formed equal to 1.8 $\times 10^{49}$ ergs M$_\odot ^{-1}$, 40$\%$ is carried away by the winds and the rest is assumed to be radiated away \citep{dallavecchia2008}. Hence a total of 0.72 $\times 10^{49} \times$ SFR ergs s $^{-1}$ is the net mechanical injection rate into the galaxy via star formation. Here, SFR is the star formation rate in M$_\odot$ s$^{-1}$.
To estimate the SFR, we used IRAS 60 $\mu$m and 100 $\mu$m fluxes from the IRAS catalog of galaxies and QSOs observed.
We calculated SFR as 4.5 $\times$ 10$^{-44}$ L$_\mathrm{FIR}$ \citep{kennicutt1998}, where L$_\mathrm{FIR}$=4$\pi$D$^2 \times$FIR$\times$L$_{\odot}$, where FIR = 1.26(2.58f$_{60}$+f$_{100}$)$\times$10$^{-14}$, L$_{\odot}$=3.8$\times$10$^{33}$ ergs s$^{-1}$ and the unit of f$_{60}$ and f$_{100}$ is  W m$^{-2}$ \citep{helou1988}.

In figure~\ref{energyop}, we have compared the energy injection rates from all the three mechanisms discussed above. The top-left panel shows that the energy injection rate from the AGN bolometric luminosity and the jet is comparable in this sample. Almost half of the sources in our sample are distributed above and below the identity line. It can also be noted that there is no correlation between the jet emission and the AGN bolometric luminosity. Several studies investigating the relationship between the jet and the accretion disk emission have shown the existence of the ``fundamental plane (FP) of black hole activity" \citep{Merloni03}. This is the correlation between the 5~GHz radio luminosity representing the jet and the BH mass scaled X-ray luminosity representing the emission related to the accretion on the central black hole. Given, that our sources do not show a one-to-one correlation between the jet and the accretion, we test whether our sources follow the ``fundamental plane of BH activity". We have plotted the fundamental plane for our sources in the top right panel of Figure~\ref{energyop}. The grey-shaded region represents the 1$\sigma$ error in the original relation. While all our sources follow this relation within the errors, the Spearman's rank correlation coefficient between the X and Y coordinates for our sources is 0.27 with a p-value of 0.49, suggesting no correlation. We note that our sample size is small, and therefore this study needs to be pursued with a larger sample. We also note however that the FP may intrinsically be flawed when it comes to jetted AGN \citep[see for example,][]{Zhang18}.

We also note that the trends seen in the top left panel of Figure~\ref{energyop} and the left panel of Figure~\ref{oiicorrelations} are similar to some extent. Both [O~{\small III}] and hard X-ray emission are used as proxies for the bolometric luminosities and hence the coincidence. \cite{saikia2018} studied the fundamental plane of black hole activity for a sample of LLAGN using radio luminosities derived using VLA A-array observations at 15~GHz. Interestingly, they study the correlation between both X-ray and [OIII] luminosities which are used as a proxy for the accretion rates and the core radio luminosity. They find that in both cases the radio luminosity correlates better with the BH mass scaled accretion rates rather than these quantities on their own. However, such a correlation would imply that the extended radio emission in Seyfert galaxies is correlated with the accretion rates in these systems. 
In radio jets, the jet power seems to be correlated with a combination of black hole mass and accretion rate. 

The plot of the energy injection rate from the star formation rate vs that from the bolometric luminosity of the AGN is plotted in the bottom-left panel of Figure~\ref{energyop}. Here as well, there are about the same number of sources above and below the identity line. It is also obvious from the plot that the star formation rate does not depend on the AGN bolometric luminosities (Spearman's rank correlation coefficient is 0.083 with a $p$-value of 0.83). Hence, in our sample of sources, we do not see any evidence for AGN feedback. 
The bottom-right panel shows the jet power scaled by the AGN bolometric luminosity vs the energy injected due to star-formation. We see that all three processes inject comparable energies into the medium, except for two sources where the jet power dominates the other sources.


\subsection{Episodic Jet Activity} The radio morphology of the Seyfert galaxy NGC~4388 in our EVLA data suggests at least two episodes of activity. Similarly, the polarized image of NGC~2639 suggested episodic activity as well \citep{sebastian2019b}. Multiple activity episodes are suggested for NGC~5506 based on the different lobe position angles (PAs) in our data and those of \citet{gallimore06}. Secondary lobes have been detected in polarized emission in the Seyfert galaxy NGC~2992 by \citet{irwin2017}. There also appears to be a sharp change in the jet PA of NGC~3516 in our data; the inner lobe itself shows an S-shaped morphology which is followed closely by the [O~{\small III}] gas. The GMRT images at 325 and 610 MHz of the Seyfert galaxy NGC~4235 by \cite{kharb2016} suggest the presence of relic lobe emission. Therefore overall, the majority of our sample Seyfert galaxies show indications of episodic AGN activity. 

We checked the spectral index images generated as a by-product of MS-MFS imaging in CASA for NGC\,4388, NGC\,3516, and NGC\,2992. We find that in both NGC\,4388 and NGC\,3516, the outer lobes show steeper spectral indices compared to the inner lobes. Using the sub-band data, we have created a poorer-resolution ($\sim5\arcsec$) two-point spectral index image of NGC\,4388 (see Figure~\ref{fignew4388}). While the image naturally appears noisy owing to the narrow range of frequencies used, the steepness of the outer lobe to the north, is clearly visible. Such a steep spectrum in the outer lobes is consistent with the presence of an older population of plasma in the outer lobes. On the other hand, in NGC\,2992, the outer lobe towards the north shows a flatter spectral index compared to the inner ones. The emission in the outer lobes is not very prominent and hence the result needs to be cross-checked with newer observations. We see a similar trend in the spectral index image of NGC\,2992 produced by the CHANG-ES survey \citep{irwin2012}, although the resolution of the images is relatively poor to make strong claims. However, if real, this may mean that some re-acceleration process is at play in the outer lobes.

From the lifetime estimates of Seyfert jets and the minimum statistical lifetime of Seyfert activity in a particular galaxy, \cite{sanders1984} have argued that there might be as many as 100 episodes of AGN activity in Seyfert galaxies, with each episode lasting less than $10^6$ years. These might be induced by several minor mergers leading to the formation of a new accretion disk each time, with an orientation dependent on the mass inflow direction. This would in turn result in multiple radio outflows, misaligned with each other. Our present study is consistent with these suggestions of short-lived episodic activity in Seyfert galaxies.


\section{Conclusions and Summary}
We have presented a polarization-sensitive EVLA study of nine Seyfert galaxies along with a comparison sample of seven starburst galaxies at 1.5 and 5.5~GHz. We summarise the main results below.

\begin{enumerate}
\item Bubble-like or lobe-like radio emission is observed only in the Seyfert galaxies whereas the radio emission is more spread out over the galactic disks in starburst galaxies without an AGN. 
    
\item Polarization is detected in four Seyfert galaxies (44$\%$ of the sample) and one starburst galaxy using regular imaging and two more using RM synthesis. A comparison of the polarization properties of our Seyfert galaxies in conjunction with those in the literature indicates parallel inferred magnetic fields at the lobe/bubble edges and mostly longitudinal fields inside the jets/lobes of the Seyfert galaxies, although signatures of RM gradients in some sources like NGC\,2639, NGC\,3079 and NGC\,4388, point towards more complex magnetic field structures. While there appear to be differences in the fractional polarization of Seyfert galaxies and starburst galaxies (with higher values in the higher resolution images for Seyfert galaxies), more data are needed to confirm this. We conclude that polarization can indeed be used as a tool to distinguish between the AGN jet/lobe emission from that related to star-formation.
    
\item Lack of excess radio emission in superwind hosting starburst galaxies compared to the radio-FIR correlation rules out the shock origin (from supersonic winds) of the excess radio emission seen in Seyfert galaxies. While the origin of the synchrotron emission itself may not be dominated by stellar-related processes, the comparable mechanical power output from AGN accretion, jets, and star formation, suggests that they all might be playing some role in the long term evolution of the radio emission. 
\item The correlation between the extended radio lobes and the [O~{\small III}] line emission suggests a connection between the radio emission and the emission lines. We favor the jet-NLR interaction scenario based on the S-shaped symmetry observed both in radio continuum and emission lines, in a few Seyfert galaxies. Jets in Seyfert galaxies appear to be stunted due to a close interplay between the jet and surrounding media. 
\item Several Seyfert galaxies in our sample show evidence for episodic AGN activity. This is supported by our in-band spectral index images, as well as other observations of the sample objects in the literature. The jet activity in Seyfert galaxies is intermittent, consistent with similar conclusions in the literature.
    
\end{enumerate}
\section{Data Availability}
The data underlying this article will be shared on reasonable request to the corresponding author.
\section*{Acknowledgements}
The National Radio Astronomy Observatory is a facility of the National Science Foundation operated under cooperative agreement by Associated Universities, Inc. This research has made use of the NASA/IPAC Extragalactic Database (NED) which is operated by the Jet Propulsion Laboratory, California Institute of Technology, under contract with the National Aeronautics and Space Administration.
S. Baum  and C. O'Dea are grateful to the Natural Sciences and Engineering Research Council of Canada (NSERC) for support.



\bibliographystyle{mnras}

%








\bsp    
\label{lastpage}
\end{document}